\documentclass[rmp,aps,floatfix,twocolumn,eqsecnum]{revtex4}
\usepackage{makeidx}
\usepackage{amssymb}
\usepackage{epsfig}

\begin{document}

\title{Granular Electronic Systems}
\author{I.~S.~Beloborodov}
\affiliation{Materials Science Division, Argonne National
Laboratory, Argonne, Illinois 60439, USA}
\author{K.~B.~Efetov}
\affiliation{Theoretische Physik~III, Ruhr-Universit\"{a}t Bochum,
44780 Bochum, Germany
and \\
L.~D.~Landau Institute for Theoretical Physics, 117940 Moscow,
Russia.}
\author{A.~V.~Lopatin}
\affiliation{Materials Science Division, Argonne National
Laboratory, Argonne, Illinois 60439, USA}
\author{V.~M.~Vinokur}
\affiliation{Materials Science Division, Argonne National
Laboratory, Argonne, Illinois 60439, USA}

\begin{abstract}

A granular metal is an array of metallic nano-particles imbedded
into an insulating matrix. Tuning the intergranular coupling
strength a granular system can be transformed into either a good
metal or an insulator and, in case of superconducting particles,
experience superconductor-insulator transition.  The ease of
adjusting electronic properties of granular metals makes them most
suitable for fundamental studies of disordered solids and assures
them a fundamental role for nanotechnological applications. This
Review discusses recent important theoretical advances in the
study of granular metals, emphasizing on the interplay of
disorder, quantum effects, fluctuations and effects of confinement
in formation of electronic transport and thermodynamic properties
of granular materials.

\end{abstract}

\maketitle \tableofcontents

\section{Introduction}

\label{INT}

\subsection{ Why are granular electronic systems interesting? }

Granular conductors form a new class of artificial materials with
tunable electronic properties controlled at the nanoscale and
composed of close-packed granules varying in size from a few to
hundred nanometers (often referred to as {\it nanocrystals}).  The
granules are large enough to possess a distinct electronic
structure, but sufficiently small to be mesoscopic in nature and
exhibit effects of quantized electronic levels of confined
electrons.  Granular conductors combine the unique properties of
individual- and the collective properties of coupled nanocrystals
opening a new route for potential novel electronic, optical, and
optoelectronic applications.  Applications range from light
emitting devices to photovoltaic cells and bio-sensors. The
intense interest is motivated not only by the important
technological promise but by the appeal of dealing with the
experimentally accessible model system that is governed by tunable
cooperative effects of disorder, electron correlations, and
quantum phenomena~\cite{Murray93,Gaponenko98,Mow05}.

\begin{figure}[tbp]
\includegraphics[width=8.8cm]{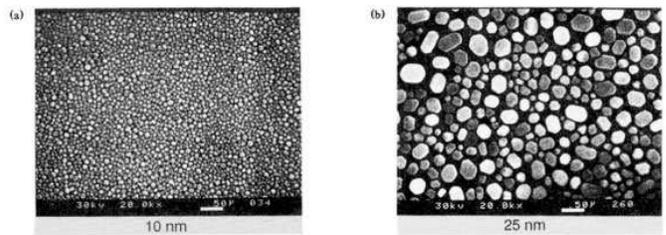}
\caption{Scanning electron microscope photographs of indium
evaporated onto $SiO_{2}$ at room temperatures~\cite{Yu91}.}
\label{gs1}
\end{figure}

Among traditional methods of preparation of such materials, the
most common are the thermal evaporation and sputtering techniques.
During those processes metallic and insulating components are
simultaneously evaporated or sputtered onto a substrate. Diffusion
of metallic component leads to the formation of small metallic
grains, usually 3-50 \textrm{nm}, see Fig.~\ref{gs1}. Variations
of grain sizes within a sample produced by these methods can be
achieved to as low as about $\sim 10\%$~\cite{Gerber97}. Depending
on the materials used for the preparation, one can obtain magnetic
systems, superconductors, insulators, etc.

Recent years have seen a remarkable progress achieved in the
design of granular conductors with the controllable structure
parameters.  Granules can be capped with organic (ligands) or
inorganic molecules which connect and regulate the coupling
between them.  Altering the size and shape of granules one can
regulate quantum confinement effects. In particular, tuning
microscopic parameters one can vary the granular materials from
being relatively good metals to pronounced insulators as a
function of the strength of electron tunnelling couplings between
conducting grains.  This makes granular conductors a perfect
exemplary system for studing metal-insulator transition and
related phenomena.

\begin{figure}[tbp]
\includegraphics[width=5.7cm]{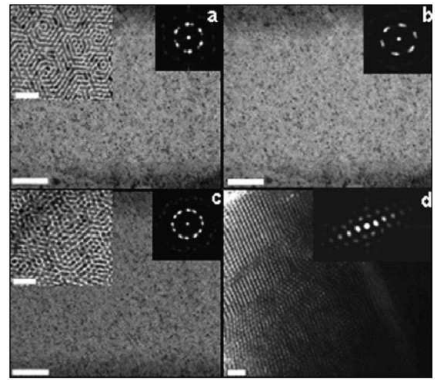}.
\caption{Transmission electron micrographs showing periodic
granular (a) bilayers, (b) trilayers, (c) tetralayers, and (d)
thick films. The insets on the left sides of panels (a) and (c)
are the zoomed-in images. The scale bars correspond to 200 $nm$
(a)-(c) and 40 $nm$ (d)~\cite{Tran05}} \label{periodic_gs}
\end{figure}

One emerging technique to create granular systems is through
self-assembling colloidal
nanocrystals~\cite{Murray2000,Collier98}. For
instance~\cite{Lin2001} describe almost perfectly periodic
two-dimensional arrays of monodisperse (with grain size variations
within $\sim 5\%$) gold nanoparticles covered by ligand molecules
that play a role of an insulating layer. Such samples are produced
via self-assembling of gold nanoparticles at the liquid-air
interface during the evaporation of a colloidal
droplet~\cite{Narayanan2004}. By changing the experimental
conditions, multilayers of nanocrystals can also be created, see
Fig.~\ref{periodic_gs}~\cite{partha01, Parthasarathy04, Tran05}.
Other important examples include Langmuir films of colloidal
Ag~\cite{Collier97,Du02} and arrays of semiconductor quantum
dots~\cite{Wehrenberg02,Yu04,Du02,Murray93}.

Of the range of techniques developed for the fabrication of
semiconductor quantum dot arrays, the most successful appeared the
self-assembled  technique of epitaxial growth of two
semiconductors having significantly different lattice constants.
For the prototype system of $InAs$ on $GaAs$, where the lattice
mismatch is 7\%, $InAs$ initially deposited on $GaAs$ grows as a
strained two-dimensional layer (referred to as the wetting
layer)~\cite{Mow05}. These materials along with $In_{2}O_{3}:Sn$
(known as indium tin oxide, ITO) are the most widely studied
systems. They exhibit a high visible transparency and a good
electrical conductance. They are used as electrodes in light
emitting diodes~\cite{kim,zhu} solar cells~\cite{gordon}, smart
windows and flat panel displays.

All these experimental achievements and technological prospects
call for a comprehensive theory able to provide quantitative
description of transport and thermodynamic properties of granular
conductors and can, therefore, serve as a ground for a clever
design of devices for emerging new generation nano-electronics.

It has been realized for quite some time that granularity can
bring new physics extending already wealthy list of remarkable
effects exhibited by disordered systems even further.  One of the
early observation was the stretched exponential temperature
behavior of the conductivity  in  the strongly disordered films
and arrays of metallic granules in the insulating regime
(see~\cite{Abeles75} for a review):
\begin{equation}
\sigma \left( T\right) \sim  \exp \left( -{\sqrt{T_0/T}}\right),
 \label{e3}
\end{equation}
with $T_0$ being a material dependent constant. This behavior
resembled the Mott-Efros-Shklovskii variable range hopping
conductivity in semiconductors~\cite{Efros,Shklovskii} and
appeared to be generic for granular arrays both metallic and
semiconducting -- either irregular~\cite{Abeles75} and strictly
periodic~\cite{Yu04,Tran05,Romero05}. Several explanations have
been advanced for this striking behavior, but real understanding
was achieved only
recently~\cite{Shklovskii04,Beloborodov05,Tran05,Feigelman05}.

A most recent incitement to intense and deep study of physics of
granular materials was given by the influential works
~\cite{Gerber97,Simon} where a logarithmic dependence of the
conductivity $\sigma \left( T\right) $
\begin{equation}
\sigma \left( T\right) = a + b \ln T  \label{e2},
\end{equation}
with $a$ and $b$ being material dependent constants, was observed
in the metallic conductivity domain. This logarithmic behavior,
has been observed in both two and three dimensional samples thus
ruling out the tempting explanation in terms of the weak
localization~\cite{Abrahams79,Khmelnitskii79} or interaction
corrections~\cite{Altshuler80,Lee_review,Belitz94} which result in
logarithmic behavior in two dimensions only.  The call for
understanding these results brought to life new
models~\cite{Efetov02,Beloborodov03,Lopatin04,Efetov02b} which
evolved into a new direction of research that will be one of the
major topics of our review.

The feature that plays a fundamental role, especially at low
temperatures, is the pronounced discreteness of the electronic
levels due to electron confinement within a single grain.  The
mesoscopic scale of the grains brings about the levels statistics
and all the wealth of the related effects~\cite{halperin,nagaev}.
The mean level spacing $\delta $ in a single grain is inversely
proportional to the volume of the grain,
\begin{equation}
\delta =\left( \nu V\right) ^{-1},  \label{e1}
\end{equation}
where $V$ is the volume of the grain and $\nu $ is the density of
states at the Fermi energy. For metal particles of size of several
nanometers, the parameter $\delta $ is typically of order of
several Kelvin. For example, for an aluminum particle with radius
$R=5$ \textrm{nm} one has $\delta \sim 1 K$. However, we
concentrate on the temperature range $T > \delta$ where the
quantum size effects are not important. In fact both types of
conductivity behavior, Eqs.~(\ref{e3}) and (\ref{e2}), were
observed at temperatures $T >\delta$.


\begin{figure}[tbp]
\includegraphics[width=5.7cm]{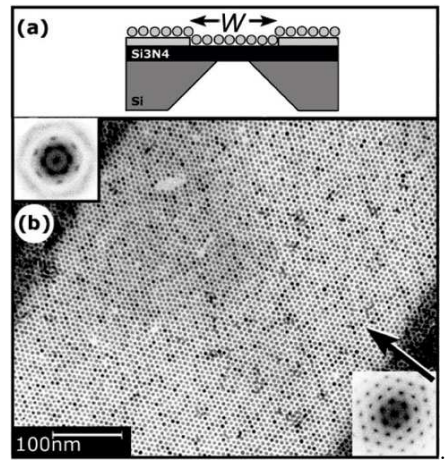}.
\caption{ a) Sketch of a nanocrystal monolayer and in-plane
electrodes, b) Highly ordered superlattice between electrodes
visible at the upper left and lower right.~\cite{partha01}}
\label{prl2001image}
\end{figure}


Superconductivity brings yet another aspect to diversity of the
interesting effects in granular
materials~\cite{Gerber97,Shapira83,Jaeger89,Hadacek04}. One of
them is a counterintuitive suppression of the conductivity due to
superconducting fluctuations.

In this review we summarize the recent theoretical progress in
understanding of the phenomena observed in granular metals and
superconductors. These are effects that do not demand ultra low
temperatures (and thus very sophisticated experimental setups) and
are thus promising from the standpoint of possible applications.
Physics of these phenomena is not particularly material specific
and our consideration will be based on the correspondingly general
models of the granular systems.

\subsection{Physical quantities characterizing granular materials}

This Review deals with the systems which we call a granular metals
and which are well modelled by an array of the identical in size
and shape but mesoscopically different metallic particles, the
intergranular electron coupling being described via the tunnelling
matrix. The grain arrangements may be either periodic or
irregular.

The effect of irregularities in the grain positions and in the
strengths of the tunneling coupling on the physical properties of
granular systems is different for metallic and insulating samples.
If the coupling between the grains is sufficiently strong and the
system is well conducting, the irregularities are not very
important. On the contrary, irregularities become crucial in the
limit of low coupling where the system is an insulator.  As a
matter of practice, recent advances in fabrication techniques now
permit very regular arrays with the fluctuations in the granule
size within the $5\%$ precision~\cite{Yu04,Tran05}. As regular
arrays promise many technological applications the further
progress in making perfect systems is expected.

At the same time, disorder related to either internal defects
inside and/or on the surface of the individual granules and to the
charged impurities in the insulating substrate/matrix is
unavoidable. Even in metallic grains with the mean free part
exceeding their size, the electrons that move ballistically within
the granules yet scatter at the grain boundaries irregularities;
the eletfron motion becomes chaotic with the resulting effect
equivalent to the action of the intragranular disorder (see,
e.g.~\cite{efetov}). This surface chaotization  would have been
absent in the atomically perfect spherical (or, say, cubical)
granules.  In such ideal granules the additional degeneracies of
the energy levels would have led to singularities in physical
quantities. However, the slightest deviations (even of the order
of $1\mathring{A}$ or of the order of the electron wavelength) of
the shape of the grains from the ideal spheres or cubes would have
lifted the accidental degeneracy of the levels (this would be
equivalent to adding an internal disorder). Thus one can justly
assume that the grains are always microscopically irregular.
Moreover, the coupling between the grains is the source of the
additional irregularities. We thus can adopt the model with the
diffusive electron motion within the each grain without any loss
of generality.  In this model the use of expression of
Eq.~(\ref{e1}) for the mean level spacing $\delta $ is fully
justified.

The key parameter that determines most of the physical properties
of the granular array is the average tunnelling conductance $G$
between the neighboring grains. It is convenient to introduce the
dimensionless conductance $g$ (corresponding to one spin
component) measured in the units of the quantum conductance
$e^{2}/\hbar $ : $g=G/(2e^{2}/\hbar)$. As we will see below, the
samples with $g\gtrsim 1$ exhibit metallic transport properties,
while those with $g\lesssim 1$ show an insulating behavior.

One of the most important energy parameters of the granular system
is the single grain Coulomb charging energy $E_{c}$. This energy
is equal to the change in the energy of the grain when adding or
removing one electron, and it plays a crucial role in the
transport properties in the insulating regime when electrons are
localized in the grains, such that the charge of each grain is
quantized. The physics of the insulating state is closely related
to the well known phenomenon of the Coulomb blockade of a single
grain connected to a metallic reservoir.

The behavior of a single grain in a contact with a reservoir has
been discussed in many articles and reviews (see
e.g.~\cite{Averin91,Averin92,Aleiner02}).  The main features of
the Coulomb blockade can be summarized as follows: (i) If the
grain is weakly coupled to the metallic contact, $g\ll 1$, the
charge on the grain is almost quantized; this is the regime of the
so-called Coulomb blockade. (ii) In the opposite limit, $g\gg 1$,
the effects of the charge quantization are negligible and the
electrons freely exchange between the granule and reservoir.

Although the systems we consider are arrays of interconnected
granules rather than a single grain coupled to a bulk metal, a
somewhat similar behavior is expected: In the regime of the strong
coupling between the grains, $g\gg 1$, electrons propagate easily
through the granular sample and the Coulomb interaction is
screened. In the opposite limit of the low coupling, $g\ll 1$, the
charge on each grain gets quantized as in the standard Coulomb
blockade behavior. In this case an electron has to overcome
electrostatic barrier of the order $E_c$ in order to hop onto the
neighboring granule. This impedes the transport at temperatures
$T$ lower than $E_{c}$.

Throughout this review we will be always assuming that the mean
level spacing $\delta $ from Eq.~(\ref{e1}) is the smallest energy
scale. In particular, in all cases we take the condition $E_{c}\gg
\delta $ be satisfied. This is the most realistic condition when
dealing with the metallic particles of a nanometer-size scale,
given that the charging energy $E_{c}$ is inversely proportional
to the radius $a$ of the grains, whereas the mean level spacing
$\delta $ is inversely proportional to the volume $V$ and also
taking into account the high density of energy states in metals.
Note that this may not be the case in arrays of semiconductor
dots, where $E_c$ and $\delta$ may appear of the same order, but
this goes beyond the scope of our review.

Another important thing to remember is that the intergranular
(tunneling)  conductance $g$ is much smaller than the intragrain
conductance $g_{0}$ by the very meaning of the notion of
``granular system":
\begin{equation}
g\ll g_{0}.  \label{granularity}
\end{equation}
The intra-grain conductance $g_{0}$ is brought by scattering on
impurities or on the boundaries of the grains and the
inequality~(\ref{granularity}) means that the grains are not very
dirty. The case $g\sim g_{0}$ can be viewed as a homogeneously
disordered system and we do dwell on this limit here. The single
grain conductance $g_{0}$ can be most easily defined as the
physical conductance of a granule of a cubic geometry measured in
the units of the quantum conductance $e^{2}/h$. In mesoscopic
physics it is customary, however, to relate the single grain
conductance $g_{0}$ to the single grain Thouless energy $E_{Th}$
as $ g_{0}=E_{Th}/\delta $, where the energy $E_{Th}$ is defined
as
\begin{equation}
E_{Th}=D_{0}/a^{2},  \label{thouless}
\end{equation}
and $D_{0}=v_{F}^{2}\tau /d$ is the classical diffusion
coefficient. Other parameters are $v_{F}$, the Fermi velocity,
$\tau $, the elastic scattering time within the grains, and $d$,
the dimensionality of the grain. The length $a$ in
Eq.~(\ref{thouless}) is the linear size (radius) of the grain. If
the grains are not very dirty, such that the electrons move inside
the grains ballistically, the mean free path $l=v_{F}\tau $ should
be replaced by the size of the grains $2a$. The Thouless energy
$E_{Th}$, Eq.~(\ref{thouless}), is proportional to the inverse
time that it takes for an electron to traverse the grain. The
energy $E_{Th}$, Eq.~(\ref{thouless}), exceeds the mean level
spacing $\delta $, Eq.~(\ref{e1}), and therefore the intragrain
conductance $g_{0}$ is always large, $g_{0}\gg 1$, whereas the
inter-granular conductance $g$ can be either larger or smaller
than unity.

The above parameters  make a full set of variables describing
properties of the normal granular metals. If the constituent
particles are made out of the superconductor material, a wealth of
new interesting phenomena arises.  The behavior of such a system
can be quantified by adding one more energy parameter, the
superconducting gap $\Delta$ of the material of a single granule.
Now, if the inter-grain coupling is sufficiently strong, the
system can turn a superconductor at sufficiently low temperatures.
Properties of such a superconductor are not very different from
those of bulk superconductors.

On the contrary, in the opposite limit of weak coupling between
the granules, an array of superconducting grains can transform
into an insulator at $T=0$. In this regime the Cooper pairs are
locally formed in each grain but remain localized inside the
grains due to the strong on-site Coulomb repulsion leading to the
Coulomb blockade.   Because of localization, the number of the
Cooper pairs in each grain is fixed and, according to the
uncertainty principle, this leads to strong phase fluctuations.
Thus the system does not develop the global coherence, and the
global macroscopic superconductivity is suppressed. One can
describe this effect using the model of superconductor grains
coupled via the Josephson junctions. Such a system is
characterized by three energy parameters: the superconductor gap,
$\Delta $, characterizing a single grain, the Josephson coupling
$J$, and the grain charging Coulomb energy $E_{c}$. Strong
Josephson coupling, $J\gg E_{c}$, suppresses the phase
fluctuations leading to the globally coherent superconducting
state at sufficiently low temperatures. If $J\ll E_{c}$, the
Coulomb blockade prevails, the Cooper pairs get localized at $T\to
0$, and the system falls into an insulating state.

Note that even in the insulator state the superconducting gap
$\Delta $ still exists in each grain and its value is close to the
gap magnitude in the bulk provided $\Delta \gg \delta $. If the
latter inequality is not fulfilled the superconducting gap in the
grains is suppressed or can even be fully
destroyed~\cite{Anderson59,Anderson64}. There are interesting
effects in this regime but we do not consider them here because
the relevant region of the parameters corresponds to either too
small grains or too weak superconductors.

At low temperatures, the Josephson coupling $J$ is expressed via
the tunnelling conductance $g$ as $J=\pi g\Delta /2 $
\cite{ambegaokar63} and, at first glance, one could have concluded
that the transition between the insulating and superconducting
states should have happened at $g \sim E_{c}/\Delta $. However,
this simple estimate holds only in the weak coupling regime,
$g<1,$ that assumes $E_c < \Delta$. At stronger coupling, the
Coulomb energy $E_{c}$ is renormalized down to the value
$E_{c}\rightarrow \tilde{E} \sim \Delta
/g$,~\cite{Larkin83,Chakravarty87} due to electron tunnelling
between the neighboring grains. Therefore, in the limit of the
strong coupling, $g \gg 1$, the effective Coulomb energy
$\tilde{E}$ is always smaller than the Josephson coupling,
$\tilde{E}<J$, implying the superconducting ground state. Shown in
Fig.~\ref{IS phase} is the schematic phase diagram summarizing the
above consideration~\cite{Chakravarty87}.
\begin{figure}[t]
\resizebox{.21\textwidth}{!}{\includegraphics{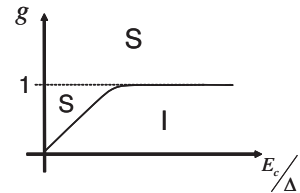}}
\vspace{0.6cm} \caption{ Schematic phase diagram for granular
superconductors at temperatures $T=0$. Symbols $S$ and $I$ stand
for superconducting and insulating phases respectively.} \label{IS
phase}
\end{figure}
For many superconducting granular samples available experimentally
the ratio $E_{c}/\Delta $ is large. In this case, as one can see
from phase diagram in Fig.~\ref{IS phase}, the transition between
the superconducting and insulating at $T \to 0$ occurs at $g \sim
1$ \cite{Orr86,Chakravarty87}.

To conclude our brief introduction to granular superconductors we
note that for the most of experimentally available samples, the
grains are much smaller than the bulk superconducting coherence
length of the granule material
\begin{equation}
a\ll \xi _{0}.  \label{k2}
\end{equation}
This allows one to neglect variations of the superconducting order
parameter $\Delta $ inside the grains and treat a single grain as
a zero dimensional object. As the condition~(\ref{k2}) is at this
point the most common experimental situation, we will further
concentrate mainly on this regime.

As we now see, depending on the parameters of the system, many
different physical situations appear and the phase diagram of a
granular material is quite reach. In the next two sections,
Sec.~\ref{normal} and Sec.~\ref{superconducting}, we will focus on
the systems consisting of the normal and superconducting grains
respectively.

\section{Normal granule arrays}

\label{normal}

\subsection{Transport properties}

\label{preliminary}

We start the discussion of the properties of the granular systems
consisting of the normal grains by presenting the main results and
their qualitative explanations. This may help the reader to
understand the basic physics of the system and to learn important
formulae suitable for a direct comparison with experiments without
going into the technical detail of theoretical models, which will
be presented at the subsequent parts of each section.

\subsubsection{Classical conductivity}

The key parameter that determines the transport properties of the
granular materials is the tunneling conductance $g$. In the strong
coupling regime, $g\gg 1$, a granular array has metallic
properties, while in the opposite case, $g\ll 1$, the array is an
insulator. The insulating state appears as a result of the strong
Coulomb correlations that block electron transport at low
temperatures. Generally speaking, apart from the Coulomb
interaction effects one has also to consider the effects of
quantum interference that may play an important role in the low
conducting samples.  In homogenously disordered systems the
interference effects play an important role leading to the
localization of electron states in the absence of
interaction~\cite{Anderson58}.

In the metallic regime and at high enough temperatures both
Coulomb correlation and interference effects are weak. In this
case the global sample conductivity  $\sigma_{0}$  is given by the
classical Drude formula, in particular, for a periodic cubic
granular array
\begin{equation}
\sigma _{0}=2e^{2}ga^{2-d},  \label{conductivity0}
\end{equation}
where $a$ is the size of the grains and $d$ is the dimensionality
of the sample. Formula~(\ref{conductivity0}) has a straightforward
meaning: in order to obtain the physical conductance of the
contact, one should multiply its dimensionless conductance, $g$,
by $2e^{2}$ (the factor $2$ is due to spin and $\hbar=1$) and,
then, multiplying the result by $a^{2-d}$ one arrives at the
conductivity per unit volume.

Although Eq.~(\ref{conductivity0}) is written for the periodic
array, the electron system is not translationally invariant,
otherwise the sample conductivity would be infinite.
Equation~(\ref{conductivity0}) is valid for grains with an
internal disorder, such that the electron motion inside the grains
is chaotic. When hopping from grain to grain, the electron
momentum is not conserved and this leads to the finite
conductivity $\sigma _{0}$, Eq.~(\ref{conductivity0}). The
conductance of the contact $g$ depends on the microscopic
properties of the contact, and we will consider it in most cases
as a phenomenological dimensionless parameter controlling behavior
of the system.  Note that the model of a periodic array assumes no
variation in tunneling conductances.

Upon decreasing temperature the Coulomb interactions become
relevant and formula Eq.~(\ref{conductivity0}) does not hold any
more.  Below we will discuss Coulomb effects in more detail and
find that their manifestation in granular systems may differ
noticeably from that in \textquotedblleft homogeneously
disordered\textquotedblright\ metals.

\subsubsection{Metallic regime}

In the metallic regime electrons tunnel easily from granule to
granule.  The time $\tau _{0}$ that the electron spends inside a
grain plays an important role in the metallic regime: the
corresponding characteristic energy $\Gamma =\tau _{0}^{-1}$ is
related to the tunneling conductance and the mean energy level
spacing as
\begin{equation}
\Gamma =g\delta .  \label{Gamma}
\end{equation}%
The energy $\Gamma $ can also be interpreted as the width of the
smearing of the energy levels in the grains. In the limit of large
conductances, $g\gg 1$, this width exceeds the energy spacing
$\delta $ and the discreteness of the levels within a single grain
ceases to be relevant.

Since the electron motion on the scales well exceeding the granule
size is always diffusive (even in the case of the ballistic
electron motion inside each grain), the electron motion on the
time scales larger than $\Gamma ^{-1}$ can be described by the
effective diffusion coefficient $D_{eff}$ related to $\Gamma $ as
\begin{equation}
D_{eff}=\Gamma a^{2}.  \label{Deffective}
\end{equation}
Then the Einstein relation gives the conductivity of the granular
sample as
\begin{equation}
\sigma _{0}=2\,e^{2} \nu  D_{eff}.  \label{e10}
\end{equation}
For periodic arrays Eq.~(\ref{e10}) is equivalent to
Eq.~(\ref{conductivity0}), which can be seen from Eqs.~(\ref{e1},
\ref{Gamma}, \ref{Deffective}). At the same time Eq.~(\ref{e10})
is more general than Eq.~(\ref{conductivity0}) since with the
properly defined diffusion constant $D_{eff}$ it applies to arrays
with the arbitrary grain arrangement as well.

The energy scale $\Gamma $ plays a very important role; many
physical quantities have qualitatively different behavior
depending on whether they are dominated by the energies higher or
lower than $\Gamma $.

From the experience with the homogeneously disordered metals one
can envision two major causes that may alter the classical
conductivity $\sigma _{0} $ in Eq.~(\ref{conductivity0}): (i)
electron-electron interactions~\cite{Altshuler85,Lee_review} and
(ii) quantum interference effects~\cite{Khmelnitskii79,
Abrahams79}. Accordingly, constructing the theory of granular
conductors with the reference of highly advanced theory of
disordered metals, one can expect two corresponding distinct
corrections to $\sigma_0$.  Since in the metallic domain electrons
effectively screen out the on-site Coulomb interactions the bare
magnitudes of which are high because of small sizes of the grains,
the notion of the interaction corrections to conductivity is well
justified.

To gain the qualitative understanding of the interaction effects,
we introduce the characteristic interaction temporal-  and the
corresponding spatial scales $\tau_{T}\sim \hbar/T$ and $L_{T}\sim
\sqrt{D_{eff}/T}$ associated with this time. One expects that the
behavior of the interaction correction is different on the
distances exceeding the granule size, $L_{T}>a$, and within the
granule, $ L_{T}<a$. Using Eq.~(\ref{Deffective}) for the
effective diffusion coefficient $D_{eff}$, one immediately sees
that these conditions imply the existence of two distinct
temperature regions $T>\Gamma$ and $T<\Gamma$ with respect to
interaction contributions. Accordingly, the correction to the
conductivity due to Coulomb interaction can be written as a sum of
contributions coming from the large, $\varepsilon >\Gamma ,$ and
the low, $\varepsilon <\Gamma $, energies. This ``contribution
separation" naturally follows, as we will see below, from the
diagrammatic approach where the two contributions in question are
represented by the two distinct sets of the diagrams. Deferring
the details for later, the result is as follows: denoting the
corrections coming from the high and low with respect to $\Gamma $
energies as $\delta \sigma _{1}$ and $\delta \sigma _{2}$,
respectively, we write:
\begin{equation}
\sigma =\sigma _{0}+\delta \sigma _{1}+\delta \sigma _{2}.
\label{mainresult1}
\end{equation}
with
\begin{equation}
\frac{\delta \sigma _{1}}{\sigma _{0}}=-{\frac{{1}}{{\ 2\pi
dg}}}\,\ln \left[ {\frac{{gE_{c}}}{\max {(T,\Gamma )}}}\right]
\label{mainresult3}
\end{equation}
~\cite{Efetov02,Efetov02b} and
\begin{equation}
\frac{\delta \sigma _{2}}{\sigma _{0}}=\left\{
\begin{array}{lr}
{\frac{{\alpha }}{{12\pi ^{2}g}}}\sqrt{{\frac{{T}}{\Gamma
}}}\hspace{1.6cm}
d=3, &  \\
-\frac{1}{4\pi ^{2}g}\ln \frac{\Gamma }{T}\hspace{1.4cm}d=2, &  \\
-{\frac{{\beta }}{{4\pi g}}}\sqrt{\frac{\Gamma
}{T}}\hspace{1.6cm}d=1, &
\end{array}
\right.  \label{mainresult44}
\end{equation}
where $\alpha \approx 1.83$ and $\beta \approx 3.13$ are numerical
constants~\cite{Beloborodov03}. The high energy contribution
$\delta \sigma _{1}$ in Eq.~(\ref{mainresult1}) contains the
dimensionality of the array $d$ merely as a coefficient and is, in
this sense, universal. On the contrary, the low energy
contribution $\delta \sigma _{2}$ in Eq.~(\ref{mainresult1})  has
a different functional form for different array dimensionalities
(note that for the 3d correction we have kept the
temperature-dependent part only).

\begin{figure}[tbp]
\includegraphics[width=8.6cm]{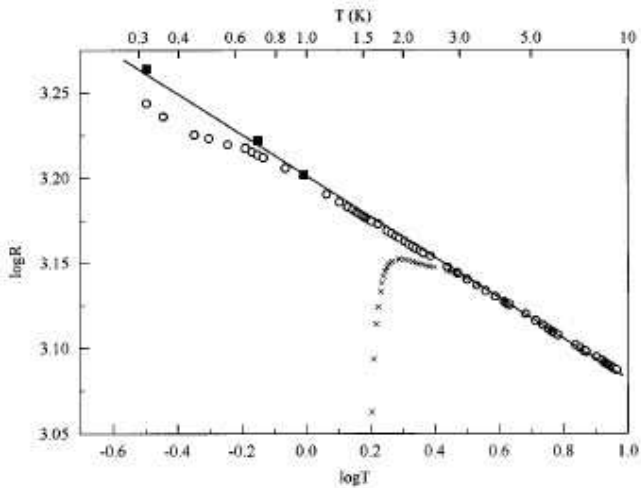}
\caption{Resistance of $3d$ granular $Al-Ge$ sample as a function
of temperature on a log-log scale, as measured at (zero)
($\times$) and 100 $ kOe $ field (open circles). Sample room
temperature resistance is 500 $\Omega $~\protect\cite{Gerber97}.}
\label{gerber97_picture}
\end{figure}

At high, $T>\Gamma$, temperatures the correction $\delta \sigma
_{1}$, Eq.~(\ref{mainresult1}), grows logarithmically with
decreasing temperature.  Upon further lowering the temperature
this correction saturates at $T\approx\Gamma$ and remains constant
at $T<\Gamma $.  Then the correction $\delta \sigma _{2}$ in
Eq.~(\ref{mainresult44}) from the low energy scales, $\varepsilon
\leq \Gamma $, where the coherent electron motion on the scales
larger than the grain size $a$ dominates the physics, and  which
is similar to that derived for homogeneous disordered
metals~\cite{Altshuler85} comes into play. In the low temperature
regime it is this term that entirely determines the temperature
dependence of the conductivity.  At the same time, the
contribution $\delta \sigma _{1}$, although being temperature
independent, still exists in this regime and, as a matter of fact,
can even be larger in magnitude than $\delta \sigma _{2}$.
Equation~(\ref{mainresult3}) can be written for $T>\Gamma $ in the
form Eq.~(\ref{e2}) that has been observed in a number of
experiments~\cite{Simon,Gerber97,fujimori,rotkina}, (see, e.g.
Fig.~\ref{gerber97_picture}).

Rewriting the low energy contribution $\delta \sigma _{2}$, in
terms of the effective diffusion coefficient $D_{eff}$ from
Eq.~(\ref{Deffective}) one reproduces the Altshuler-Aronov
corrections~\cite{Altshuler85}. This reflects a universal
character of a large scale behavior of a disordered system and we
will show it rigorously in Sec.~\ref{very.low}, that indeed the
granular metal model can be reduced to an effective disordered
medium on distances much larger than the single grain size (i.e.
coming from the low energy, $T<\Gamma$ excitations). The
contribution $\delta \sigma _{1}$, which is dominated by the
energies $\varepsilon
>\Gamma $, is the consequence of and specific to the granularity
-- and does not exist in the homogeneously disordered metals.

Now we turn to quantum interference (weak localization) effects
that exist also in the systems without any electron-electron
interaction. In the metallic regime where the perturbation theory
with respect to inverse tunneling conductance, $1/g$, the
interaction and weak localization corrections can be considered
independently in the leading order.

The weak localization correction is of a purely quantum origin: it
stems from the quantum interference of electrons moving along the
self-intersecting trajectories and is proportional to the return
probability of an electron diffusing in a disordered medium. In
one- or two-dimensional conductors the probability of infinite
times returns is unity and returning trajectories can be
infinitely long.  Thus a fully coherent electron propagation would
lead to a divergent weak localization correction. Finite phase
relaxation (dephasing) time disables long trajectories and limits
the correction. Adapting here the concept of the effective
disordered medium we can assume that the results
of~\cite{Altshuler85} for homogeneously disordered metals apply to
granular metals with the proper renormalization of the diffusion
constant, we introduce the effective dephasing length $L_{\phi
}=\sqrt{D_{eff}\tau _{\phi }}\sim a\sqrt{\Gamma\tau_{\phi}}$,
where $\tau _{\phi }$ is the dephasing time, which dtermines the
scale for the interference effects. At low temperatures the
dephasing time $\tau _{\phi }$ is large and the decoherence length
$L_{\phi }$ can exceed the size $a$ of a single grain, $L_{\phi
}>a$. In this regime the relevant electron trajectories pass
through many granules and quantum interference effects are similar
to those in homogeneously disordered metals and contribute
essentially to the conductivity.  With the increasing temperature
the decoherence length $L_{\phi }$ decreases and as soon as it
drops to the grain size, $L_{\phi }\sim a$, the trajectories
contributing to weak localizations traverse to only one neighbor
and the weak localization correction is suppressed. Using
Eq.~(\ref{Deffective}) we write the last condition separating the
domains of ``relevance" and ``irrelevance" of weak localization
effects as $\tau _{\phi }\sim \Gamma ^{-1}$. For $\tau _{\phi
}>\Gamma ^{-1}$ (or $L_{\phi }>a$) the quantum interference
effects are important, while for $\tau _{\phi }<\Gamma ^{-1}$ (or
$L_{\phi }<a$) the corresponding correction drops rapidly with
temperature.

The final result for the quantum interference corrections
reads~\cite{BLV_2004,Varlamov05}
\begin{equation}
\frac{\delta \sigma _{\scriptscriptstyle WL}}{\sigma
_{0}}=-\frac{1}{4\pi ^{2}g}\ln \left( \tau _{\phi }\Gamma \right)
,  \label{wl1}
\end{equation}
for granular films, and
\begin{equation}
\frac{\delta \sigma _{\scriptscriptstyle WL}}{\sigma
_{0}}=-\frac{1}{2\pi g} \left( \tau _{\phi }\Gamma \right) ^{1/2},
\label{wl2}
\end{equation}
for granular wires. In both equations $\tau _{\phi }$ is the
dephasing time. Within this time the wave function retains its
coherence. The most common mechanism of dephasing, the
electron-electron interactions gives for the dephasing
time~\cite{Altshuler82}:
\begin{equation}
\tau _{\phi }^{-1}=\left\{
\begin{array}{lr}
\frac{T}{g}\hspace{2.6cm}d=2, &  \\
\left( \frac{T^{2}\delta }{g}\right) ^{1/3}\hspace{1.4cm}d=1. &
\end{array}
\right.  \label{tauphi}
\end{equation}
In this case the condition $\Gamma\approx\tau_{\phi}^{-1}$ defines
(in 2d case) yet another characteristic energy scale,
$T^*=g^2\delta$, that marks the interval of relevance of weak
localization effects.

The quantum interference correction $\delta \sigma
_{\scriptscriptstyle WL}$ are suppressed by applying even the
relatively weak magnetic field; the dependence upon the magnetic
field can serve as a test for identifying the weak localization
effects.  At sufficient fields thus the main temperature
dependence of the conductivity will come from the
electron-electron interaction effects, Eqs.~(\ref{mainresult3})
and (\ref{mainresult44}).

Both, the electron-electron interactions and quantum interference
effects decrease conductivity of a granular system at low
temperatures similarly to the same effect in homogeneously
disordered systems. The novel important feature is that the
granularity restrains screening thus enhancing the role of Coulomb
interaction: this is reflected by the contribution $\delta \sigma
_{1}$ to conductivity which is specific to granular conductors but
absent in the homogeneously disordered metals.  As a result, in
$3d$ [and, to some extent, in $2d$, see below], granular systems
the Coulomb interactions can become a main driving force of a
metal-insulator transition.

This question worth more detailed discussion.  The ``granular"
contribution, $\delta \sigma _{1}$, comes from the short distances
and is actually the renormalization of the tunneling conductance
between the grains:
\begin{equation}
g\rightarrow \tilde{g}=g-{\frac{{1}}{{\ 2\pi d}}}\,\ln \left[
{\frac{{gE_{c}}}{\max {(T,\Gamma )}}}\right].  \label{e55}
\end{equation}
Using renormalization group methods one can show that
Eq.~(\ref{e55}) represnts in fact the solution of the
renormalization group equation for the effective conductance
$\tilde{g}$ rather than a merely  perturbative correction. As
such, it holds not only till the second term in the r.~h.~s. of
Eq.~(\ref{e55}) is much smaller than the first one but in a
broader temperature region - as long as the renormalized
conductance is large, $\tilde{g}>1.$  It is important that the
logarithm in Eq.~(\ref{e55}) saturates at temperatures of the
order of $\Gamma $.  Then one sees from Eq.~(\ref{e55}) that the
renormalized conductance $\tilde{g}$ may remain large in the limit
$T\rightarrow 0$ only provided the original (bare) conductance $g$
is larger than its critical value
\begin{equation}
g_{c}=(1/2\pi d)\ln (E_{c}/\delta ).  \label{gC}
\end{equation}%
If $g<g_{c}$ the effective conductance $\tilde{g}$ renormalizes
down to zero at finite temperature. Of course, as soon as
$\tilde{g}$ becomes of the order of unity, Eq.~(\ref{e55}) is no
longer valid, but it is generally accepted -- in the spirit of the
renormalization group approach -- that the conductance flow to low
values signals the electron localization (a recent exact solution
for a model equivalent to a single grain connected to a metallic
contact lends confidence to this conclusion~\cite{zamolodchikov}).
The result~(\ref{gC}) can be also obtained by the analysis of the
stability of the insulating state that we discuss below.

The physical meaning of the critical conductance $g_{c}$ is most
transparent for a $3d $ system: in this case, there are no
infrared divergencies and both the localization and
Altshuler-Aronov corrections are not important even in the limit
$T\rightarrow 0$. This implies that so long as $g>g_{c}$ the
granular conductor remains metallic.

At $g<g_{c}$ the conductance renormalizes to low values at
temperatures exceeding $\Gamma$ signaling thus the development of
the Coulomb blockade. Upon the further temperature decrease the
system resistance begins to grow exponentially at certain
characteristic temperature $T_{ch}$ indicating the onset of the
insulating behavior. The temperature $T_{ch}$ tends to zero in the
limit $g\rightarrow g_{c}$.  Thus, in $3d$ the value $g_{c}$ marks
the boundary between the insulating and metallic states at
$T\rightarrow 0$.

The interpretation of the critical value $g_{c}$ is less
straightforward in the case of granular films, since at $g>g_{c}$
the low temperature conductivity corrections due to interaction
and localization effects diverge logarithmically. This means
apparently, that the system turns an insulator without a sharp
transition. Yet the notion of $g_{c}$ still makes sense as a mark
distinguishing between the systems that are strong Coulomb
insulators at low temperatures ($g<g_{c}$) and those that are weak
insulators ($g>g_{c}$).

All the above results have been obtained within the model of a
periodic granular array (cubic lattice) neglecting the dispersion
of the tunneling conductances. In reality, the typical granular
samples are disordered. It is then important to understand what
effect the irregularities in the granule's arrangement may have.
It is plausible and intuitive that the universal regime of low
energies is hardly affected by the irregularities since all the
physical characteristics in this regime can be expressed through
the effective diffusion coefficient of the medium. At the same
time, the physical results that are controlled by the local
physics, in particular, the critical value $g_{c}$ are sensitive
to the grain arrangements. In some cases the irregularities can be
incorporated at almost no expense. For example, breaking some
finite fractions of the junctions between the granules merely
changes the average coordination number $z$, which appears as the
$1/2d$ factor in the expressions~(\ref{e55}, \ref{gC}) for the
critical conductance. This means that the dispersion in tunneling
conductances is not expected to change noticeably, for example,
the position of the metal-insulator transition (which may acquire
the percolation character) and can be accounted for via replacing
$2d$ in Eqs.~(\ref{e55}, \ref{gC})  by the effective coordination
number $z_{eff}<z$.

In general the proper treatment of disorder in the grain
arrangement may appear more tedious. Yet we do not expect that it
can change the physics of the metallic state qualitatively.
Irregularities of the grain displacements and of other quantities
characterizing the system play much more important role in the
insulating regime which we briefly review in the following
subsection.

\subsubsection{Insulating regime}

We begin with a consideration of a periodic granular array which,
in the regime of a weak coupling between the grains, is an
exemplary Mott insulator at low temperatures. The electron
transport is mediated by the electron hopping from grain to grain.
However, leaving a neutral grain and entering its neighbor costs a
considerable electrostatic energy, and the electron transport is
blocked at low temperatures by the Coulomb gap in the electron
excitation spectrum $\Delta _{M}$. At very small tunneling
conductances this gap is simply the Coulomb charging energy of the
grain, $\Delta _{M}=E_{c}$.  Virtual electron tunneling to
neighboring grains leads to a reduction of the Mott gap $\Delta
_{M}$ and, in the limit of noticeable tunneling, it decays
exponentially in $g$~\cite{Beloborodov05} until it reaches the
inverse escape rate from a single grain $\Gamma $. At $\Delta
_{M}\sim \Gamma $ the system falls into a regime of weak Coulomb
correlations and the insulator-metal transition occurs in $3d$.
Using the estimate $\Delta _{M}\sim \Gamma $ for estimating the
transition point one arrives, within the logarithmic accuracy, at
the same result for the critical conductance $g_{c}$ as that
derived from a renormalization group considerations (see for
details Section~\ref{insulating}).

The presence of the hard gap in the excitation spectrum leads to
the activation dependence (Arrhenius law) of the conductivity on
temperature:
\begin{equation}
\sigma \sim e^{-\Delta _{M}/T},\hspace{1cm}T\ll \Delta _{M}.
\label{activation1}
\end{equation}
Indeed, the finite temperature conductivity is due to the
electrons and holes that are present in the system as real
excitations. Their density is given by the Gibbs distribution that
results in the exponential dependence of the conductivity,
Eq.~(\ref{activation1}).

However, the activation behavior is usually not observed in real
granular samples at low temperatures. Instead, the experimentally
observed resistivity follows the law, Eq.~(\ref{e3}), that
resembles the Efros-Shklovskii law derived for doped
semiconductors. The fact that the observed conductivity behavior
cannot be explained in terms of the periodic model suggests that
disorder  plays the crucial role in formation the low temperature
conductivity in the insulator state.

\begin{figure}[tbp]
\includegraphics[width=7.0cm]{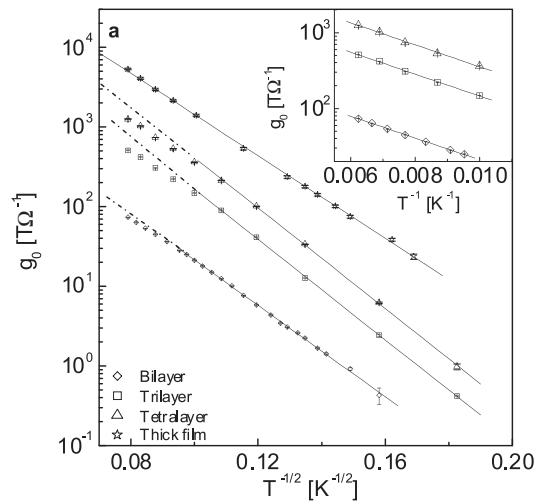}
\caption{Conductance $g_0$ vs inverse temperature $T^{-1/2}$ for
periodic granular multilayer and thick-film shown in
Fig.~\protect\ref{periodic_gs}~\cite{Tran05}. Inset: for the
high-temperature range $g_0$ has been replotted as a function of
$T^{-1}$, indicating Arrhenius behavior from $100 - 160$ K.}
\label{Tran05_picture}
\end{figure}

The stretched exponential Shklovskii-Efros-like conductivity
behavior in granular conductors remained a challenging puzzle for
a long time.  Several explanations had been advanced, in
particular, it was proposed~\cite{Abeles75}, that the capacitance
random variations resulting from grain size dispersion could
provide the dependence, Eq.~(\ref{e3}). However, as was pointed
out in Refs.~\cite{pollak, Shklovskii04}, the capacitance disorder
can never lift the Coulomb blockade in a single grain completely
and therefore cannot give rise to the finite density of states at
the Fermi level. Furthermore, the stretched exponential
dependence, Eq.~(\ref{e3}), was recently observed in the periodic
arrays of quantum dots~\cite{Yakimov03} and artificially
manufactured metallic periodic granular systems~\cite{Tran05}, see
Fig.~\ref{Tran05_picture}, where the size of granules and the
periodicity in the dots arrangement were controlled within a few
percent accuracy. Either of those systems does not posses a
noticeable capacitance disorder at all, yet the
dependence~(\ref{e3}) has been observed. These indicates that it
should be electrostatic disorder unrelated to the grain size
variations but caused most probably by charged defects in the
insulating matrix/substrate that is responsible for lifting
Coulomb blockade and formation the finite density of states near
the Fermi level resulting in the dependence~(\ref{e3}).

There is however another ingredient necessary for the
variable-range-hopping type (VRH) conductivity to occur (apart the
finite density of states at energies close to Fermi level): the
finite, although exponentially decaying with the distance,
probability, for tunneling to the spatially {\it remote} -- and
not only to the adjacent -- states close to the Fermi level.
Accordingly, the problem of the hopping transport in the granular
conductors is two-fold and should contain: (i) explanation the
origin of the finite density of states near the Fermi-level and
the role of the Coulomb correlations in forming this density of
states and (ii) constructing and quantitative description of the
mechanism of tunneling over long distances through the dense array
of metallic grains.

The behavior of the density of states, as we have mentioned above,
can form due to the onsite random potential, which in its turn is
induced by the carrier traps in the insulating matrix in the
granular conductors. The traps with energies lower than the Fermi
level are charged and induce the potential of the order of
$e^{2}/\kappa r$ on the closest granule, where $\kappa $ is the
dielectric constant of the insulator and $r$ is the distance from
the granule to the trap. This compares with the Coulomb blockade
energies due to charging metallic granules during the transport
process. Such a mechanism was considered by~\citet{Shklovskii04}.
Speaking about the $2d$ granular arrays and/or arrays of quantum
dots, one can expect that the induced random potential originates
also from imperfections and the charged defects in the substrate.

We model the electrostatic disorder via the random potential
$V_{i}$, where $i$ is the grain index. Such a potential gives rise
to a flat bare density of states at the Fermi level. In a complete
analogy with semiconductors, the bare density of states should be
suppressed by the long-rang Coulomb
interaction~\cite{Efros,Shklovskii}.

Next we have to consider the electron hopping over the distances
well exceeding the average granule size in a dense granular array.
This processes can be realized as tunneling via the virtual
electron levels in a sequence of grains. The virtual electron
tunneling through a \emph{single} granule or a quantum dot (the so
called co-tunneling) was first considered by~\citet{Averin} where
two different mechanisms for a charge transport through a single
quantum dot in the Coulomb blockade regime were identified.
Namely, there are elastic and inelastic co-tunneling mechanisms.
In the course of elastic co-tunneling the charge is transferred
via the tunneling of an electron through an intermediate virtual
state in the dot such that the electron leaves the dot with the
same energy as it came in. In the latter mechanism (inelastic
co-tunneling) an electron that comes out of the dot has a
different energy from that of the incoming one. After inelastic
co-tunneling the electrons leaves behind in the granule the
electron-hole excitation absorbing the in- and out- energy
differences.  Note that both these processes are realized via
classically inaccessible intermediate states, i.e. both mechanisms
occur in the form of the charge transfer via a virtual state. The
inelastic cotunneling dominates at temperatures larger than
$T_{1}\sim \sqrt{E_{c} \, \delta }$~\cite{Averin}.

These two co-tunneling mechanisms can be generalized to the case
of the multiple co-tunneling through several grains. The tunneling
probability should fall off exponentially with the distance (or
the number of granules left behind
$N$)~\cite{Beloborodov05,Feigelman05,Tran05} and this is
equivalent to the exponentially decaying probability of the
tunneling between the states near the Fermi surface in the theory
of Mott-Efros-Shklovskii~\cite{Efros,Shklovskii}.

Thus the hopping processes in the amorphous semiconductors and
granular materials are alike -- up to the specific expressions for
the localization lengths -- and minimizing the hopping probability
in the same manner  as in the classic Efros-Shklovskii work, one
ends up with the hopping conductivity $\sigma $ in a form
\begin{equation}
\sigma \sim \exp \left[ -\left( T_{0}/T\right) ^{1/2}{}\right],
\label{e6}
\end{equation}
where $T_{0}$ is a characteristic temperature depending on the
particular microscopic characteristics. Explicit expressions for
this temperature in different regimes are given in Section
\ref{hopping}. Eq.~(\ref{e6}) explains the experimentally observed
conductivity behavior in poorly conducting granular materials (see
Eq.~(\ref{e3})).

Our discussion of the variable range hopping via virtual electron
tunneling through many grains is based on the assumption that the
hopping length $r^{\ast }$ exceeds the size of a single grain $a$.
This length decays when the temperature increase and reaches the
grain size at some characteristic temperature $\tilde{T}$. Then
the VRH picture does not apply any longer, the hops occur between
the adjacent granules only.  Once the probability of a single jump
is defined, the quantum effects can be neglected and one can use a
classical approach. In particular, at temperatures $T\geq
\tilde{T}$ one expects the conductivity to follow the simple
Arrhenius law.

The classical approach to the transport in granular metals was
developed in Refs.~\cite{Middleton93,Jha05}. One of the results of
this study is the presence of the threshold voltage below which
the conductivity is exactly zero at $T=0$. The existence of such a
threshold voltage is a consequence of the classical approach where
multiple cotunneling processes are not taken into account.

In order to match the results of the classical and hopping
theories one has to generalize the approach of~\citet{Middleton93}
to include the multiple cotunneling processes. Development of such
a theory, in our opinion, is an interesting and important task.

\subsection{Model and main theoretical tools}

\label{model}

From the theoretical point of view a granular conductor is an
appealing exemplary system whose behavior is governed by a
non-trivial interplay of electron-electron interactions, disorder
and quantum fluctuations.  The powerful approaches that allowed
for recent breakthroughs in our understanding the physics of
granular media are based on the effective field theories; in
particular, the phase action technique appeared especially
suitable.  For the situations that fall outside its range of
applicability, the appropriate diagrammatic techniques,
generalizing those for the homogeneously disordered systems, can
be developed.  In this subsection we present in detail a model for
description of granular metals and introduce both complimentary
methods that serve as a foundation for quantitative description of
granular conductors.

\subsubsection{Hamiltonian}

We model the granular system as an array of metallic particles
connected via tunneling contacts. The Hamiltonian $\hat{H}$
describing a granular conductor has the form
\begin{equation}
\hat{H}=\sum_{i}\hat{H}_{0,i}+\hat{H}_{I}+\hat{H}_{t}, \label{a02}
\end{equation}%
where $\hat{H}_{0,i}$ stands for the Hamiltonian of
non-interacting electrons in a grain $i,$ $\hat{H}_{c}$ represents
the interactions, and $\hat{H}_{t}$ describes the electron
tunneling between the grains. We now discuss each term in
Eq.~(\ref{a02}).

The $\hat{H}_{0,i}$ term describes the free electrons within the
each grain in the presence of impurities
\begin{equation}
\hat{H}_{0,i}=\int \hat{\psi}_{i}^{+}\left( \mathbf{r}\right)
\left( -\frac{\mathbf{\nabla }^{2}}{2m}+u_{i}(\mathbf{r})-\mu
\right) \hat{\psi}_{i}\left( \mathbf{r}\right) d\mathbf{r},
\label{a01}
\end{equation}
where $\hat{\psi}_i^{+}(\mathbf{r),\hat{\psi}_{i}(r)}$ are the
electron creation and annihilation operators, $\mu $ is the
chemical potential and $u\left( \mathbf{r}_{i}\right) $ represents
disorder responsible for the electron scattering inside the $i$-th
grain. We adapt the Gaussian distribution for $ u\left(
\mathbf{r}_{i}\right) $ with pair correlations
\begin{equation}
\langle u_{i}(\mathbf{r})\,u_{j}(\mathbf{r}^{\prime })\rangle
={\frac{1}{2\pi \nu \tau _{\mathrm{imp}}}}\delta
(\mathbf{r}-\mathbf{r}^{\prime })\,\delta _{ij}.  \label{a30}
\end{equation}
Throughout our review we assume that all the grains are in the
diffusive limit, i.e. the electron mean free path $l$ within each
grain is smaller than the grain size $a$. This assumption
simplifies our calculations because it allows to avoid considering
the electron scattering from the grain boundaries that becomes the
main ``disorder" mechanism in the case of ballistic grains. At the
same time, most of the results obtained in the diffusive limit are
expected to hold also for the ballistic grains with an irregular
surface as long as the single grain diffusion coefficient $D_{0}$
does not enter the final result. This happens in the normal grains
provided all relevant energies are smaller than the Thouless
energy $E_{Th}$ of a single grain, Eq.~(\ref{thouless}). Actually,
for typical grain sizes of the order $100\mathring{A}$ the mean
free path $l$ is comparable with the granule size $a$.

A single grain may be considered within the standard diagrammatic
approach developed for disordered metals~\cite{Abrikosov65}. The
main building block for the diagrams is a single electron Green
function averaged over disorder and can be derived by the self
consistent Born approximation (SCBA). The diagram shown in
Fig.~\ref{selfenergy}a represents the relevant contribution to the
self energy:
\begin{equation}
G_{0\varepsilon }\left( \mathbf{p}\right) =\left( i\varepsilon
-\xi (\mathbf{ \ p})+i\frac{\mbox{sgn}\left( \varepsilon \right)
}{2\tau _{\mathrm{imp}}}\right) ^{-1}.  \label{a15}
\end{equation}
Another important block for the diagrammatic technique is the
diffusion propagator (diffusion) which is just the impurity
averaged particle-hole propagator. In bulk disordered metals it is
given by
\begin{equation}
D(\omega , \mathbf{q})={\frac{1}{{D}_{0}{\mathbf{q}^{2}+|\omega
|}}}, \label{e15}
\end{equation}
where $D_{0}$ is the classical diffusion coefficient and $\omega $
is the Matsubara frequency.

In a grain, however, the term $\mathbf{q}^{2}$ has to be replaced
by the Laplace operator with the proper boundary
conditions~\cite{efetov83,efetov,Aleiner02}. This procedure leads
to the quantization of the diffusion modes and to the appearance
of the lowest excitation energy of the order of $D_{0}/a^{2}$ with
$a$ being the grain size. This is just the Thouless energy
$E_{Th}$ that has been defined in Eq.~(\ref{thouless}).

The Thouless energy of a nanoscale grain is large and this allows
us to simplify calculations neglecting all non-zero space
harmonics in the diffusion propagator, Eq.~(\ref{e15}). Throughout
our review we will assume that $E_{Th}$ is the largest energy
scale associated with our system, and will use the zero
dimensional approximation for the diffusion propagator
\begin{equation}
D(\omega )=1/|\omega |.  \label{Zero_dimensional_diffusion}
\end{equation}

Now we turn to the description of the electron-electron
interaction, the second term $\hat{H}_I$ in the Hamiltonian,
Eq.~(\ref{a02}). We begin with the consideration of the
interaction effects in an isolated grain and then we will turn to
the discussion of the intergranular terms that are important
because of the long range character of the Coulomb interaction.

The most general form of the electron-electron interaction
$\hat{H}_I$ in an isolated grain is
\begin{equation}
\hat{H}_{I}^{(0)}=\frac{1}{2}\sum_{p,q,r,s}\mathcal{H}_{pqrs} \,
\psi _{p,\alpha }^{+}\psi _{q,\beta }^{+}\psi _{r,\beta }\psi
_{s,\alpha },  \label{e19}
\end{equation}
where the subscripts  $p$, $q$, $r$, $s$ stand for the states in
the grains and $\alpha ,$ $\beta $ label electron spins.

The matrix $\mathcal{H}_{pqrs}$ is a complicated object, but in
the low energy region of the parameters, not all of the matrix
elements are of the same order. In the case of disordered grain
that we consider, only the elements with the equal in pairs
indices survive \cite{Aleiner02}, all others are small in the
parameter $1/g_0,$ where $g_0$ is the single grain conductance.
Thus in the leading order in $1/g_0$ the in-grain interaction term
is \cite{Aleiner02}
\begin{equation}
H_I^{(0)} = E_c\,  \hat n^2 + J_S \hat{ \vec S^2} +\lambda \, \hat
T^\dagger \hat T,
\end{equation}
where $\hat{n}=\int \hat{\psi}^{+}\left( \mathbf{r}\right)
\hat{\psi} _{i}\left( \mathbf{r}\right) d\mathbf{r}-N_{0}$ is the
number of excessive (with respect to the electron number in the
charge neutral state $N_{0}$) electrons in the grain,  $\hat{\vec
S}$ is the total spin of the grain and $\hat T^\dagger, \hat T$
are the Cooper pair creation and annihilation operators:
$T=\sum_{p} \hat\psi_{p,\uparrow} \hat \psi_{p,\downarrow}.$ The
interaction strengths in these three channels are controlled by
coupling constants $ E_c, J_S, \lambda$ respectively. The coupling
constants in the spin and BCS channels, $J_S$ and $\lambda$,
cannot be large and are, at most, of the order of $\delta$. At the
same time the charging energy $E_c$ is usually much larger than
$\delta$. For this reason, in the absence of the
superconductivity, the most interesting and noticeable effects
come from the Coulomb correlations.

The above considerations of the interactions in a single grain can
be easily generalized to the case of a granular array. It is clear
that the bare interactions in the spin and Cooper pair channels
are short-range and thus they have to be diagonal in the granular
indices. At the same time the Coulomb interaction is long-range
and it's of-diagonal components cannot be neglected. Thus, we
arrive at the following Hamiltonian that describes the Coulomb
correlations
\begin{equation}
\hat{H}_{c}=\frac{e^{2}}{2}\sum_{ij}\hat{n}_{i}C_{ij}^{-1}\hat{n}_{j},
\label{Coulomb__Interacation}
\end{equation}
where $C_{ij}^{-1}$ is the capacitance matrix which can be found
by solving a classical electrostatic problem of metallic particles
embedded into the insulating matrix. Note that since metallic
grains have infinite dielectric constant, the effective dielectric
constant of the whole sample can be considerably larger than the
dielectric constant of its insulating component. Thus the
effective single grain charging energy can be much less than the
electrostatic energy of a single grain in a vacuum.

The Hamiltonian $\hat{H}_{c}$ describing the long range part of
the Coulomb interaction, (\ref{Coulomb__Interacation}), has been
derived for granular superconductors in the early
work~\cite{Efetov80a}. As far as the normal grains are concerned,
the proper derivation is given in the review~\cite{Aleiner02}. The
term $\hat{H}_{c}$, Eq.~(\ref{Coulomb__Interacation}), commutes
with the free part $\hat{H}_{0,i}$, Eq.~(\ref{a01}) and therefore
does not describe dynamics in a single insulated granule. For the
macroscopic transport to occur the electron trasfer from grain to
grain has to be switched on, and we turn now to the third term
$\hat{H}_{t} $ in Eq.~(\ref{a02}) describing the electron
tunneling between the neighboring grains. We write this term in
the form
\begin{equation}
\hat{H}_{t}=\sum_{i,j;\,p,q}t_{ij;pq}\hat{\psi}_{pi}^{+}\hat{\psi}_{qj},
\label{a2a}
\end{equation}
where the summation is performed over the states $p$, $q$ of each
grain and over the neighboring grains $i$ and
$j$.~\cite{cohen,ambegaokar63,Abrikosovbook}. It is assumed that
the linear size of the inter-grain contact area well exceeds
atomic distances, i.e. there is a large number of conducting
channels between the grains. At the same time, the magnitude of
the potential barrier between the grains can be large, this can
easily be achieved experimentally by the process of oxidation.
Therefore, the tunneling conductance $g$, which is roughly
proportional to the area of the contact and to the square of the
tunneling matrix element $t_{ij}$, can be set both larger and
smaller than unity.

While the dependence of the tunneling elements $t_{ij;pq}$ on the
state numbers $p,q$ is not important for transport through the
conventional point contacts and is usually
neglected~\cite{cohen,ambegaokar63,Abrikosovbook}, it may become
relevant for granular arrays. Namely, the importance of this
dependence is related to whether the electron motion in the grains
is chaotic or can be viewed as an integrable billiard. In order to
clarify this point we write the matrix elements $t_{ij;pq}$ as
\begin{equation}
t_{ij;pq}=\int t_{ij}\left( \mathbf{s}_{ij}\right) \phi _{p}^{\ast
}\left( \mathbf{s}_{ij}\right) \phi _{q}\left(
\mathbf{s}_{ij}\right) d\mathbf{s}_{ij},  \label{e21}
\end{equation}
where $\mathbf{s}_{ij}$ is the coordinate on the junction between
the grains $i$ and $j$ and $\phi _{p}\left( \mathbf{r}_{i}\right)
$ are the wave functions. The integration over $s_{ij}$ in
Eq.~(\ref{e21}) extends over the area of the junction.

If the grains do not contain any disorder and have a perfect shape
(e.g. cubes or spheres), the array is a completely periodic system
and according to the Bloch theorem the total resistivity is zero.
The presence of the internal disorder or irregularities of the
shape of the grains change the situation giving rise to a finite
resistivity. In the latter case one may employ the random matrix
theory~(RMT) \cite{mehta,Beenakker97,Alhassid00} for an isolated
single grain or, which is equivalent, the zero dimensional $\sigma
$-model~\cite{efetov}. It is well known~\cite{mehta} that this
theory, when applied to the eigenfunctions leads to a Gaussian
distribution $W\left\{ \left\vert \phi_{p}\left( \mathbf{r}\right)
\right\vert \right\} $ of their amplitudes
\begin{equation}
W\left\{ \left\vert \phi _{p}\left( \mathbf{r}\right) \right\vert
\right\} =\exp \left( -\left\vert \phi _{p}\left(
\mathbf{r}\right) \right\vert ^{2}V\right),  \label{gaussian}
\end{equation}
where $V$ is the volume of the grain. Further, the Gaussian
distribution of the amplitudes of the wave functions $\phi
_{p}\left( \mathbf{r}_{i}\right) $ and the fact that their
correlations decay rapidly with the increasing both spatial
distance and the energy levels spacing (the characteristic scales
are the wavelength and the mean level spacing, respectively) leads
to a Gaussian distribution of the matrix elements $t_{ij;pq}$.

A simple transformation brings these correlators to:
\begin{eqnarray}
\langle t_{p_{1}q_{1}}\,t_{p_{2}q_{2}}\rangle
&=&\frac{g_{ij}\,\delta _{i}\delta _{j}}{2\pi }\,(\delta
_{p_{1}p_{2}}\,\delta _{q_{1}q_{2}}+\delta
_{p_{1}q_{2}}\,\delta _{p_{2}q_{1}}),  \nonumber \\
\left\langle t_{pq}\right\rangle &=&0.  \label{e16}
\end{eqnarray}%
In Eq.~(\ref{e16}) the matrix elements $t_{pq}$ are taken for the
same contact between the grains $i$ and $j$, otherwise the
correlations vanish. As we will see, the constant $g_{ij}$ is
nothing but the tunneling conductance of the contact.

Equation~(\ref{e16}) is written for time reversal invariant
systems (orthogonal ensemble). This, in particular, means that
there is no magnetic field and/or and no magnetic impurities
present and therefore all eigenfunctions of an insulated grain can
be chosen real. In the limit of comparatively strong magnetic
fields, one arrives at the unitary ensemble and the first term in
the correlation function of the matrix elements in Eq.~(\ref{e16})
vanishes.  However, for the metallic grains of the size of the
order of $5$-$10nm$ the characteristic magnetic field that require
the description in terms of the unitary ensemble is of the order
of several Tesla. Thus in the subsequent discussion we will be
using Eq.~(\ref{e16}), assuming that the applied magnetic fields
are not that high.

To conclude this part, we have reformulated the initial theory
with impurities and regular tunneling matrix elements in terms of
the model with random tunneling elements. This equivalence holds
for the zero dimensional grains, or in other words, for the
situations when all the characteristic energies are smaller than
the Thouless energy $E_{Th}$, Eq.~(\ref{thouless}). Working with
the random tunneling elements turns out the more convenient
approach.

The model introduced in this subsection, Eqs.~(\ref{a02},
\ref{a01}, \ref{a30}, \ref{Coulomb__Interacation}, \ref{a2a},
\ref{e16}) describes a strongly correlated electronic disordered
system which cannot be solved exactly. Below we introduce two
complimentary approaches to explore this problem in different
regimes. One of these approaches -- the diagrammatic technique --
is especially useful provided the perturbative  expansion with
respect to one of the relevant parameters is possible. The second
method is based on the gauge transformation which allows to get
rid of the explicit Coulomb interaction term in the Hamiltonian at
the expense of an appearing phase field. As we will show below,
this approach offers a very powerful tool in its domain of
applicability.

\subsubsection{ Diagrammatic technique for granular metals}

\label{diagrmamatic_technique}

\begin{figure}[t]
\hspace{-0.4cm} \includegraphics[width=2.7in]{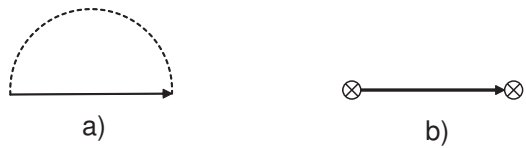}
\caption{ Self energy of the election Green function averaged over
the impurity potential inside the grains and over tunnelling
elements between the grains. Averaging over the impurity potential
is represented by the dotted line (a) while tunnelling elements
are represented by crossed circles (b). } \label{selfenergy}
\end{figure}

There are two routes to construct diagrammatic technique for
granular metallic systems. First, one can work in the basis of the
exact single grain eigenfunctions and use the distribution of the
tunneling elements~(\ref{e16}) for the description of the
scattering between the different states of the neighboring grains.
In this case the impurities inside each grain are treated
``exactly," and, by definition, the diagrams that represent
impurity scattering within each grain do not appear in this
representation. The alternative approach is, in essence,
equivalent to the conventional cross technique: one begins with
the momentum representation and then carries out the standard
averaging over impurities within the each grain. The tunneling
between the grains can be then viewed as an intergranular
scattering. The corresponding matrix elements, as in the first
method can, be viewed as Gaussian random variables. We prefer to
follow the latter approach since it is straightforwardly related
to the standard diagrammatic technique for homogeneously
disordered metals making it easier to make comparisons to well
known situations in disordered metals when possible.

Following the guidelines of~\citet{Abrikosov65} we construct the
action expansions with respect to both types of disorder:
potential disorder within the each grain $u\left( \mathbf{r}
\right) $ and the intergranular scattering matrix elements
$t_{pq}$. Shown in Fig.~\ref{selfenergy} are the lowest order
diagrams representing both contributions to the self energy of the
single electron Green function in the granular metals obtained by
averaging over $u\left( \mathbf{r}\right) $ and $t_{pq}$. The
diagram a) describes the potential scattering within a single
grain, while the diagram b) is due to the intergranular
scattering. Both processes result in a similar contribution
proportional to $\mathrm{sgn} \omega $ to the electron
self-energy. This shows, that on the level of the single electron
Green function; the intergranular scattering results merely in the
renormalization of the relaxation time $\tau $
\begin{equation}
\tau ^{-1}=\tau _{0}^{-1}+2\Gamma d,  \label{self}
\end{equation}%
where $\tau _{0}$ is the electron mean free time in a single
grain.
\begin{figure}[t]
\hspace{-0.5cm}
\includegraphics[width=2.8in]{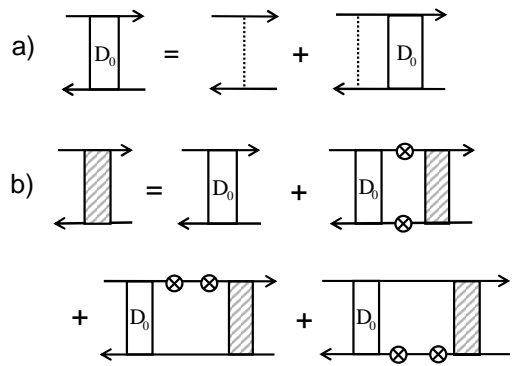} \caption{
Diagrams represent the Dyson equation for a single grain diffusion
propagator Eq.~\protect\ref{Zero_dimensional_diffusion} (a) and
for the whole granular system~Eq.~\protect\ref{dif_prop} (b).
Dotted lines represent the impurity scattering while crossed
circles stay for intergranular tunnelling elements. }
\label{Diffuson_Diagram}
\end{figure}

The next step is to consider the diffusion motion of electron
through the granular metal. The diffusion motion inside a single
grain is given by the usual ladder diagram that results in the
diffusion propagator~$D\left( \omega \right) $,
Eq.~(\ref{Zero_dimensional_diffusion}). The tunneling between the
grains does not change the selection rules for the diagrams.
Typical diagrams are shown in Fig.~\ref{Diffuson_Diagram} b). The
diagrams are generated by connection the tunneling vertices, and
only the diagrams without the intersection are to be kept. This is
similar to what one has in the standard impurities
technique~\cite{Abrikosov65}. In order to derive the expression
for the total diffusion propagator one should sum up the ladder
diagrams shown in Fig.~\ref{Diffuson_Diagram} b). For the periodic
array of the grains we arrive at the following formula for the
diffusion propagator $D(\omega ,\mathbf{q})$
\begin{equation}
D(\omega ,\mathbf{q})=\tau ^{-1}\,(|\omega |+\Gamma \lambda
_{\mathbf{q} })^{-1},  \label{dif_prop}
\end{equation}
where
\begin{equation}
\lambda _{\mathbf{q}}=2\sum_{\mathbf{a}}(1-\cos \mathbf{qa}),
\label{e240}
\end{equation}
$\mathbf{a}$ is the (super) lattice vectors, and $\mathbf{q}$ is
the quasi-momentum. In Eq.~(\ref{dif_prop}) only the zero space
harmonics of the diffusion motion in the grain is taken into
account. This approximation is valid in the limit
\begin{equation}
\Gamma \ll E_{Th}  \label{e24}
\end{equation}%
where the energies $\Gamma $ and $E_{Th}$ are given by
Eqs.~(\ref{Gamma}) and (\ref{thouless}), respectively. In the
limit of small quasimomenta, $q\ll a^{-1}$, we have $\lambda
_{q}\rightarrow a^{2}q^{2}$, such that the
propagator~(\ref{dif_prop}) describes the diffusion motion on the
scales much larger than the size of a single grain $a$ with an
effective diffusion coefficient
\begin{equation}
D_{eff}=a^{2}\Gamma .  \label{e25}
\end{equation}
The other necessary blocks of the diagrammatic technique can be
constructed analogously to the diffusion propagator.

\subsubsection{Coulomb interaction and gauge transformation}

\label{gauge}

Taking an alternative route, one can eliminate the explicit
Coulomb term in the model for the granular metals,
Eqs.~(\ref{a02}, \ref{a01}, \ref{a30},
\ref{Coulomb__Interacation}, \ref{a2a}, \ref{e16}), by introducing
a new phase field via a gauge transformation. This approach proves
to be especially useful in the regime of strong Coulomb
correlations, where the diagrammatic technique developed in the
previous subsection may break down.

To introduce this approach it is convenient to adapt the formalism
of the functional integration that allows replacing the
calculations with the Hamiltonian $\hat{H}$, Eq.~(\ref{a02}), by
computation of a functional integral over classical fermion fields
$\psi _{i}\left( X\right) $ and their complex conjugate $\psi
_{i}^{\ast }\left( X\right) $, where $X=\left( \mathbf{r},\tau
\right) $. These fields must satisfy the fermionic antiperiodicity
condition
\begin{equation}
\psi _{i}\left( \tau \right) =-\psi _{i}\left( \tau +\beta
\right), \label{e28}
\end{equation}
where $\beta =1/T$ is the inverse temperature. We define the field $%
\psi _{i}\left( X\right) $ such that it is different from zero
only in the $i$th grain.

The Lagrangian $L\left[ \psi \right] $ entering the functional
integral for the partition function $Z$
\begin{equation}
Z=\int \exp \left( -\int_{0}^{\beta }L\left[ \psi \right] d\tau
\right) D\psi \label{e25a}
\end{equation}
takes the form
\begin{equation}
L\left[ \psi \right] =\sum_{i}L_{0i}\left[ \psi \right]
+L_{t}\left[ \psi \right] + L_{c}\left[ \psi \right]\;\; ,
\label{e25b}
\end{equation}
where
\begin{equation}
L_{0i}\left[ \psi \right] =\int \psi _{i}^{\ast }\left( X\right)
\left( \frac{\partial }{\partial \tau }-\frac{\mathbf{\nabla
}^{2}}{2m}+u(\mathbf{r} )-\mu \right) \psi _{i}\left( X\right)
d\mathbf{r},  \label{e25c}
\end{equation}%
\begin{equation}
L_{t}\left[ \psi \right] =\sum_{i,j;\,p,q}t_{ij;pq}\psi
_{pi}^{\ast }\left( \tau \right) \psi _{qj}\left( \tau \right) ,
\label{e25e}
\end{equation}
and
\begin{equation}
L_{c}\left[ \psi \right] =\frac{e^{2}}{2}\sum_{ij}n_{i}(\tau
)C_{ij}^{-1}n_{j}(\tau )\;\; ,  \label{e25d}
\end{equation}
where $n_{i}(\tau )=\int \psi _{i}^{\ast }(X)\psi
_{i}(X)d\mathbf{r}$ is the density field of the grain $i$.

The Coulomb interaction $L_{c}\left[ \psi \right] $,
Eq.~(\ref{e25d}), can be decoupled using a Gaussian integration
over the auxiliary field $\bar{V}_{i}\left( \tau \right)
$~\cite{Efetov02b}
\begin{eqnarray}
&&\exp \left[ -\frac{e^{2}}{2}\sum_{ij}\int_{0}^{\beta }n_{i}(\tau
)C_{ij}^{-1}n_{j}(\tau )d\tau \right]  \nonumber \\
&=&\int \exp \left[ i\sum_{i}\,\int_{0}^{\beta }\bar{V}_{i}(\tau
)\,\psi
_{i}^{+}(X)\psi _{i}(X)dX\right]  \nonumber \\
&&\times Z_{\bar{V}}^{-1}\exp \left[ -S\left( \bar{V}\right)
\right] d{\bar V}_{i}, \label{e26}
\end{eqnarray}
where $\bar{V}_{i}(\tau )$ is the bosonic field obeying the
boundary condition $\bar{V}_{i}\left( \tau \right)
=\bar{V}_{i}\left( \tau +\beta \right) $ and $Z_{\bar{V}}$ is the
partition function,
\[
Z_{\bar{V}}=\int \exp \left[ -S\left( \bar{V}\right) \right]
d\bar{V}_{i}.
\]
The action $S\left( \bar{V}\right) $ has the following form
\begin{equation}
S\left[ \bar{V}\right] =\frac{1}{2e^{2}}\int_{0}^{\beta }d\tau
\sum_{ij}\bar{V}_{i}(\tau )C_{ij}\bar{V}_{j}(\tau ).  \label{a7}
\end{equation}
We see from Eqs.~(\ref{e25} - \ref{e26}) that the Lagrangian
becomes quadratic in fields $\psi $ after the decoupling,
Eq.~(\ref{e26}). Instead of dealing with the Coulomb interaction,
$L_{c}\left[ \psi \right] $, Eq.~(\ref{e25d}), one should consider
now an effective Lagrangian $L_{0i}^{eff}\left[ \psi
,\bar{V}\right] $ for the grain $i$:
\begin{equation}
L_{0i}^{eff}\left[ \psi ,\bar{V}\right] =L_{0i}\left[ \psi \right]
-i\int_{0}^{\beta }\bar{V}_{i}\left( \tau \right) \psi
_{i}^{+}\left( \tau \right) \psi _{i}\left( \tau \right) d\tau .
\label{e27}
\end{equation}
The effective action $L_{0i}^{eff}\left[ \psi ,\bar{V}\right] $ is
now expressed in terms of electron motion in a granular matter in
the presence of the fluctuating potential $\bar{V}_{i}\left( \tau
\right) $ of the grains.

We remove the field $\bar{V}_{i}(\tau )$ from the Lagrangian $L_{0i}^{eff}%
\left[ \psi ,\bar{V}\right] $, Eq.~(\ref{e27}), using a gauge
transformation of the fermionic fields
\begin{equation}
\psi _{i}\left( \mathbf{r},\tau \right) \rightarrow e^{-i\varphi
_{i}\left( \tau \right) }\psi _{i}\left( \mathbf{r},\tau \right)
,\text{ \ \ \ }\dot{\varphi}_{i}\left( \tau \right) =\bar
V_{i}\left( \tau \right), \label{a4}
\end{equation}
where the phases $\varphi _{i}\left( \tau \right) $ depend on the
imaginary time $\tau $ but not on the coordinates inside the
grains. This is a consequence of the form of the Coulomb
interaction, Eqs.~(\ref{Coulomb__Interacation}, \ref{e25d}). Since
the action of an isolated grain is gauge invariant, the phases
$\varphi _{i}(\tau )$ enter the whole Lagrangian of the system
only through the tunneling matrix elements
\begin{equation}
t_{ij}\rightarrow t_{ij}\,e^{i\varphi _{ij}(\tau )},
\label{tilde_t}
\end{equation}%
where $\varphi _{ij}(\tau )=\varphi _{i}(\tau )-\varphi _{j}(\tau
)$ is the phase difference of the $i-$th and $j-$th grains.

At the first glance, the transformation, Eq.~(\ref{a4}), have
removed completely the potentials $\bar{V}\left( \tau \right) $
from the effective Lagrangian $L_{0i}^{eff}\left[ \psi
,\bar{V}\right] $, Eq.~(\ref{e27}). In fact, this is not the case
since the gauge transformation as defined in Eq.~(\ref{a4}),
violates the antiperiodicity condition, Eq.~(\ref{e28}) and the
resulting effective Lagrangian for the single grain changes.  In
order to preserve the boundary condition, Eq.~(\ref{e28}), the
certain constraints should be imposed  on the phases $\varphi
_{i}\left( \tau \right) $ and potentials $\bar{V}_{i}\left( \tau
\right) $. Namely, the phases should obey the condition
\begin{equation}
\varphi _{i}\left( \tau \right) =\varphi _{i}\left( \tau +\beta
\right) +2\pi k_{i},  \label{e29}
\end{equation}
which leads to the following constraint for the potential
$\bar{V}_{i}\left( \tau \right) $
\begin{equation}
\int_{0}^{\beta }\bar{V}_{i}(\tau )d\tau =2\pi k_{i},
\label{constraint}
\end{equation}
where $k_{i}=0,\pm 1,\pm 2,\pm 3...$.

The constraint on the phases $\varphi _{i}\left( \tau \right) $,
Eq.~(\ref{e29}), can be reformulated via introducing a function
$\phi _{i}\left( \tau \right) $ assuming arbitrary real values
from $-\infty $ to $\infty $ and obeying a periodicity condition
\begin{equation}
\phi \left( \tau \right) =\phi \left( \tau +\beta \right).
\label{e30}
\end{equation}
With the aid of this function the phase $\varphi _{i}\left( \tau
\right) $ can be written in the form:
\begin{equation}
\varphi _{i}\left( \tau \right) =\phi _{i}\left( \tau \right)
+2\pi Tk_{i}\tau ,  \label{a5}
\end{equation}
which satisfies Eq.~(\ref{e29}). The potential $\bar{V}_{i}\left(
\tau \right) $, in its turn, is taken in a form
\begin{equation}
\bar{V}_{i}(\tau )=\rho _{i}+\tilde{V}_{i}(\tau ),
\label{potential}
\end{equation}%
where the static variable $\rho _{i}$ varies in the interval
\begin{equation}
-\pi T<\rho _{i}<\pi T,\text{\ }  \label{b7}
\end{equation}%
and the dynamic variable $\tilde{V}_{i}(\tau )$ satisfies the
constraint (\ref{constraint}). Note that the static part of the
potential $\rho _{i}$ cannot be gauged out, contrary to dynamic
contribution, $\tilde{V}_{i}(\tau )$, and the effective Lagrangian
assumes the form $L_{0i}^{eff}\left[ \psi ,\rho \right] $. In the
limit of not very low temperatures,
\begin{equation}
T\gg \delta,  \label{f1}
\end{equation}
the static potential $\rho _{i}$ drops out from the fermionic
Green functions and does not influence the system behavior. This
can be understood by noticing that at moderately high
temperatures~(\ref{f1}) the discreteness of the levels in the
grains is not important. Then, in the Green functions, one may
replace  the variable $\varepsilon _{p}-\mu \rightarrow \xi ,$
where $\varepsilon _{p}$ are eigenenergies and $\xi $ is a
continuous variable varying from $-\infty $ to $\infty $.
Integrating over $\xi $ one may shift the contour of the
integration into the complex plane and remove $\rho _{i}$ provided
it obeys the inequality~(\ref{b7}). More details can be found
in~\citet{Efetov02b}. Note that the variable $\rho _{i}$ cannot be
neglected in the Lagrangian $L_{0i}^{eff}\left[ \psi ,\rho \right]
$ in the limit of very low temperatures.

An integer $k_{i}$ in Eqs.~(\ref{e29}, \ref{constraint}, \ref{a5})
represents an extra degree of freedom related to the charge
quantization and is usually called the \textquotedblleft winding
number\textquotedblright . The physical meaning of the winding
numbers becomes especially clear in the insulating regime where
they simply represent static classical electron charges.

Thus, at not very low temperatures, (\ref{f1}), one can write the
partition function $Z$ in the form
\begin{equation}
Z=\int \exp \left[ -\int \left( L_{0}\left[ \psi \right]
+L_{1}\left[ \psi ,\varphi \right] +L_{2}\left[ \varphi \right]
\right) d\tau \right] D\psi D\varphi ,  \label{f2}
\end{equation}
where
\begin{equation}
L_{0}\left[ \psi \right] =\sum_{i}L_{0i}\left[ \psi \right] ,
\label{f3}
\end{equation}
with $L_{0i}\left[ \psi \right] $ from Eq.~(\ref{e25c}). The
tunneling term $L_{1}\left[ \psi ,\varphi \right] $ is
\begin{equation}
L_{1}[\psi ,\varphi ]=\sum_{ij}\int_{0}^{\beta }d\tau
t_{ij;pq}\psi _{ip}^{\ast }(\tau )\psi _{jq}(\tau )\exp \left[
i\varphi _{ij}\left( \tau \right) \right] ,  \label{f4}
\end{equation}
and the term $L_{2}\left[ \varphi \right] $ describing the
charging effects
reads%
\begin{equation}
L_{2}[\varphi ]=\sum_{ij}\frac{C_{ij}}{2e^{2}}\frac{d\varphi
_{i}(\tau )}{d\tau }\frac{d\varphi _{j}(\tau )}{d\tau }.
\label{f5}
\end{equation}
One should integrate in Eq.~(\ref{f2}) over the anticommuting
variables $\psi $ with the antiperiodicity condition
Eq.~(\ref{e28}). The integration over $\varphi $ includes the
integration over the variable $\phi _{i}$ (see Eq.~(\ref{a5})) and
summation over $k_{i}.$ Equations~(\ref{f2}- \ref{f5}, \ref{e30},
\ref{a5}) completely specify the model that will be studied in the
subsequent sections.

\subsubsection{Ambegaokar-Eckern-Sch{\"o}n functional}

\label{phaseaction}

The partition function $Z,$~Eqs.~(\ref{f2} - \ref{f5}), can be
further simplified by integration over the fermion fields $\psi $.
This can hardly be done exactly but one can use a cumulant
expansion in the tunneling term $L_{1}[\psi ,\varphi ]$,
Eq.~(\ref{f4}). Of course, even in the absence of the tunneling
the random potential in $L_{0}\left[ \psi \right] $,
Eqs.~(\ref{f3}, \ref{e25c}), can make the problem highly
non-trivial. Yet in the limit of not very low temperatures,
Eq.~(\ref{f1}), the problem can be simplified because of the
possibility of neglecting interference effects and considering the
disorder within the SCBA taking Green functions from
Eq.~(\ref{a15}).  We perform the cumulant expansion in the
tunneling term $S_{1}[\psi ,\varphi ]$ up to the lowest
non-vanishing second order, integrating over fermionic degrees of
freedom and averaging over the tunneling matrix elements with the
help of Eq.~(\ref{e16}).

The resulting action, originally derived in~\citet{AES} for a
system of two weakly coupling superconductors, contains only the
phases $\varphi _{i}\left( \tau \right) $ but no fermionic degrees
of freedom left. Then for the granular metals the partition
function $Z$ can be written as
\begin{equation}
Z=\int \exp \left( -S\right) D\varphi ,\text{
}\hspace{0.5cm}S=S_{c}+S_{t} \label{a9}
\end{equation}
where $S_{c}$ describes the charging energy
\begin{equation}
S_{c}=\frac{1}{2e^{2}}\sum_{ij}\int_{0}^{\beta }d\tau
C_{ij}\frac{d\varphi _{i}\left( \tau \right) }{d\tau
}\frac{d\varphi _{j}\left( \tau \right) }{d\tau },  \label{a10}
\end{equation}
and $S_{t}$ stands for tunneling between the grains
\begin{equation}
S_{t}=\pi g\sum_{\langle ij\rangle }\int_{0}^{\beta }d\tau d\tau
^{\prime }\alpha \left( \tau -\tau ^{\prime }\right) \sin
^{2}\left[ \frac{\varphi _{ij}(\tau )-\varphi _{ij}(\tau ^{\prime
})}{2}\right] .  \label{a11}
\end{equation}
The function $\alpha \left( \tau -\tau ^{\prime }\right) $ in
Eq.~(\ref{a11}) has the form
\begin{equation}
\alpha \left( \tau -\tau ^{\prime }\right) =T^{2}\mathrm{Re}\sin
^{-2}\left[ \pi T(\tau -\tau ^{\prime }+i\eta )\right] .
\label{f7}
\end{equation}
where $\eta \rightarrow +0$ and one should take the real part in
Eq.~(\ref{f7}).

Despite the fact that the above functional was obtained via an
expansion in the tunneling term $S_{1}[\psi ,\varphi ]$, its
validity is not limited by the insulating  regime,$g\ll 1$, only.
The functional can be used in the metallic regime at temperatures
$T\gg \Gamma ,$ where $\Gamma $ is given by Eq.~(\ref{Gamma}).

Of course, the phase action looses some information about the
original model, Eqs.~(\ref{e25a} - \ref{e25d}), and its
applicability in each particular case has to be carefully
analyzed. In general, the functional $S$, Eqs.~(\ref{a9} -
\ref{f7}) does not apply in cases where the coherent diffusive
motion of an electron on the scale of many grains is important.
For example, the weak localization correction cannot be obtained
within this approach.

Moreover, one has to be careful in using this approach even in the
weak coupling regime, $g\ll 1$. For example, the hopping
conductivity in the low temperature elastic regime is also beyond
the accuracy of the phase action, Eqs.~(\ref{a9}- \ref{f7}), since
it requires consideration of the elastic multiple co-tunneling
processes which the phase action misses.

Yet the phase functional approach is extremely powerful for many
applications. For example, it enables obtaining non-perturbative
results for the conductivity in the metallic regime at
temperatures $T\gg \Gamma.$

\subsection{Metallic properties of granular arrays at not very low
temperatures}

\label{not.low}

In this section, we will derive the logarithmic correction to the
conductivity, Eq.~(\ref{mainresult3}), in the temperature regime
$T\gg \Gamma $. While this correction has the same origin as the
Althsuler-Aronov (AA) result derived for the homogeneously
disordered metals~\cite{Altshuler85,Lee_review}, its form is
specific to granular metals because the AA correction. In contrast
to Eq.~(\ref{mainresult3}), AA corrections are sensitive to the
sample dimensionally and exhibit the logarithmic behavior only in
two dimensions.

The logarithmic correction, Eq.~(\ref{mainresult3}), as we will
show below, can be obtained as a result of the renormalization of
the tunneling coupling $g$ due to the Coulomb correlations. It
comes from the short distances, and this explains its
insensitivity to the sample dimensionally. At temperatures smaller
than $\Gamma $ the coherent electron motion on the scales of many
grains becomes important, and the conductivity correction acquires
the form similar to that of the homogeneously disordered metals.

For the sake of simplicity we will consider a periodic system
assuming that the granular array forms a cubic lattice. This
implies that all the grain sizes and intergranular conductances
are equal to each other. At the end of this section we will
discuss the applicability of the result (\ref{mainresult3}) to
systems containing irregularities. We will start with perturbation
theory in $1/g$, then generalize our derivation of
Eq.~(\ref{mainresult3}) using the renormalization group approach.

\subsubsection{Perturbation theory}

\label{Perturbation_theory}

In the limit of large tunneling conductances $g\gg 1$ the
tunneling term $S_{t}$, Eq.~(\ref{a11}), suppresses large
fluctuations of the phase $\varphi (\tau )$. In this regime the
tunneling term $S_{t}$ can be expanded in phases $\varphi
_{ij}(\tau )$ because they are small. The charge quantization
effects in this regime are not pronounced allowing to neglect all
non-zero winding numbers $k_{i}$. At the same time, the phase
fluctuations can change considerably the classical result,
Eq.~(\ref{conductivity0}).

\begin{figure}[tbp]
\includegraphics[width=2.5in]{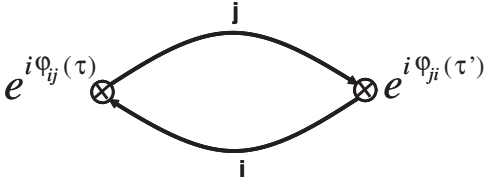} \vspace{-0.3cm}
\caption{ This diagram represents the conductivity in the leading order in $%
1/g$. The crossed circles represent the tunnelling matrix elements $%
t_{kk^{\prime }}^{ij}\,e^{i\protect\varphi _{ij}(\protect\tau )}$
where phase factors appear from the gauge transformation.}
\label{AES_conductivity}
\end{figure}

The quadratic in $\varphi _{i}(\tau )$ part of the action $S$ in
Eq.~(\ref{a9} - \ref{a11}) will serve as the bare action for the
perturbation theory. Keeping terms of the second order in $\varphi
_{i}(\tau )$ in Eqs.~(\ref{a9} - \ref{a11}) and performing the
Fourier transformation in both the coordinates of the grains and
the imaginary time we reduce the action $S$ in Eq.~(\ref{a9}) to
the form
\begin{equation}
S_{0}=T\sum_{\mathbf{q},n}\varphi _{\mathbf{q},n}G_{\mathbf{q}
,n}^{-1}\varphi _{-\mathbf{q},-n},  \label{a13}
\end{equation}%
where
\begin{equation}  \label{Green_fluct}
G_{\mathbf{q}n}^{-1}=\omega _{n}^{2}/4E(\mathbf{q})+2g\left\vert
\omega _{n}\right\vert \lambda _{\mathbf{q}}.  \label{a13a}
\end{equation}
where $\lambda _{\mathbf{q}}$ is given by Eq.~(\ref{e240}). Here
$\mathbf{q}$ are the quasi-momenta of a periodic array,
$E(\mathbf{q})$ is the Fourier transform of the charging energy
related to the Fourier transform of the capacitance matrix
$C\left( \mathbf{q}\right) $ as $E(\mathbf{q})=e^{2}/2C(
\mathbf{q}).$ The conductivity $\sigma \left( \omega \right) $ of
the array of the grains can be obtained from the current-current
correlation function via the standard Kubo formula. The
conductivity in the leading order in $1/g$ is shown in
Fig.~\ref{AES_conductivity}. Its analytical expression is given
by~\cite{Efetov02b}

\begin{equation}
\sigma (\omega )={\frac{{2\pi e^{2}gT^{2}ia^{2-d}}}{{\omega }}}
\,\int_{0}^{\beta }d\tau \,{\frac{{1-e^{i\Omega \tau }}}{{\sin
^{2}(\pi T\tau )}}\exp }\left( -\tilde{G}_{\mathbf{a}}\left( \tau
\right) \right) , \label{conductivity1}
\end{equation}
where the analytic continuation to the real frequency is assumed
as $\Omega \rightarrow -i\omega $ and the function
$\tilde{G}_{\mathbf{a}}\left( \tau \right) $ is
\begin{equation}
\tilde{G}_{\mathbf{a}}\left( \tau \right) =4Ta^{d}\sum_{\omega
_{n}>0}\int \frac{d^{d}\mathbf{q}}{\left( 2\pi \right)
^{d}}G_{\mathbf{q}n}\sin ^{2}\frac{\mathbf{qa}}{2}\sin
^{2}\frac{\omega _{n}\tau }{2}. \label{a14}
\end{equation}
where $d$ is the dimensionality of the array. The factor ${\exp
}\left( -\tilde{G}_{\mathbf{a}}\left( \tau \right) \right) $ in
Eq.~(\ref{conductivity1}) is responsible for the interaction
effects and appears as a result of the averaging of the phase
exponents $e^{i\phi(\tau) }$ emerging after the gauge
transformation, Eq.~(\ref{a4}). One can see from Eqs.~(\ref{a13a},
\ref{a14} ) that the function $\tilde{G}_{\mathbf{a} }\left( \tau
\right) $ behaves as a logarithm $\ln \left( gE_{c}\tau \right) $
at the values of $\tau $ of the order $1/T$ that are essential for
calculation. With the logarithmic accuracy one can neglect the
$\omega _{n}^{2}$ term in $G_{\mathbf{q,}n}^{-1}$,
Eq.~(\ref{a13a}), and reduce Eq.~(\ref{a14}) to
\begin{equation}
\tilde{G}_{\mathbf{a}}\left( \tau \right)
=\frac{T}{dg}\sum_{\omega _{n}>0}^{\omega _{c}}\frac{1-\cos \left(
\omega _{n}\tau \right) }{\omega _{n}}.  \label{b12}
\end{equation}
In Eq.~(\ref{b12}) one should sum over positive Matsubara
frequencies up to the cutoff $\omega _{c}\sim gE_{c}$. For larger
frequencies the first term in Eq.~(\ref{a13a}) is no longer small
and there is no logarithmic contribution from these large
frequencies. Equation~(\ref{b12}) shows a remarkable independence
of the result on the structure of the lattice. What is also
important, there are no \textquotedblleft
infrared\textquotedblright\ divergencies in the integral over
$\mathbf{q}$ in any dimensionality including $2d$ and $1d$.

With the logarithmic accuracy, one can replace $\tau $ by $1/T$ in
the function $\tilde{G}_{\mathbf{a}}(\tau)$ and calculate the
remaining integral over $\tau $ in Eq.~(\ref{conductivity1})
ignoring the dependence of the function $\tilde{G}_{\mathbf{a}}$
on $\tau $. Taking the limit $\omega \rightarrow 0$ we obtain the
result (\ref{mainresult3}) in the temperature interval $T\gg
\Gamma :$
\begin{equation}
\sigma =\sigma _{0}\left( 1-\frac{1}{2\pi dg}\ln \left[
\frac{gE_{c}}{T}\right] \right) .  \label{b18}
\end{equation}
It was shown in~\citet{Efetov02b} that the terms of the order $
(1/g)^{2}\ln ^{2}\left( gE_{c}/T\right) $ are cancelled out in the
expansion of the conductivity correction, which means that the
accuracy of Eq.~(\ref{b18}) exceeds the accuracy of the first
order correction. Furthermore, this cancellation is not accidental
and the result~(\ref{b18}) turns out to be applicable even at
temperatures at which the conductivity correction becomes of the
same order as $\sigma _{0}$ itself. This fact is explained in the
next subsection where the temperature dependence of the
conductivity is considered within the RG scheme.

\subsubsection{Renormalization Group}

\label{RG}

In order to sum up all the logarithmic corrections to the
conductivity we use RG scheme suggested for a one-dimensional
$XY$-model long ago~\cite{kosterlitz} and used later in a number
of works~\cite{bulgadaev85,guinea,falci}. As the starting
functional we take the tunneling action $S_{t}$
\begin{equation}
S_{t}=\pi g\sum_{\langle i,j\rangle }\int_{0}^{\beta
}\int_{0}^{\beta }d\tau d\tau ^{\prime }\alpha \left( \tau -\tau
^{\prime }\right) \sin ^{2}\left[ \frac{\varphi _{ij}\left( \tau
\right) -\varphi _{ij}\left( \tau ^{\prime }\right) }{2}\right] .
\label{b19}
\end{equation}
The charging part $S_{c}$ is not important for the renormalization
group because it determines only the upper cutoff of integrations
over frequencies. In the limit $T\rightarrow 0$ the function
$\alpha (\tau -\tau ^{\prime })$ is proportional to $\left( \tau
-\tau ^{\prime }\right) ^{-2}$ and the action is dimensionless. We
neglect in this subsection the non-zero winding numbers $k_{i}$
and replace the phases $\varphi _{i}\left( \tau \right) $ by the
variables $\phi _{i}\left( \tau \right) $, Eq.~(\ref{a5}).

Following standard RG arguments we want to find how the form of
the action $S_{t}$ changes under changing the cutoff. Generally
speaking, it is not guaranteed that after integrating over the
phases $\phi $ in an interval of the frequencies one comes to the
same function $\sin ^{2}\phi $ in the action. The form of the
functional may change, which would lead to a functional
renormalization group.

In the present case appearance of terms $\sin ^{2}2\phi $, $\sin
^{2}4\phi $, etc. is not excluded and, indeed, they are generated
in many loop approximations of the RG. Fortunately, the one loop
approximation is simpler and the renormalization in this order
results in a change of the effective coupling constant $g$ only.

To derive the RG equation we represent the phase $\phi $ in the
form
\begin{equation}
\phi _{ij\omega }=\phi _{ij\omega }^{\left( 0\right)
}+\overline{\phi }_{ij\omega },  \label{b20}
\end{equation}
where the function $\phi _{ij\mathbf{\omega }}^{(0)}$ is the slow
variable and it is not equal to zero in an interval of the
frequencies $0<\omega <\lambda \omega _{c}$, while the function
$\overline{\phi }_{ij\omega }$ is finite in the interval $\
\lambda \omega _{c}<\omega <\omega _{c}$, where $\lambda $ is in
the interval $0<\lambda <1$. Substituting Eq.~(\ref{b20}) into
Eq.~(\ref{b19}) we expand the action $S_{t}$ up to terms quadratic
in $\overline{\phi }_{ij\omega }$. Integrating in the expression
for the partition function
\[
Z=\int \exp \left( -S_{t}\left[ \phi \right] \right) D\phi
\]%
over the fast variable $\overline{\phi }_{ij\omega }$ \ with the
logarithmic accuracy we come to a new renormalized effective
action $\tilde{S}_{t}$
\begin{eqnarray}
\tilde{S}_{t} &=&2\pi g\sum_{\langle i,j\rangle }\int_{0}^{\beta
}\int_{0}^{\beta }d\tau d\tau ^{\prime }\alpha \left( \tau -\tau
^{\prime
}\right)  \nonumber \\
&\times &\sin ^{2}\left[ \frac{\phi _{ij}\left( \tau \right) -\phi
_{ij}\left( \tau ^{\prime }\right) }{2}\right] \left( 1-\frac{\xi
}{2\pi gd}\right) ,  \label{b21}
\end{eqnarray}
where $\xi =-\ln \lambda $. It follows from Eq.~(\ref{b21}) that
the form of the action is reproduced for any dimensionality $d$ of
the lattice of the grains. This allows us to write immediately the
following renormalization group equation
\begin{equation}
\frac{\partial g\left( \xi \right) }{\partial \xi }=-\frac{1}{2\pi
d}. \label{b22}
\end{equation}
The solution of Eq.~(\ref{b22}) is simple. Neglecting the Coulomb
interaction in the action $S_{t}$, Eq.~(\ref{b19}), is justified
only for energies smaller than $gE_{c}$ and this gives the upper
cutoff. Then, the renormalized conductance $g\left( T\right) $
takes the form
\begin{equation}
g\left( T\right) =g-\frac{1}{2\pi d}\ln \frac{gE_{c}}{T}.
\label{b23}
\end{equation}
and we come using Eq. (\ref{conductivity0}) to Eq.~(\ref{b18}) for
the conductivity. Both the quantities depend on the temperature
logarithmically.

Equation~(\ref{b23}) is obtained in the one loop approximation and
should be valid so long as the effective conductance $g\left(
T\right) $ remains much larger than unity. This assumes that the
result of the perturbation theory, Eq.~(\ref{b18}) can be extended
to a wider temperate interval and, in fact, it is valid as long as
\begin{equation}
g-\left( 2\pi d\right) ^{-1}\ln \left( gE_{c}/T\right) \gg 1.
\label{b24}
\end{equation}
At high temperatures, when the inequality (\ref{b24}) is
fulfilled, one can speak of metallic behavior of the system. At
lower temperatures $T<T_{c}$, where
\begin{equation}
T_{c}=gE_{c}\exp \left( -2\pi gd\right),  \label{b24a}
\end{equation}
the system is expected to show insulating properties.

The result for the conductivity correction~(\ref{b18}) was
obtained within a cubic lattice model. It is important to
understand how it will change in a more realistic case of an
irregular array. First, we note that Eq.~(\ref{b18}) can easily be
generalized to the case of an arbitrary periodic lattice with the
result
\begin{equation}
\sigma =\sigma _{0}\left( 1-\frac{1}{\pi zg}\ln \left[
\frac{gE_{c}}{T}\right] \right) ,  \label{cond_general}
\end{equation}
where $z$ is the coordination number of the arbitrary periodic
lattice.

The role of the dispersion of the tunneling conductance was
studied in~\citet{Feigelman05} within the $2d$ model that assumed
regular periodic positions and equal sizes of grains but random
tunneling conductances. It was shown that the
dependence~(\ref{cond_general}) holds in a wide temperature range
in the case of a moderately strong conductance dispersion.
However, the distribution of the conductances broadens and this
effect becomes especially important close to the metal-insulator
transition where it was suggested to describe the transition in
terms of percolation,~\cite{Feigelman05}. One may expect that in
$3d$ samples the effect of the conductance dispersion on the
macroscopic transport will be less important than in $2d$ ones,
while, on the contrary, it will be more dramatic in $1d$ samples.

Finally, we would like to comment on the validity of
Eq.~(\ref{cond_general}) in the case of an arbitrary irregular
array. First, we note that in the RG approach the capacitance
disorder leads to a local renormalization of the tunneling
conductances. Then, the general irregular system can be viewed as
an irregular array of equal size grains with random tunneling
conductances. The main difference of such a system from that
considered in~\cite{Feigelman05} is that the coordination number
of each grain varies from grain to grain. In such a case one
expects that the coefficient in front of the second term in
Eq.~(\ref{cond_general}) will be determined by an effective
coordination number in a system that is expected to be close to
the average number of neighbors of a grain.

Following the RG scheme we did not take into account non-zero
winding numbers $k_{i}$, Eqs.~(\ref{e29} - \ref{a5}). This is a
natural approximation because the contribution of such
configurations should be exponentially small in $g$ ($\sim \exp
(-cg),$ where $c$ is of the order of unity). So long as the
effective conductance $g$ is large in the process of the
renormalization, the contribution of the non-zero winding numbers
$k_{i}$ can be neglected. They become important when $g$ becomes
of the order of unity. This is the region of the transition into
the insulating state and apparently the non-zero winding numbers
play the crucial role in forming this state. (However, the authors
of recent publications~\cite{Altland05,kamenev,Jul'ka} argued that
the non-zero winding numbers might become important in $1d$
samples at temperatures $T^{\ast }$ parametrically exceeding
$T_{c}$, Eq.~(\ref{b24a})).

\subsection{Metallic properties of granular arrays at low temperatures}

\label{very.low}

The phase functional technique may not be used for obtaining the
conductivity corrections at temperatures $T\leq \Gamma $ since in
this regime the coherent electron motion over a large number of
grains, which is missed in the functional approach~\cite{AES},
becomes important. In this section we will derive the conductivity
correction within the diagrammatic perturbation theory described
in the Sec.~\ref{Perturbation_theory}. Unlike the phase functional
approach, this technique does not allow us to obtain the non
perturbative results in an easy way but it has an advantage of
being applicable at arbitrary temperatures. In particular, we will
see below that it reproduces at $T\gg \Gamma $ the perturbative
result, Eq.~(\ref{b18}). As we will show, the conductivity
correction agrees in the low temperature regime with the one
obtained in Refs.~\cite{Altshuler85,Lee_review,Belitz94} for
homogeneously disordered samples
\begin{figure}[t]
\hspace{-0.5cm} \includegraphics[width=2.8in]{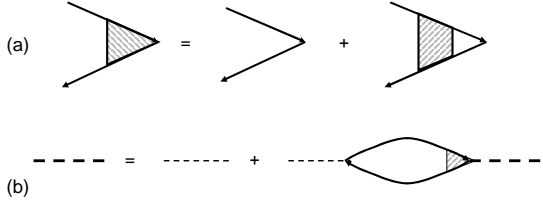}
\caption{ Diagrams representing the vertex correction (a) and the
renormalized Coulomb interaction (b). } \label{Vertex}
\end{figure}

\begin{figure}[tbp]
\hspace{-0.5cm} \includegraphics[width=3.0in]{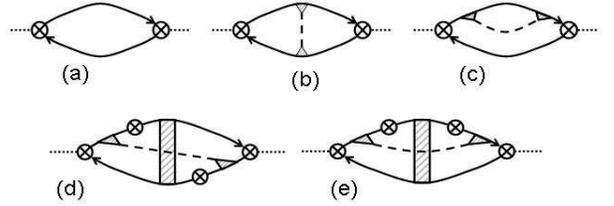}
\caption{Diagrams describing the conductivity of granular metals:
the diagram (a) corresponds to $\protect\sigma _{0}$ in
Eq.~(\protect\ref{mainresult1}) and it is the analog of Drude
conductivity. Diagrams (b)-(e) describing first order correction
to the conductivity of granular metals due to electron-electron
interaction. The solid lines denote the propagator of electrons
and the dashed lines describe effective screened electron-electron
propagator. The sum of the diagrams (b) and (c) results in the
conductivity correction $\protect\delta \protect\sigma _{1}$ in
Eq.~(\protect\ref{mainresult1}). The other two diagrams, (d) and
(e) result in the correction $\protect\delta \protect\sigma _{2}$.
} \label{fig:2}
\end{figure}

In Section~\ref{Perturbation_theory}, we have already described
the main building blocks of the diagrammatic technique including,
in particular, the
diffusion propagator defined by the ladder diagram in Fig.~\ref%
{Diffuson_Diagram}. The same ladder diagram describes the dressing
of the interaction vertex as shown in Fig.~\ref{Vertex}a. The
dressed vertex can be used to obtain the polarization operator
that defines the effective dynamically screened Coulomb
interaction (Fig.~\ref{Vertex}b):
\begin{equation}
V(\Omega ,\mathbf{q})=\left[
{\frac{{C(\mathbf{q})}}{{e^{2}}}}+{\frac{{\ 2\lambda
_{\mathbf{q}}}}{{|\Omega |+\lambda _{\mathbf{q}}\Gamma }}}\right]
^{-1}.  \label{f8}
\end{equation}%
The conductivity of the granular metals is given by the analytical
continuation of the Matsubara current-current correlation
function. In the absence of the electron-electron interaction the
conductivity is represented by the diagram (a) in Fig.~\ref{fig:2}
that results in high temperature (Drude) conductivity $\sigma
_{0}$ defined in Eq.~(\ref{conductivity0}). First order
interaction corrections to the conductivity are given by the
diagrams (b-e) in Fig.~\ref{fig:2}. These diagrams are analogous
to those considered in~\citet{Altshuler85} for the correction to
the conductivity of homogeneous metals.

We consider the contributions from diagrams (b,c) and (d,e)
separately: The sum of the diagrams (b,c) results in the following
correction to the conductivity
\begin{equation}
\frac{\delta \sigma _{1}}{\sigma _{0}}=-{\frac{1}{{2\pi
dg}}}\mathrm{Im} \sum_{\mathbf{q}}\int d\omega \,\gamma (\omega
)\,\lambda _{\mathbf{q}}\,\, \tilde{V}(\omega ,\mathbf{q}),
\label{Diagrams12}
\end{equation}
where $\gamma (\omega )={\frac{{d}}{{d\omega }}}\omega \coth
{\frac{{\omega }}{{2T}}},$ and the potential $\tilde{V}(\omega
,\mathbf{q})$ is the analytic continuation of the screened Coulomb
potential $V(\Omega ,\mathbf{q})$ with dressed interaction
vertices including those attached at the both ends
\begin{equation}
\tilde{V}(\omega ,\mathbf{q})={\frac{{\
2\,E_{c}(\mathbf{q})}}{{(\lambda _{ \mathbf{q}}\Gamma -i\omega
)\,(4\,\lambda _{\mathbf{q}}E_{c}(\mathbf{q})-i\omega )}}}.
\label{effectivinteraction}
\end{equation}%
The expression for the screened Coulomb interaction
$\tilde{V}(\omega ,\mathbf{q})$, Eq.~(\ref{effectivinteraction}),
is written in a simplified form using the fact that the charging
energy $E_{c}(\mathbf{q})=e^{2}/2C(\mathbf{q}),$ expressed in
terms of the Fourier transform of the capacitance matrix
$C(\mathbf{q})$, is much larger than the escape rate $\Gamma .$

Performing the integration over the frequency and summing over the
quasimomentum $\mathbf{q}$ in Eq.~(\ref{Diagrams12}) with the
logarithmic accuracy we obtain the correction to the conductivity,
Eq.~(\ref{mainresult3}). One can see from Eq.~(\ref{Diagrams12})
that the contribution $\delta \sigma _{1}$ in
Eq.~(\ref{mainresult3}) comes from the large energy scales, $
\varepsilon \geq \Gamma $. At low temperatures $T\leq \Gamma $,
the logarithm is cut off by the energy $\Gamma $ and is no longer
temperature dependent.

To obtain the total correction to the conductivity of a granular
sample the two other diagrams, (d) and (e) in Fig.~\ref{fig:2}
should be taken into account. These diagrams result in the
following contribution to the conductivity~\cite{Beloborodov03}
\begin{equation}
{\frac{{\delta \sigma _{2}}}{{\ \sigma _{0}}}}=-{\frac{{\ 2g\delta
}}{{\ \pi d}}}\sum_{\mathbf{q}}\int d\omega \,\gamma (\omega
)\,{\text{Im}\frac{{\ \tilde{V}(\omega ,\mathbf{{q})\sum_{a}}}\sin
{\mathbf{^{2}({q}a)}}}{\lambda{_{\mathbf{q}}\Gamma -i\omega }}}.
\label{sigma2}
\end{equation}
In contrast to the contribution $\delta \sigma _{1}$,
Eq.~(\ref{Diagrams12}), the main contribution to the sum over the
quasimomentum $\mathbf{q}$ in Eq.~(\ref{sigma2}) comes from the
low momenta, $q\ll 1/a$. In this regime the capacitance matrix
$C(\mathbf{q)}$ in Eqs.~(\ref{effectivinteraction}) and
(\ref{sigma2}) has the asymptotic form
\begin{equation}
C^{-1}(\mathbf{q})=\frac{2}{a^{d}}\,\left\{
\begin{array}{rl}
& \ln (1/qa)\hspace{0.7cm}d=1, \\
& \pi /q\hspace{1.4cm}d=2, \\
& 2\pi /q^{2}\hspace{1.1cm}d=3.
\end{array}
\right.  \label{capacitance}
\end{equation}
Using Eqs.~(\ref{effectivinteraction} - \ref{capacitance}), we
obtain the result for the correction $\delta \sigma _{2}$ in
Eq.~(\ref{mainresult44}). This correction has a physical meaning
similar to that of the Altshuler-Aronov
correction~\cite{Altshuler85} derived for homogeneously disordered
metals.

Comparing results in Eqs.~(\ref{mainresult1}) -
(\ref{mainresult44}) with those obtained in the previous
subsection using the AES functional one can see that the
correction to the conductivity obtained in Sec.~\ref{not.low} is
equivalent to the correction $\delta \sigma _{1}$ in Eq.~(\ref
{mainresult1}), which corresponds in the diagrammatic approach to
the sum of diagrams (b) and (c) in Fig.~\ref{fig:2}. The
correction $\delta \sigma _{2}$ in Eq.~(\ref{mainresult1}) becomes
important only at low temperatures, $T\leq \Gamma $ where AES
functional is not applicable.

Using the final results of the calculations,
Eqs.~(\ref{mainresult3}) and (\ref{mainresult44}), we can make a
very important statement about the existence of a metal-insulator
transition in three dimensions. It follows from
Eq.~(\ref{mainresult44}) that for a $3d$ granular array there are
no essential corrections to the conductivity at low temperatures
$T\ll \Gamma $ coming from the low energies $\omega \leq \Gamma $,
since the correction $\delta \sigma _{2}$ is always small. This
means that the result for the renormalized conductance,
Eq.~(\ref{b23}), for $3d$ samples can be written within the
logarithmic accuracy in the form
\begin{equation}
\tilde{g}(T)=g-{\frac{{1}}{{6\pi }}}\ln \left[
{\frac{{gE_{C}}}{\max {(\tilde{g}\delta ,T)}}}\right] ,
\label{RGgen}
\end{equation}
such that it is valid for \textit{all} temperatures as long as the
renormalized conductance is large, $\tilde{g}(T)\gg 1$.

One can see from Eq.~(\ref{RGgen}) that for a large bare
conductance, $g\gg (1/6\pi )\ln (gE_{C}/\delta )$, the
renormalized conductance $\tilde{g}$ is always large and the
system remains metallic down to zero temperatures. In the opposite
limit $g<(1/6\pi )\ln (gE_{C}/\delta )$, the system flows when
decreasing the temperature to the strong coupling regime,
$\tilde{g}\sim 1,$ which indicates the onset of the insulating
phase. One can see that with the logarithmic accuracy the critical
value of the conductance $g_{c}$ is given by Eq.~(\ref{gC}), and
this value separates the metallic and insulating
states~\cite{Beloborodov03}.

One can see from Eq.~(\ref{mainresult44}) that in $1d$ and $2d$
samples the correction to the conductivity, being negative, grows
with decreasing the temperature and diverges in the limit
$T\rightarrow 0$. Such a behavior is usually attributed to a
localization in the limit $T\rightarrow 0$.

However, the situation is more interesting. In a recent
paper~\cite{basko}, it was demonstrated that there must be a
metal-insulator transition at finite temperature $T_{k}$ in
systems with a weak repulsion provided all one-particle states are
localized. This means that the conductivity is strictly zero at
$T<T_{k}$ and becomes finite at $T>T_{k}.$ The $1d$ and $2d$
granular systems with the Coulomb interaction discussed in the
present review should belong to the class of the models considered
in~\cite{basko} (all one-particle states should be localized in
$1d$ and $2d$ systems for any disorder) and one can expect such a
transition. Clearly, the results of Eq.~(\ref{mainresult44})
should hold for $T\gg T_{k}$.

At the same time, the $3d$ samples do not fall into the class of
the models studied by~\citet{basko} because there exist both
localized and extended one-particle states. Therefore, for $3d$
granular systems, one can speak of the metal-insulator transition
at $T=0$ only. The transition point can be varied by changing,
e.g. the tunneling conductance between the grains or their size.

Finally, we would like to note that the low temperature
conductivity correction $\delta \sigma _{2}$ is less sensitive to
a particular model of a granular sample than the high temperature
result, Eq.~(\ref{b18}). As we will discus in the next section,
the low temperature conductivity correction is determined by the
scales larger than the single grain size and can be expressed in
terms of the effective diffusion coefficient $D_{eff}$. This
assumes that the underlying structure of the model is not
important in this regime.

\subsection{Universal description of granular materials}

\label{universal}

One can see from the previous subsection that at low temperatures,
$T\ll \Gamma $, the dependence of the correction $\delta\sigma_2$,
Eq.~(\ref{mainresult44}), to the conductivity of granular metals
coincides exactly with the corresponding result for the
conductivity of the homogeneously disordered samples. A question
immediately arises: is it an accidental coincidence that the two
different physical systems exhibit in the main approximation the
identical low temperature transport behavior, or there is an
underlying deep connection between the two? Can on describe both
the region $T\leq \Gamma $ and $T\geq \Gamma $ in an unified
manner? The main goal of this subsection is to answer these
important questions.

A very convenient description of the low temperature behavior of
homogeneously disordered metals is based on the so-called $\sigma
$-model. There are several formulations of the $\sigma $-model
based on the replica trick~\cite{Wegner79,Efetov80},
supersymmetry~\cite{efetov83,efetov}, and Keldysh Green
functions~\cite{Kamenev99}. The approach of~\citet{Efetov80} has
been generalized by~\citet{Finkelstain_review} to include the
interaction effects. It is clear that the $\sigma $-model
of~\citet{Finkelstain_review} cannot be used for the granular
materials because it does not contain the scale $\Gamma $.

Fortunately, the approach of~\citet{Finkelstain_review} can rather
easily extended to include the granularity. This can be done using
Eqs.~(\ref{a02}) - (\ref{Coulomb__Interacation}) and following the
usual way~\cite{Beloborodov01,Andreev04}: 1) we write a generating
functional in terms of functional integrals over anticommuting
Grassmann variables, 2) average over the disorder using a replica
trick, 3) we decouple a $\psi ^{4}$ effective interaction
appearing after the averaging by a Gaussian integration over a
$Q$- matrix field and do the same for the Coulomb interaction term
in Eq.~(\ref{Coulomb__Interacation}) using axillary fields
$\bar{V}$~\cite{Efetov80,Finkelstain_review}, 4) we integrate thus
obtained exponential of a quadratic form over the $\psi $ fields
and finally, 5) find a saddle point in the free energy functional
containing the fields $Q$ and $\bar{V}$ only and expand around
this saddle point. The final expression for the effective
low-energy action reads:
\begin{eqnarray}
S\left[ Q,V\right] &=&-\frac{\pi }{2\delta
}\sum\limits_{i}\mathrm{Tr}\left[
({\hat{\varepsilon}} + \bar{V}_{i})Q_{i}\right]  \label{action} \\
&&-{\frac{{\pi g}}{8}}\sum_{\langle i,j\rangle
}\mathrm{Tr}[Q_{i}Q_{j}]+ \frac{1}{2e^{2}}\sum_{i,j}C_{ij}\left(
\bar{V}_{i}\,\bar{V}_{j}\right) .  \nonumber
\end{eqnarray}
Here the sums are performed over the grain indices, the symbol
$\langle ...\rangle $ means summation over the nearest neighbors,
$C_{ij}$ is the capacitance matrix, $\hat{\varepsilon}=i\partial
_{\tau }$ and the symbol \textquotedblleft
$\mathrm{Tr}$\textquotedblright\ means the trace over spin and
replica indices and integration over $\tau $. The field $\bar{V}$
in Eq.~(\ref{action}) is a time dependent vector in the replica
space and the corresponding \textquotedblleft scalar
product\textquotedblright\ is implied: $\left(
\bar{V}_{i}\bar{V}_{j}\right) =\sum_{\alpha
}\int_{0}^{1/T}\bar{V}_{i\alpha }\left( \tau \right)
\,\bar{V}_{j\alpha }\left( \tau \right) d\tau $, where $ \alpha $
is the replica index ($0\leq \alpha \leq N$). The $Q$- matrix in
Eq.~(\ref{action}) is the matrix in the time (it depends on two
times $\tau $ and $\tau ^{\prime }$), spin, and replica spaces
subject to the constraints $ Q^{2}=1$, $\mathrm{Tr}Q=0$. The
variable $\tau $ enters $Q$ in the same way as the replica and
spin ones, which means that $\left( Q_{1}Q_{2}\right) _{\tau ,\tau
^{\prime }}\equiv \int Q_{1}\left( \tau ,\tau ^{\prime \prime
}\right) Q_{2}\left( \tau ^{\prime \prime },\tau ^{\prime }\right)
d\tau ^{\prime \prime }$.

For energies smaller than the Thouless energy $E_{Th}$ of one
grain, Eq.~( \ref{thouless}), the $Q$-matrices in
Eq.~(\ref{action}) are coordinate independent within each grain.
They can be written in the form
\begin{equation}
Q_{i}=U_{i}\Lambda U_{i}^{-1},  \label{f9}
\end{equation}
where the function $\Lambda $ equals
\begin{equation}
\Lambda _{\tau ,\tau ^{\prime }}=\frac{iT}{\sin \pi T\left( \tau
-\tau ^{\prime }\right) },  \label{f10}
\end{equation}
and $U_{i}$ is an unitary matrix. The action $S[Q,\bar{V}]$ in
Eq.~(\ref{action}) describes the entire region of both low, $T\leq
\Gamma ,$ and not very low, $T\geq \Gamma,$ temperatures discussed
in the previous subsections. For $T\gg \Gamma ,$ an essential
contribution comes from $U\left( \tau ,\tau ^{\prime }\right) $ in
Eq.~(\ref{f9}) diagonal in time, spin and replica spaces. This
means that we write this function as
\begin{equation}
U_{i}\left( \tau ,\tau ^{\prime }\right) =\delta _{\alpha \beta
}\delta \left( \tau -\tau ^{\prime }\right) \exp \left[ i\varphi
_{\alpha i}\left( \tau \right) \right],  \label{f11}
\end{equation}
where $\alpha $\textbf{,}$\beta $ stand for both replica and spin
indices.

Substituting Eqs.~(\ref{f9}-\ref{f11}) into Eq.~(\ref{action}),
calculating the integral
\[
\int \exp \left( -S\left[ Q,\bar{V}\right] \right) D\bar{V}
\]%
and taking the limit $N=0$, which is trivial in this limit, we
come to the AES action, Eqs.~(\ref{a9}, \ref{f7}).

The $\sigma $-model, Eq.~(\ref{action}), has a well-defined
continuum limit. It can be obtained for slow spatial variations of
the $Q$-matrix by expanding the second term in Eq.~(\ref{action})
to the second order in gradients of $Q$, and replacing the
summation over $i$ by the integration over $\mathbf{r}$. The
tunneling term in Eq.~(\ref{action}) describes in the continuum
limit the diffusion and we reduce the action, Eq.~(\ref{action}),
to the form
\begin{eqnarray}
S &=&-\pi \nu \int \mathrm{Tr}\left[ (\hat{\varepsilon}+ \bar{V}
)Q-\frac{D}{4}
(\nabla Q)^{2}\right] d\mathbf{r}  \nonumber \\
&+&\int \frac{d\mathbf{r}d\mathbf{r^{\prime
}}}{a^{2d}}\mathrm{Tr}\left[ \bar{V}_{
\mathbf{r}}\frac{C_{\mathbf{r}\mathbf{r^{\prime
}}}}{2e^{2}}\bar{V}_{\mathbf{ r^{\prime }}}\right] .
\label{effective}
\end{eqnarray}%
where $\nu $ is the density of states. The coefficient $D$ in the
second term of Eq.~(\ref{effective}) is in this approximation just
the classical diffusion coefficient determined by
Eq.~(\ref{Deffective}).

However, we can derive Eq.~(\ref{effective}) more accurately, not
just neglecting the contribution of energies exceeding $\Gamma $
but taking them into account. Calculating the contribution of the
energies exceeding $\Gamma $ we can use Eq.~(\ref{f11}) for the
excitations with such energies. As a result, we come again to
Eq.~(\ref{effective}) but with a renormalized coefficient $D$ that
can be written in the form
\begin{equation}
D = g_{eff}a^{2}\delta,  \label{f110}
\end{equation}
where $g_{eff}$ equals (c.f. with Eq.~(\ref{b23}))
\begin{equation}
g_{eff}=g-\frac{1}{2\pi d}\ln \frac{E_{c}}{\delta }.  \label{f12}
\end{equation}

Since the effective model~(\ref{effective}) operates with matrix
fields $Q$ that have only long range degrees of freedom, it
applies after an appropriate high energy renormalization to any
disordered metal including a homogeneously disordered one.
Integrating over the potentials $\bar{V}$ one can come to the
$\sigma $-model of~\citet{Finkelstain_review} (the limit of a long
range interaction is implied).

Thus, all the information about the granularity of the sample is
hidden in the low temperature limit in the coefficients of the
effective model~(\ref{effective}). The conductivity of the sample
is related to the effective diffusion coefficient through the
usual Einstein relation
\begin{equation}
\sigma =2e^{2}D(a^{d}\delta )^{-1}.  \label{f13}
\end{equation}%
The effective model, Eq.~(\ref{effective}), together with
equation~(\ref{b23}) for the renormalized conductance naturally
explains the results for the low temperature, $T\ll \Gamma $,
conductivity described in section~\ref{very.low}. The interaction
correction to conductivity consists of two terms: 1) The first
contribution is temperature independent and comes from the high
energy renormalization of the diffusion coefficient $D$,
Eqs.~(\ref{f13}, \ref{sigma1}). It can be written in the form
(c.f. with Eq.~(\ref{mainresult3}))
\begin{equation}
\delta \sigma _{1}=-\sigma _{0}\frac{1}{2\pi dg}\,\ln \left[
\frac{E_{c}}{\delta }\right] ,  \label{sigma1}
\end{equation}
where $\sigma _{0}$ is the classical Drude conductivity defined in
Eq.~(\ref{conductivity0}). This contribution is specific for the
granular materials.

2) The second contribution to conductivity,
Eq.~(\ref{mainresult44}), is temperature dependent and comes from
the low energy renormalization of the diffusion coefficient $D$ in
the effective model~(\ref{effective}). It coincides with the
corresponding correction to conductivity obtained for
homogeneously disordered metals~\cite{Altshuler85}.

Making use of the effective description of the granular metals in
terms of effective the $\sigma $-model, Eq.~(\ref{effective}),
applicable at low temperatures we conclude: all the phenomena that
are described in terms this model including localization effects,
magnetoresistance, Hall conductivity, etc., are universal.

Using the $\sigma $-model, Eq.~(\ref{effective}), we also describe
quantum interference (weak localization) corrections to the
conductivity,~\cite{Khmelnitskii79}. In the leading order in the
inverse tunneling conductance, $1/g$, the interaction and weak
localization corrections can be considered independently. The
quantum interference corrections may be obtained from the
effective $\sigma $-model, Eq.~(\ref{effective}), using directly
the corresponding results for homogeneously disordered
metals~\cite{Efetov80} with the proper effective diffusion
coefficient, $D=ga^{2}\delta $. For $2D$ and $1D$ samples it is
important to take into account dephasing effects since the weak
localization correction diverges. The dephasing time $\tau _{\phi
}$ can also be extracted directly from the results for
homogeneously disordered metals~\cite{Altshuler82} with the proper
effective diffusion coefficient, $D$. The final result for the
weak localization corrections is given by Eqs.~(\ref{wl1}) and
(\ref{wl2}).

\subsection{Insulating Properties of Granular Metals: Periodic model}

\label{insulating}

In the previous section we discussed the transport properties of metallic
granular arrays with strong intergranular tunneling coupling. Now we turn to
the opposite limit of weakly coupled grains, $g\ll 1.$ We start our
consideration with a detailed description of a periodic granular array model
that is presented in this section. To remind, the periodic model assumes
periodic arrangements of equal size grains as well as the absence of the
disorder in conductances and grain potentials. At the same time, the
electron motion inside the grain is chaotic and this brings a disorder into
the model.

As we discussed in the Introduction, the periodic granular array model
predicts the insulating conductivity behavior with a hard gap in the
electron excitation spectrum and, for this reason, it cannot describe the
transport properties of realistic granular systems where the conductivity
governed by the variable range hopping mechanism. Nevertheless, it is
instructive to consider this model for two reasons: first, it is important
to be see explicitly that such a model may indeed produce only the
activation conductivity dependence; second, this model illustrates a general
effect of the charge renormalization of the Coulomb energy due to the
intergranular coupling.

\subsubsection{Activation conductivity behavior}

First, we will consider a periodic granular array with a small
intergranular coupling, $g\ll 1$. The conductivity in this case
can be found via the gauge transformation technique described in
Sec.~\ref{gauge}. The phase action in the lowest order in
tunneling conductance $g$ is simply given by the Coulomb term
$S_{c}$, Eq.~(\ref{a10}). With the help of the Kubo formula the
conductivity in the first nonvanishing order in $g$ may be written
as~\cite{Efetov02b}
\begin{equation}
\sigma (\omega )={\frac{{2\pi e^{2}gT^{2}ia^{2-d}}}{{\omega }}}
\,\int_{0}^{\beta }d\tau \,{\frac{{1-\exp {i\Omega \tau }}}{{\sin
^{2}(\pi T\tau )}}}\;\Pi (\tau ).  \label{p1}
\end{equation}%
Here the analytic continuation to the real frequency is assumed as $\Omega
\rightarrow -i\omega $ and the function $\Pi (\tau )$ represents the phase
correlation function
\begin{equation}
\Pi (\tau _{1}-\tau _{2})=\langle \exp \left( i\left( \varphi \left( \tau
_{1}\right) -\varphi \left( \tau _{2}\right) \right) \right) \rangle ,
\label{p2}
\end{equation}
where the averaging is performed with the Coulomb action $S_{c}$
in Eq.~(\ref{a10}).

At first glance, calculation of the correlation function $\Pi
\left( \tau _{1}-\tau _{2}\right) $ with the action $S_{c}$
reduces to computation of a Gaussian integral. However, it is not
so because at finite temperatures it is necessary to take into
account all winding numbers, Eq.~(\ref{a5}), in order to obtain
the correlation function $\Pi (\tau )$ correctly. At low
temperature the straightforward calculation results
in~\cite{Efetov02b}
\begin{equation}
\sigma =2\sigma _{0}\exp \left( -E_{c}/T\right) ,  \label{activation}
\end{equation}
where $\sigma _{0}$ is the Drude high temperature conductivity and
$ E_{c}$ is the single grain charging energy. This result has a
clear physical meaning: The conductivity of the insulating
granular array is mediated by electrons and holes that are present
in the system as real excitations. Their concentration is given by
the Gibbs law and the factor $2$ appears due to the presence of
both electrons and holes. From the result (\ref{activation}) we
conclude that the activation exponent is determined by the Mott
gap - that is the energy one pays to add or remove an electron
from the system.

One can extend the calculation of the conductivity to higher
orders in $g$ expanding in the tunneling term $S_{t}$,
Eq.~(\ref{a11}). It was shown by~\citet{Loh} that the activation
behavior persists at least in the next order in $g$. Apparently,
the activation behavior, Eq.~(\ref{activation1}) is quite
universal for the periodic arrays and what one should clarify is
the dependence of the Mott gap $\Delta _{M}$ on the conductance
$g.$ Now we turn to this question.

\subsubsection{ Mott gap at small tunneling conductances}

At finite intergranular coupling the Mott gap turns out to be
reduced due to the processes of virtual tunneling of electrons to
the neighboring grains. This effect can be most easily studied in
the limit of the low coupling between the grains within the
straightforward perturbation theory in the tunneling conductance.
The gap $\Delta _{M}$ can be formally defined as
\begin{equation}
\Delta _{M}=E_{N=1}-\mu -E_{N=0},  \label{Delta}
\end{equation}%
where $\mu $ is the chemical potential, $E_{N=0}$ is the ground
state energy of the charge neutral array and $E_{N=1}$ is the
minimal energy of the system with an extra electron added to the
neutral state. The correction to the Mott gap due the to the
electron tunneling to the neighboring grains can be found in the
second order of the perturbation theory
\begin{equation}
\Delta E=-\sum_{k}{\frac{{\ |V_{k,0}|^{2}}}{{\ E_{k}-E_{0}}}},
\label{second_order}
\end{equation}
where the matrix elements of a perturbation $V$ are taken between the ground
state $0$ and excited states $k$.

The correction to the energy due to the finite intergranular
coupling should be included in both the terms $E_{N=1}$ and
$E_{N=0}$ in Eq.~(\ref{Delta}). We consider in the zero
approximation isolated grains and the tunneling Hamiltonian
$\hat{H}_{t},$ Eq.~(\ref{a2a}), is our small perturbation. Using
Eq.~(\ref{second_order}) we have to calculate matrix elements of
the tunneling between neigboring grains. As all grains are
equivalent, we consider the tunneling between the grains $1$ and
$2$ assuming that the charge $N$ on the grain $1$ can be $0$ or
$1,$ while the grain $2$ is initially neutral. Contribution of the
hops between the grain $1$ and all its other neighbors leads
merely to the factor $z$ (coordination number) in the final
result. Using Eq.~(\ref{second_order}) we should calculate the
corrections to the energies $E_{N=0}$ and $E_{N=1}.$

First, let us consider the correction to the ground state energy $E_{N=0}$.
In this case the matrix elements $V_{k,0}$ correspond to electron tunneling
between the initially neutral neighboring grains. The excitation energy of
this process equals $\varepsilon _{k_{1}}+\varepsilon _{k_{2}}+E_{eh},$
where $\varepsilon _{k_{2}}$ is the bare (with no Coulomb energy included)
energy of an electron excitation in grain $2$, $\varepsilon _{k_{1}}$ is the
bare energy of a hole excitation in grain $1$, and $E_{eh}$ is the
electrostatic energy the electron-hole excitation that should be determined
from Eq.~(\ref{Coulomb__Interacation}). It is equal to $%
E_{eh}=E_{11}^{c}+E_{22}^{c}-2E_{12}^{c},$ where $E_{ij}$ can be
obtained from Eq.~(\ref{Coulomb__Interacation}) putting $n_{i}=1$
and $n_{j}=-1$. For the periodic array of the grains under
consideration, the energy $E_{eh}$ reduces to
\begin{equation}
E_{eh}=2E_{c}-2E_{12}^{c}.  \label{electron_hole}
\end{equation}
The energy correction corresponding to such a process is
\begin{equation}
-\Delta E_{N=0}=2\sum_{k_{1},k_{2}}\,{\frac{{\
|t_{k_{1}k_{2}}|^{2}}}{{\varepsilon _{k_{1}}+\varepsilon
_{k_{2}}+E_{eh}}}},  \label{Delta0}
\end{equation}%
where the factor $2$ takes into account the equivalent process of the
electron hopping from grain $2$ to grain $1$.

Analogously, we find the correction to the energy $E_{N=1}$. The
excitation energy of the process of the electron tunneling from
the grain $2$ to the grain $1$ equals $\varepsilon
_{k1}+\varepsilon
_{k_{2}}+3E_{11}^{c}+E_{22}^{c}-4E_{12}^{c}=\varepsilon
_{k_{1}}+\varepsilon _{k_{2}}+2E_{eh}$, while the excitation
energy of the process of electron tunneling from the grain $1$ to
the grain $2$ is $\varepsilon _{k_{1}}+\varepsilon
_{k_{2}}+E_{11}^{c}-E_{22}^{c}=\varepsilon _{k_{1}}+\varepsilon
_{k_{2}}.$ The corresponding correction to the energy $E_{N=1}$
reads
\begin{equation}
-\Delta E_{N=1}=\sum_{k_{1},k_{2}}\,{\frac{{\
|t_{k_{1}k_{2}}|^{2}}}{{\varepsilon _{k_{1}}+\varepsilon
_{k_{2}}+2E_{eh}}}}+{\frac{{\ |t_{k_{1}k_{2}}|^{2}}}{{\varepsilon
_{k_{1}}+\varepsilon _{k_{2}}}}}. \label{Delta1}
\end{equation}
One can see that corrections to the energy levels,
Eqs.~(\ref{Delta0}, \ref{Delta1}), are ultraviolet divergent.
However, their difference is finite and it gives the correction to
the Mott gap, $\Delta_M$. Subtracting Eq.~(\ref{Delta0}) from
Eq.~(\ref{Delta1}), replacing the summation over the states by
integrals over a continuous variable $\varepsilon $, such that
$\varepsilon _{k_{1}(k_{2})}\rightarrow \varepsilon _{1(2)},$ we
can easily caclulate the Mott gap, $\Delta _{M}$. Taking into
account the spin degeneracy we obtain the following expression for
the Mott gap~\cite{Beloborodov05}
\begin{equation}
\Delta _{M}=E_{c}-{\frac{{2z}}{{\pi }}}\,g\,E_{eh}\,\ln 2,
\label{Delta_pert}
\end{equation}
where $z$ is the coordination number of the array of the grains
and $E_{eh}$ is the energy necessary to create an electron-hole
excitation in the system by removing an electron from a given
grain and putting it on a neighboring one. In the case of a
diagonal Coulomb interaction $E_{ij}=E_{c}\,\delta _{ij}$ the
energy of an electron-hole excitation $E_{eh}$ is simply twice the
Coulomb charging energy $E_{c}$.

Derivation of Eq.~(\ref{Delta_pert}) was carried out neglecting
the fact that an extra electron added to the neutral system in the
presence of a finite intergranular coupling can move diffusively
over the sample. Contributions corresponding to such processes are
suppressed by an extra small factor $\delta /E_{c}\ll 1$ and,
thus, can be neglected. More details of the calculations can be
found in Ref.~\cite{Beloborodov05}.

\subsubsection{Mott gap at large tunneling conductances}

From Eq.~(\ref{Delta_pert}) one can see that the Mott gap $\Delta
_{M}$ is significantly reduced at values $gz\sim 1$, where the
perturbation theory becomes inapplicable. Suppression of the gap
$\Delta _{M}$ at large tunneling conductances $g$ can be found
with the help of the phase action, Eqs.~(\ref{a9} - \ref{a11}). In
fact, the renormalization of the charging energy at $gz\gg 1$ can
be obtained with the renormalization group approach that we have
described in subsection~\ref{RG}.

Indeed, the RG Eq.~(\ref{b22}) is written assuming that $g$ is a
function of the independent variable $\xi $. We can invert this
equation and assume that $\xi $ is a function of $g$. The running
variable $\xi $ determines an effective charging energy
$E_{c}^{eff}$ as $\xi =\ln E_{c}^{eff}$ and we obtain solving the
latter equation
\begin{equation}
E_{c}^{eff}\sim A\exp \left( -\pi gz\right)  \label{p3}
\end{equation}
where $A$ is a constant.

In order to determine the Mott gap $\Delta _{M}$ we notice that at
conductances $gz\sim 1$ the charging energy is of the order of the
Mott gap $ \Delta _{M},$ as one can see from
Eq.~(\ref{Delta_pert}). Matching Eq.~(\ref{p3}) with
Eq.~(\ref{Delta_pert}) we conclude that the Mott gap $\Delta _{M}$
at large $gz$ is reduced exponentially and can be written as
\begin{equation}
\Delta _{M}\sim E_{c}\;\exp \left( -\pi gz\right) .
\label{Delta_M}
\end{equation}
The RG approach used to obtain Eq.~(\ref{Delta_M}) is applicable
only at energy scales larger than the inverse escape rate from a
single grains $\Gamma ,$ which means that Eq.~(\ref{Delta_M}) is
valid as long as the Mott gap $\Delta _{M}$ is larger than $\Gamma
.$ In this region the system should be the insulator. In the
opposite limit one can expect that the granular material becomes a
metal.

Taking $\Delta _{M}\sim \Gamma $ and resolving Eq.~(\ref{Delta_M})
with respect to the conductance $g$ we obtain the critical
conductance $g_{c}$, Eq.~(\ref{gC}), that marks the boundary
between the insulating and metallic phases at low temperature.
Thus, we come to the conclusion that the Mott gap $\Delta _{M}$
remains finite as long as $g<g_{c}$, which assumes the activation
behavior, Eq.~(\ref{activation1}). The Arrhenius law,
Eq.~(\ref{activation1}), is typical for crystalline insulators.

\subsubsection{Metal - Insulator transition in periodic granular arrays}

In the previous subsection we have shown that in the case of three
dimensional arrays there is a critical conductance $g_{c},$ such
that the samples with $g < g_{c}$ are insulators at $T\rightarrow
0$, while those with $g > g_{c}$ are metals. The analysis was
performed from both metallic and insulator sides and both the
approaches agree as concerns the value of the critical conductance
$g_{c}$, Eq.~(\ref{gC}). Unfortunately, they are not applicable in
the vicinity of the phase transition and do not allow us to find
the critical behavior of the Mott gap $\Delta _{M}$ near $g_{c}$.

The order of the Mott transition in $3d$ periodic granular array
is also not known, since one cannot exclude a weakly first order
phase transition. For, example the Mott transition in the Hubbard
model in dimensionality $d \geq 3$ is believed to be of the first
order, and since physics of the transition is similar, the same
can be expected for the periodic granular array. We note however
that in spite of the similarities between the periodic granular
array model and the Hubbard model, there are essential differences
between the two. Namely, in granular metals, even in the case of
the periodic samples, the electron motion within a grain as well
as on the scales exceeding the intergranular distance is
diffusive, while in the Hubbard model electrons move through a
periodic lattice that allows to label their states by
quasimomenta.

The model of the granular metals has also an additional physical
parameter $\delta $- the mean energy level spacing in a single
grain, that has no analog in the Hubbard model. Yet, the physics
of the transition looks similar in both the cases, and one may
expect that the methods developed for the study of the Mott
transition in the system of strongly interacting electrons on the
lattice.

In the case of $1d$ and $2d$ granular samples the question of the
Mott transition at $T\rightarrow 0$ cannot be formulated since,
even in the metallic phase, the conductivity corrections are
divergent at low temperatures, see
Eqs.~(\ref{mainresult3})-(\ref{wl2}). However, the critical
conductance $g_{c}$ still has a meaning of the boundary between
the hard gap insulator behavior at $g<g_{c}$ and a weak insulating
behavior $g>g_{c}$ that essentially coincide with insulating
properties of the ordinary disordered metals. We note that the
fate of the ordinary disordered metals with interaction at low
temperature is not quite clear due to difficulties related to the
divergence of the conductivity
corrections~\cite{Finkelstain_review}. In particular, the metal to
insulator transition recently observed in disordered
films~\cite{Abrahams_review,Kravchenko04} has not been yet
theoretically understood, although some plausible scenarios have
been suggested recently (see, e.g. \cite{Finkelsteinscience}).

It is clear that understanding the metal-insulator transition in the
granular materials is one of the most difficult theoretical problems and
considerable efforts are necessary to solve it.

\subsection{Hopping conductivity in Granular Materials}

\label{hopping}

In the previous section we have shown that a strictly periodic
granular model predicts and explains the activation conductivity
behavior only. Experimentally observed temperature behavior of the
conductivity of granular metals and arrays of quantum dots at low
temperatures, however, is usually not of the activation type but
resembles the Efros-Shklovskii law,
Eq.~(\ref{e3}),~\cite{Yakimov03,Yu04,Tran05,Liao05,Romero05}. As
we have already discussed in the introduction, it is crucial for
understanding the hopping transport in granular metals to take
into account an electrostatic disorder that seems to be present in
any realistic system. This disorder can be viewed as a random
potential $V_{i}$ applied to each grain that lifts the Coulomb
blockade on a part of grains in the sample such that the
conductivity is mediated by the electron hopping between the sites
where the Coulomb blockade is almost removed. Below, we consider
in more details the two essential ingredients of the hopping
conductivity - the density of states and the mechanisms of the
electron tunneling through the dense granular system in different
regimes.

\subsubsection{Density of states}

\label{sum_el_cot}

The electrostatic disorder causes fluctuations in the electrostatic energy
of granules and can thus lift the Coulomb blockade at some sites of the
granular sample. This results, in its turn, in the finite density of states
at the Fermi level and makes the variable range hopping the dominating
mechanism of conductivity. The bare density of states induced by the random
potential can be substantially suppressed due to the presence of the
long-rang Coulomb interaction in the same way as it happens in
semiconductors where the density of states (DOS) $\nu _{g}(\varepsilon )$ is
given by the Efros-Shklovskii expression~\cite{Efros,Shklovskii}
\begin{equation}
\nu _{g}(\varepsilon )\sim (\tilde{\kappa}/e^{2})^{d}\;|\varepsilon |^{d-1},
\label{ES_density_of_states}
\end{equation}
with $e$ being the electron charge and $\tilde{\kappa}$ being the
dielectric constant. This result has recently been confirmed
by~\citet{Ioffe,Pankov} analytically using a locator
approximation.

As we have explained in Sec.~\ref{model}, the Coulomb interaction
in the granular matter may be considered writing classical
electrostatic formulae for the electron-electron interaction,
Eq.~(\ref{Coulomb__Interacation}). This approach can also be used
in the limit of a weak coupling between the grains, $g\ll 1$. So,
we describe the Coulomb interaction in the presence of the
electrostatic disorder writing the following expression for the
Hamiltonian $H$
\begin{equation}
H=\sum_{i}V_{i}n_{i}+\sum_{ij}n_{i}\;E_{ij}^{c}n_{j},
\label{classical_model}
\end{equation}
where $V_{i}$ is the random external potential, $n_{i}$ is the
number of excess electrons on the grain $i,$ and $E_{ij}^{c}$ is
the Coulomb interaction between the grains $i$ and $j.$ Taking
into account the asymptotic behavior of the Coulomb interaction
matrix $E_{ij}^{c}\sim e^{2}/2 r_{ij}\tilde{\kappa},$ where
$r_{ij}$ is the distance between the grains $i$ and $j$ and
$\tilde{\kappa}$ is the effective dielectric constant of the
granular sample we see that the classical model
(\ref{classical_model}) for the granular materials is essentially
equivalent to the one studied by Efros and Shklovskii. Therefore
their results have to be applicable to the model of the granular
array, Eq.~(\ref{classical_model}), as well.

At the same time, one expects that the DOS in the array of the metallic
granules is larger than that in a semiconductor since each metallic grain
has a dense electron spectrum. Indeed, one has to remember that there are
many electron states in a grain that correspond to the same grain charge,
while in the model of impurity levels in the semiconductors the charge is
uniquely (up to the spin) identified with the electron state.

The energy of an unoccupied state $\varepsilon _{i}$ in the
model~specified by Eq.~(\ref{classical_model}) is by definition
the energy of an electron placed on this state. In the granular
metal, any state with the energy larger than $\varepsilon _{i}$ is
also available for the electron. Thus, in order to translate the
ES result, Eq.~(\ref{ES_density_of_states}), to the density of
electron spectrum in granular metals one has to integrate the
dependence~(\ref{ES_density_of_states}) over the energy
$\varepsilon $ and multiply it by the bare DOS in a single grain.
As a result, we obtain
\begin{equation}
\nu (\varepsilon )\sim \nu _{0}\;(|\varepsilon |\tilde{\kappa}/\,e^{2})^{d},
\label{Granular_ES}
\end{equation}
where $\nu _{0}$ is the average DOS in a single grain (defined as
the number of states per energy). However, the above expression
cannot be used in the Mott argument for the hopping conductivity
where one needs to estimate the distance to the first available
state $r$ within the energy shell $\varepsilon $ via the relation
\begin{equation}
r^{d}\;\int_{0}^{\varepsilon }d\varepsilon ^{\prime }\nu _{g}(\varepsilon
^{\prime })\sim 1.  \label{est}
\end{equation}
The problem with using the expression for DOS,
Eq.~(\ref{Granular_ES}), in Eq.~(\ref{est}) is that the
expression~(\ref{Granular_ES}) takes into account the fact that if
there is a state available for placement of an electron with a
given energy then typically \textit{on the same grain,} there will
be plenty of other states available for electron placement.
However, for application to the hopping conductivity we should not
count different electron states that belong to the same grain
since it is enough to find at least one state to insure the
transport.

Thus, when finding DOS relevant for the hopping transport, the
lowest energy states within each grain are to be counted only.
Then, we arrive at the conclusion that, although the electron DOS
in granular metals is modified according to
Eq.~(\ref{Granular_ES}), one has to use the ES expression for DOS
in its form~(\ref{ES_density_of_states}) even in granular metals
in order to find the distance to the first available state within
a given energy shell via Eq.~(\ref{est}).

Similar considerations were presented in~\citet{Shklovskii04}.
Following this reference we call the DOS that counts only the
lowest excited states in each grain and that is relevant for the
hopping conductivity \textquotedblleft the density of
\textit{ground } states\textquotedblright . In order to
distinguish this quantity from the conventional density of states
from Eq.~(\ref{Granular_ES}) we ascribe to it the subscript $g$ as
in Eq.~(\ref{ES_density_of_states}).

The presented arguments demonstrate that one can obtain a finite
density of states in a granular material in the insulating regime.
Although this is the necessary ingredient for establishing the
mechanism of the hopping conductivity, it alone is not sufficient
for this type of transport. The problem is that en electron has to
hop over several grains, which, at first glance, does not look
probable in the case of closely packed granular array. For quite a
long time, this fact did not allow to apply the Efros-Shklovskii
theory for explanation of the behavior, Eq.~(\ref{e3}), in the
granular materials and a rather artificial model
of~\citet{Abeles75} was used.

Only recently a resolution of this puzzle has been suggested
independently in publications~\cite{Beloborodov05,Feigelman05}.
The authors of these works demonstrated that a well known
phenomenon of co-tunneling (virtual
hops)~\cite{Averin,Averin91,Averin92} through a grain might be
responsible for the long range hops of the electrons. One has to
distinguish between elastic and inelastic co-tunneling processes.
The elastic co-tunneling is the dominant mechanism for the hopping
conductivity at low temperatures $T < T_{cross}$, while at larger
temperatures $T > T_{cross}$ the electron transport goes via
inelastic co-tunneling processes. The characteristic temperature
$T_{cross}$ of the crossover from the elastic to the inelastic
tunneling is given by the following formula
\begin{equation}
T_{cross} = \bar{c} \sqrt{E_{c}\delta },  \label{crossover}
\end{equation}
where $\bar{c} \approx 0.1$ is the numerical coefficient. In the
next subsections we discuss the elastic and inelastic
contributions to the variable range hopping.

\subsubsection{Hopping via elastic co-tunneling}

\label{sum_el_cot}

Considering the probability for an electron to tunnel from the
site $i_{0}$ to the site $i_{N+1}$ it is convenient to put the
Coulomb interaction energy at these sites to zero and count the
initial, $\xi _{0}$, and final, $\xi _{N+1}$, electron energies
from the Fermi level. The presence of the electrostatic disorder
on the grains is modelled by the random potential $V_{i}.$ The
energy of the electron (hole) excitation on the site $i$ is
\begin{equation}
E_{i}^{\pm }=E_{i}^{c}\pm \mu _{i},  \label{eh}
\end{equation}
where $\mu_i=V_i + 2\sum_j E_c^{ij} n_j$ is the local potential
that along with the bare potential $V_i$ includes the potential
induced by the neighboring grains. The probability $P_{el}$ of a
tunneling process via the elastic cotunneling can most easily be
found for the case of the diagonal Coulomb interaction
$E_{ij}^{c}=E_{i}^{c}\,\delta _{ij}.$ We leave calculational
details for the Appendix A and use here only the final result of
the derivation.

The probability of the elastic cotunneling through $N$ grains can be written
as
\begin{equation}
P_{el}=wg_{0}\,\left( {\frac{\bar{\Gamma}}{{\pi
\bar{E}}}}\right)^{N}\delta (\xi _{N+1}-\xi _{0}),
\label{P_el_result}
\end{equation}
where the factor $w = n(\xi _{0})[1-n(\xi _{N+1})]$ takes into
account the occupations $n(\xi _{0})$ and $n(\xi _{N+1})$ of the
initial and final states respectively (the initial state is filled
and the final one is empty). The bar in Eq.~(\ref{P_el_result})
denotes the geometrical average of the physical quantity along the
tunneling path. For example, the average energy $\bar{\Gamma}$ is
defined as
\begin{equation}
\ln \bar{\Gamma}={\frac{1}{{N+1}}}\,\sum_{k=0}^{N}\,\ln \Gamma _{k},
\label{geoave}
\end{equation}
where the summation extends over the tunneling path, $\Gamma
_{k}=g_{k}\delta _{k}$ and $g_{k}$ is the tunneling conductance
between the $k$-th and $(k+1)$-st grains. [We note that the
meaning of $\Gamma_i$ as an escape rate from a grain does not hold
in the insulating regime under consideration.] Yet, we use this
notation even in the insulating sate.
 The energy $\bar{E}$ is the
geometrical average $\ln
\bar{E}={\frac{1}{{N}}}\,\sum_{k=1}^{N}\,\ln \tilde{E}_{k},$ of
the following combination of the electron and hole excitation
energies
\begin{equation}
\tilde{E}_{k}=2\,\left( 1/E_{k}^{+}+1/E_{k}^{-}\right) ^{-1}.
\label{tilde_E}
\end{equation}
The presence of the delta function in Eq.~(\ref{P_el_result}) reflects the
fact that the tunneling process is elastic.

The form of Eq.~(\ref{P_el_result}) for the tunneling probability
$P_{el}$ corresponds to an independent sequential tunneling from
grain to grain. This equation means that on average the
probability falls off exponentially with the distance $s$ along
the path:
\begin{equation}
P_{el}\sim \exp \left( -2s/\xi _{el}\right),  \label{p5}
\end{equation}
where the dimensionless localization length $\xi _{el}$ can be written as
\begin{equation}
\xi _{el}={\frac{{\ 2}}{{\ \ln (\,\bar{E}\,\pi
/c\bar{g}\delta)}}}. \label{localization}
\end{equation}
Both the distance $s$ and the length $\xi _{el}$ in
Eqs.~(\ref{p5}, \ref{localization}) are measured in the units of
the grain size $a$. The numerical constant $c$ in
Eq.~(\ref{localization}) is equal to unity for the model of the
diagonal Coulomb interaction, $E_{ij}^{c}=E_{c}\,\delta_{ij}.$ We
show in Appendix A that the inclusion of the off-diagonal part of
the Coulomb interaction results in a renormalization of the
constant $c$ to a certain value $0.5\lesssim c<1$.

Applying the conventional Mott-Efros-Shklovskii arguments, i.e. optimizing
the full hopping probability that is proportional to
\begin{equation}
\exp [-(2s/\xi _{el})-(e^{2}/\tilde{\kappa}Tas)],  \label{e100}
\end{equation}%
one obtains Eq.~(\ref{e6}) for the hopping conductivity with the
characteristic temperature
\begin{equation}
T_{0} \sim e^{2}/{\ a\tilde{\kappa}}\,\xi _{el},  \label{T0}
\end{equation}
where $\tilde{\kappa} $ is the effective dielectric constant of a
granular sample, $a$ is the average grain size, and $\xi _{el}$ is
given by Eq.~({\ref{localization}}) (when deriving Eq.~(\ref{T0})
we considered the tunneling path as nearly straight).

In the presence of a strong electric field $\mathcal{E}$, a direct
application of the results of~\citet{Shklovskii73} gives in the
limit of low temperatures
\[
{\frac{{\ T}}{{\ e\xi _{el}a}}}\ll \mathcal{E}\ll {\frac{{\
\sqrt{E_{c}\, \delta }}}{{\ ea}}},
\]
the following expression for the nonlinear current dependence
\begin{equation}
j\sim j_{0}\;\exp \left[
-(\mathcal{E}_{0}/\mathcal{E})^{1/2}\right], \label{current_el}
\end{equation}
where the characteristic electric field $\mathcal{E}_{0}$ is given by the
expression
\begin{equation}
\mathcal{E}_{0}\sim T_{0}/e\,a\xi _{el}.
\end{equation}

In a full analogy with the hopping conductivity in
semiconductors~\cite{Shklovskii}, also in granular metals
inelastic processes are required to allow an electron to tunnel to
a state with a higher energy. In the granular metals such
processes occur due to interaction with phonons as well as due to
inelastic collisions with other electrons. It is clear that these
processes were not considered when deriving
Eq.~(\ref{P_el_result}). Therefore, the results presented in this
subsection are valid as long as the contribution of the inelastic
cotunneling to the hopping conductivity can be neglected. This
corresponds to the limit of low temperatures and electric fields
$T,\mathcal{E}ea\ll T_{cross}.$ Below, we present the results for
the opposite limit when the inelastic processes give the main
contribution to the hopping conductivity.

\subsubsection{Hopping via inelastic co-tunneling}

The inelastic cotunneling is the process when the charge is transferred by
different electrons on each elementary hop. During such a process the energy
transferred to a grain by the incoming electron differs from the energy
taken away by the outgoing one.

As in the case of the elastic cotunneling we assume that the
electron tunnels from the grain $i_{0}$ with the energy $\xi _{0}$
to the grain $ i_{N+1}$ with the energy $\xi _{N+1}.$ For the
probability of such a tunneling process through a chain of the
grains we find the following expression~[see Appendix~B and
Ref.~\cite{Beloborodov05}]
\begin{equation}
P_{in}={\frac{w}{{\ 4\pi T}}}{\frac{{\ \,\bar{g}^{N+1}}}{{\pi
^{N+1}}}}\left[ {\frac{{\ 4\pi T}}{{\ {\bar{E}}}}}\right]
^{2N}{\frac{{\ |\Gamma (N+{\frac{{ \ i\Delta }}{{2\pi
T}}})|^{2}}}{{\Gamma (2N)}}\exp }\left( -{\frac{{\Delta }
}{{2T}}}\right) ,  \label{result_in1}
\end{equation}
where $\Gamma (x)$ is the Gamma function and $\Delta =\xi _{N+1}-\xi _{0}$
is the energy difference between the final and initial states. The
appearance of the factor ${\exp }\left( -\Delta /2T\right) $ is consistent
with the detailed balance principle. Indeed, at finite temperatures an
electron can tunnel in two ways: either with an increase or with a decrease
of its energy. According to the detailed balance principle the ratio of such
probabilities must be exp$\left( \Delta /T\right) ,$ that is indeed the case
for the function $P_{in}$,~(\ref{result_in1}), since apart from the factor
exp$\left( -\Delta /2T\right) $ the rest of the equation is even in $\Delta
. $

In order to obtain the expression for the hopping conductivity one has to
optimize Eq.~(\ref{result_in1}) with respect to the hopping distance $N$
under the constraint
\begin{equation}
N\,a\,\tilde{\kappa}\,\Delta /e^{2}\sim 1,  \label{constraint1}
\end{equation}%
that follows from the ES expression for the density of the ground
states~(\ref{ES_density_of_states}). Optimization of
Eq.~(\ref{result_in1}) is a bit more involved procedure than the
standard derivation of Mott-Efros-Shklovskii law based on the
Gibbs energy distribution function (see Appendix B). Nevertheless,
it leads to essentially the same result, i.e. the ES law,
Eq.~(\ref{e6}), where the characteristic temperature $T_{0}\left(
T\right) $ takes the form
\begin{equation}
T_{0}(T) \sim e^{2}/a\tilde{\kappa}\,\xi _{in}(T),  \label{p6}
\end{equation}
with a dimensionless weakly temperature dependent localization
length $\xi _{in}(T)$,
\begin{equation}
\xi _{in}(T)={\frac{2}{{\ln [\,\bar{E}^{2}/16c\pi T^{2}\bar{g}\,]}}},
\label{inelastic_loc_len}
\end{equation}
where the coefficient $c$ equals unity for a model with a diagonal Coulomb
interaction matrix.

Comparing the characteristic lengths $\xi _{el}\left( T\right) $
and $\xi _{in}\left( T\right) $ for the elastic and inelastic
processes, respectively, we see easily that $\xi _{el}\left(
T\right) > \xi _{in}\left( T\right) $ at temperatures $T <
T_{cross}$, where the temperature $T_{cross}$ is given by
Eq.~(\ref{crossover}). This means that the mechanism of the
inelastic scattering is more efficient at not very low
temperatures exceeding $T_{cross}.$

As in the case of the elastic cotunneling, the long range electrostatic
interaction results in the reduction of the coefficient to some $1/4\lesssim
c<1$.

At zero temperature, Equation~(\ref{result_in1}) for the inelastic
probability $P_{in}$ can be simplified for $\Delta <0$ to the form
\begin{equation}
P_{in}(T=0)={\frac{{\ w\,2^{2N}\pi }}{{\ (2N-1)!}}}{\frac{{\
|\Delta |^{2N-1} }}{{\ {\bar{E}}^{2N}}}}\left( {\frac{{\
\,\bar{g}}}{{\pi }}}\right) ^{N+1}, \label{inel}
\end{equation}
while for $\Delta >0,$ Eq.~(\ref{result_in1}) gives zero since the electron
tunneling process with an increase of the electron energy is prohibited at
zero temperature.

In order to obtain the hopping conductivity in the regime of a
strong electric field $\mathcal{E}$ and low temperature, we use,
following~\citet{Shklovskii73}, the condition~(\ref{constraint1})
that defines the distance to the first available electron site
within the energy shell $\Delta $. We use also the equation
\begin{equation}
e\mathcal{E}r\sim \Delta ,  \label{Electric_Field}
\end{equation}
that relates $\Delta $ to the electric field $\mathcal{E},$ and
the fact that the distance between the initial and final tunneling
sites is $r\sim Na. $ Equations~(\ref{constraint1}) and
(\ref{Electric_Field}) allow us to determine the quantities
$\Delta \sim \sqrt{\mathcal{E}e^{3}/\tilde{\kappa}}$ and $N\sim
\sqrt{e/\tilde{\kappa}\mathcal{E}a^{2}}.$ Inserting their values
into Eq.~(\ref{inel}) we come to the following expression for the
current
\begin{equation}
j\sim j_{0}\;\exp \left[
-(\mathcal{E}_{0}/\mathcal{E})^{1/2}\right]. \label{current_inel}
\end{equation}
where the characteristic electric field $\mathcal{E}_{0}$ is a weak function
of the applied field $\mathcal{E}$
\begin{equation}
\mathcal{E}_{0}(\mathcal{E})\sim {\frac{{\
e}}{{\tilde{\kappa}\,a^{2}}}} \;\ln
^{2}[\bar{E}^{2}/e^{2}\mathcal{E}^{2}a^{2}\bar{g}].
\end{equation}
The results obtained in the insulating regime show that the
hopping conductivity is the main mechanism of transport in the low
temperature regime. The temperature dependence of the conductivity
is determined by the Efros-Shklovskii law, Eq.~(\ref{e6}), and is
similar to the one in amorphous semiconductors. The exponential
dependence of the current on a strong electric field is given by
Eqs.~(\ref{current_el}) and (\ref{current_inel})$. $ Again, this
is the same dependence as in the amorphous semiconductors. Of
course, the characteristic temperature $T_{0}$ and electric field
$\mathcal{E}_{0}$ are model dependent.

The theory for the variable range hopping (VRH) via virtual
electron tunneling through many grains is based on the assumption
that the hopping length $r^{\ast }$ exceeds the size of a single
grain $a$. This hopping length $r^{\ast }$ should be found
minimizing the exponent in Eq.~(\ref{e100}). From this condition
we see that the characteristic length $r^{\ast }\left( T\right) $
equals
\begin{equation}
r^{\ast }(T) = \sqrt{\frac{e^{2}\xi a}{2T \bar{\kappa}}},
\label{e101}
\end{equation}
where the dimensionless localization length $\xi $ is given by
either Eq.~(\ref{localization}) or Eq.~(\ref{inelastic_loc_len}),
and that it decreases with increasing the temperature. At
temperatures ${\tilde{T}}\sim e^2\xi/2a \tilde{\kappa}$ the
hopping length becomes of the order of the grain size, $r^{\ast
}(\tilde{T})\sim a$. At such distances the VRH picture does not
work anymore because the electrons hope between the neighboring
grains only. This means that at comparatively high temperatures
$\tilde{T}\ll T\ll \Delta _{M}$ one should use the Arrhenius law,
Eq.~(\ref{activation1}) instead of the Efros-Shklovskii law,
Eq.~(\ref{e6}). Such a crossover has been observed
experimentally~\cite{Tran05}.

Thus, we have obtained a remarkable result: Even though the array
of the grains may look very regular, i.e. have a periodic
structure and equal size and shape of the grains, it is almost
inevitable that at low temperatures the main mechanism of the
conduction is the variable range electron hopping and the
conductivity is determined by Eq.~(\ref{e6}) as if the system were
amorphous. This concerns also other formulae like
Eqs.~(\ref{current_el}) and (\ref{current_inel}).

\section{Arrays composed of superconducting grains}

\label{superconducting}

In this Section we consider properties of granular materials
consisting of superconducting particles. As in the
Section~\ref{normal} devoted to study of normal metals we start
our presentation in the next subsection~\label{general} with a
qualitative discussion of the most interesting effects that occur
in these systems. A quantitative analysis of these effects will be
given in the subsequent subsections.

\subsection{General properties of granular superconductors}

\subsubsection{Single grain}

A granular superconductor, similarly to a granular metal, can be viewed as
an array of superconducting granules coupled via electron tunneling.
Superconducting properties of an array in many ways are determined by the
properties of granules which it is formed of. For this reason we begin our
discussion with reviewing properties of superconductivity in a single small
isolated grain.

This question was first addressed by Anderson~\cite{Anderson59}
who realized that the s-wave superconductivity is almost
insensitive to disorder of general kind as long as it does not
break the time reversal invariance. This includes, in particular,
the diffusive scattering by the grain boundaries that is in many
ways similar to the scattering by potential impurities in the
bulk. As long as the time reversal invariance is not broken, one
can choose the basis of exact eigenfunctions and then apply the
Bardeen-Cooper-Schrieffer (BCS) theory in the usual way. The
critical temperature can still be expressed via the BCS effective
coupling constant and the density of states in the vicinity of the
chemical potential. The BCS coupling constant does not change
considerably when changing the basis and the average density of
states is not sensitive to disorder. Thus, one
concludes~\cite{Anderson59} that the critical temperature of the
single grain has to be close to the bulk value.

Anderson has also pointed out that these arguments hold even for
small grains provided the average distance between the energy
levels $\delta $, Eq.~(\ref{e1}) is still smaller than the
superconducting gap $\Delta ,$ in the bulk~\cite{Anderson59}. As
soon as the energy level spacing $\delta $ reaches the
superconducting gap, the conventional BCS theory is no longer
applicable.

This fact can easily be understood from the BCS equation for the order
parameter that can be written as
\begin{equation}
\Delta =\delta \lambda _{0}T\sum_{\omega }\sum_{\xi
_{k}}\frac{\Delta }{\omega^{2}+\xi _{k}^{2}+\left\vert \Delta
\right\vert ^{2}},  \label{s0}
\end{equation}
where $\lambda _{0}$ is the standard dimensionless constant
describing the attraction between the electrons, $\omega =2\pi
T\left( n+1/2\right) $ are Matsubara frequencies and $\delta $ is
the average mean energy level spacing in a grain determined by
Eq.~(\ref{e1}). The discrete variable $\xi _{k}$ equals
\begin{equation}
\xi _{k}=\epsilon _{k}-\mu,  \label{s0a}
\end{equation}
where $\epsilon _{k}$ are energies of the electron states in the
grains and $\mu $ is the chemical potential.

If the grain size is very large, such that the mean level spacing
$\delta$ is much smaller than the superconducting critical
temperature $T_{c0},$ the sum over the states in Eq.~(\ref{s0})
can be replaced by the integral
\begin{equation}
\delta \sum_{\xi _{k}}\rightarrow \int_{-\infty }^{\infty }d\xi,
\label{s01a}
\end{equation}
and we come to the conventional BCS equation.

At the same time, it is clear from Eq.~(\ref{s0}) that a finite
difference between the energy levels $\xi _{k}$ plays a similar
role as a finite temperature, namely, it reduces the
superconducting gap $\left\vert \Delta \right\vert $. As soon as
the mean level spacing $\delta $ exceeds $T_{c0}$ (or $\Delta
\left( T=0\right) $ for the bulk superconductors), the
superconducting gap $\left\vert \Delta \right\vert $ vanishes. To
be more precise, it remains finite due to fluctuations but its
value is sufficiently reduced and in this regime the grain cannot
be considered as a "true" superconductor. Nevertheless, the
superconducting pairing is still present manifesting, for example,
in the parity effect,~\cite{Matveev97,VonDelft96,Smith96}.

The Anderson theory agrees well with experiments. In particular,
electron spectrum of ultrasmall superconducting $Al$ grains was
studied by~\citet{Ralph95,Black96,Davidovic99} where a well
defined superconducting gap was shown to survive down to grain
sizes $\sim 10nm$, while smaller grains $\leq 5nm$ did not show
any superconducting features in their electron spectrum.

It is worth noting that, while in the diffusive/chaotic grains the
density of states in the vicinity of the chemical potential indeed
well agrees with the corresponding bulk values, this is not
necessarily the case for the ballistic grains possessing certain
geometric symmetries. For example, in a spherical grain one
naturally expects that the energy levels are strongly degenerate.
This may change the density of states on the Fermi level
significantly resulting in a considerable change of the transition
temperature. The possibility to utilize the enhanced density of
states that may appear from the orbital degeneracy in order to
increase the single grain critical temperature was recently
suggested in Refs.~\cite{Ovchinnikov05a,Ovchinnikov05b}.

Although all these effects are very interesting, in this review we
follow our line of studying models with parameters that can easily
be achieved in existing granular materials. So, we assume
everywhere that granules are sufficiently large to maintain the
local superconductivity within each grain, i.e. that the Anderson
criterion is satisfied
\begin{equation}
\delta \ll \Delta .  \label{s1}
\end{equation}
We assume also that the grains are diffusive or do not have an ideal shape
like spheres or cubes, which excludes the possibility of an accidental
degeneracy of the electron spectrum.

As a final remark in our discussion of superconductivity in a
single grain, we would like to point out that the thermodynamics
of a single grain is not affected by the presence of the Coulomb
interaction as long as the latter can be expressed in a standard
way through the operator of the total electron number. The reason
is that the number of particles in an isolated grain is conserved
and thus, the Coulomb term in the Hamiltonian commutes with the
rest of the Hamiltonian and does not influence the thermodynamic
quantities.

At the same time, phase fluctuations of the order parameter $\Delta \left(
\tau \right) $
\begin{equation}
\Delta \left( \tau \right) =\left\vert \Delta _{0}\right\vert \exp
\left[ 2i\varphi \left( \tau \right) \right],  \label{s2}
\end{equation}
are very sensitive to the presence of the charging term,
Eq.~(\ref{Coulomb__Interacation}). These fluctuations can be
conveniently characterized by the correlation function $\Pi
_{s}\left( \tau \right) $
\begin{equation}
\Pi _{s}\left( \tau \right) =\left\langle \exp \left( 2i\left[
\varphi \left( \tau \right) -\varphi \left( 0\right) \right]
\right) \right\rangle, \label{s3}
\end{equation}
where the averaging should be performed over the states of the
Hamiltonian including the Coulomb interaction,
Eq.~(\ref{Coulomb__Interacation}).

Actually, the average over the states of the Hamiltonian in
Eq.~(\ref{s3}) reduces to an average with the action $S_{c}$,
Eq.~(\ref{a10}), and can rather easily be
calculated~\cite{Efetov80a} leading to the result at $T=0$
\begin{equation}
\Pi _{s}\left( \tau \right) =\exp \left( - 4 E_c \tau \right)
\label{s4}
\end{equation}
where $E_c = E_{ii} $ is the Coulomb energy of a single grain and
a factor 4 appears because of the doubled Cooper pair charge.
Equation~(\ref{s4}) is written for the imaginary time $\tau $.
Changing to the real time, $\tau \rightarrow it$, one can conclude
that the correlation function $\Pi_s \left( \tau \right) $
oscillates in time. Although the exponential form of the function
$\Pi _{s}\left( \tau \right) $ does not lead to changes in the
thermodynamics, it is crucial for the problem of the coherence in
an array of many grains.

\subsubsection{Macroscopic superconductivity}

Having listed the superconducting properties of a single grain we
turn now to discussing the superconductivity in the whole sample.
In order to simplify the presentation we first restrict ourselves
to the zero temperature case.

It is clear that an array with sufficiently strongly coupled
grains should be able to maintain the superconducting coherence in
a whole sample because the coupling reduces the phase
fluctuations. On the contrary, in the opposite limit of a weak
coupling, one expects that the strong Coulomb interaction should
lead to the Coulomb blockade of the Cooper pairs in analogy with
the Coulomb blockade of electrons in granular metals in the low
coupling regime.

In order to quantify this intuitive statement Abeles~\cite{Abeles77},
following the earlier idea of Anderson~\cite{Anderson64}, suggested to
compare the energy of the Josephson coupling of the neighboring grains with
the Coulomb energy. Indeed, the Josephson coupling tends to lock the phases
of the neighboring grains and to delocalize the Cooper pairs, while the
Coulomb interaction tends to localize the Cooper pairs and thus enhance the
quantum phase fluctuations.

Comparing the Josephson energy $J$,
\begin{equation}
J=\pi g\Delta/2,  \label{s5a}
\end{equation}
with the Coulomb energy one would come then to the conclusion that
samples with $g>g_s\sim E_{c}/\Delta $ should be superconductors,
while those with $g<g_s $ - insulators. This result was later
derived within a model for a superconducting granular array
including both Coulomb and Josephson interactions in a number of
theoretical works~\cite{McLean79, Efetov80a, Simanek79} using
different methods. We will review the main results of these
studies in the Sec.~\ref{phase_gs}.

\begin{figure}[t]
\hspace{-0.4cm}
\includegraphics[width=2.7in]{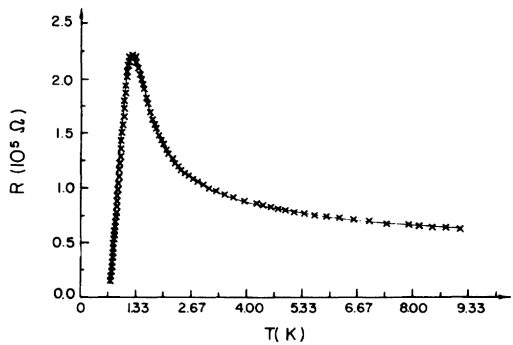} \caption{
Resistance as a function of temperature of one of the granular
samples of~\citet{Shapira83}. In spite of the fact that the sample
exhibits strong insulating behavior in the intermediate
temperature range it finally turns to the superconducting state. }
\label{Shapira_Deutscher}
\end{figure}

The Anderson-Abeles criterion predicts that the superconductivity
may survive even in such samples that would be insulators if they
were made from normal grains under equivalent conditions. Indeed,
for samples with a small ratio $E_{c}/\Delta \ll 1$, there is a
parametrically large interval of conductances $E_{c}/\Delta \ll g
\ll 1$ where superconductivity should exist in spite of the fact
that the corresponding normal array would be an insulator.

Qualitatively, this prediction was supported by
experiments~\cite{Shapira83} where some superconducting samples
were found to show clear insulating behavior above the critical
temperature. (We note, however, that the ratio $E_{c}/\Delta $ in
the samples used in this experiment was certainly large, this
fact, as we will show below, may be explained via renormalization
of the Coulomb energy). The corresponding dependence of the
resistance on temperature according to~\citet{Shapira83} is shown
in Fig.~\ref{Shapira_Deutscher}.

However, the Anderson-Abeles criterion was found later to be in
conflict with many other experiments. After the pioneering
works~\cite{Orr_PRB85,Orr86,Jaeger86,Jaeger89} it became clear
that the boundary between the superconducting and insulating
states at $T=0$ is determined by the normal state conductance of
the film rather then by the ratio of the Josephson and Coulomb
energies.

\begin{figure}[t]
\hspace{-0.4cm} \includegraphics[width=2.7in]{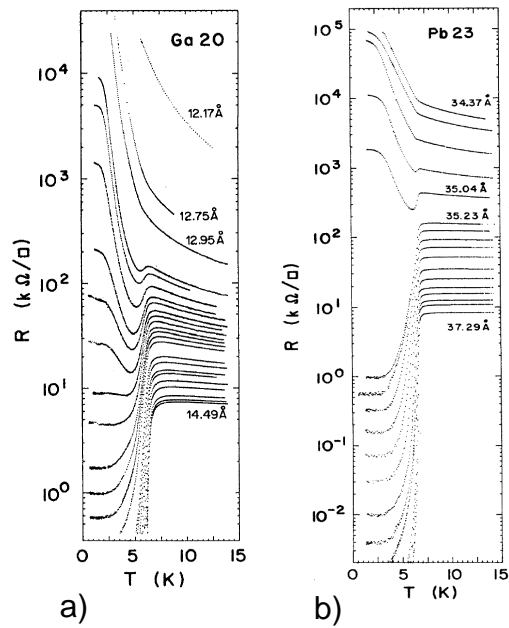}
\caption{ Temperature dependence of resistivity of granular (a) Ga
and (b) Pb ultra thin films according to~\citet{Jaeger89}.
Different curves correspond to the films with different normal
state resistances that is tuned by the sample thickness. Low
temperature state is controlled by the film normal state
resistance. } \label{Figure_Jaeger89}
\end{figure}

This fact is illustrated by the experimental resistivity
temperature dependencies of samples with different thicknesses
shown in Fig.~\ref{Figure_Jaeger89} $Ga$ and $Pb$ ultrathin films.
One can see that there exists a critical value of the normal state
resistance $R_{0}$ close to the value of quantum resistance
$h/(2e)^{2}\approx \mathrm{6.4k\,\Omega }$ such that samples with
resistance $R<R_{0}$ eventually become superconducting under
temperature decrease, while those with $R>R_{0}$ show tendency to
the insulating behavior. Such a behavior turned out to be generic
and was observed in a number of granular (as well as on
homogeneously disordered,~\cite{Liu93,Markovic99,Frydman02}) films
made from different materials.

In this respect, the superconductor-insulator transition in thin
granular films resembles the analogous transition in a resistively
shunted Josephson junction~\cite{Schmid83,bulgadaev85} where the
state of the system is known to be controlled by the value of the
shunt resistance only, no matter what the ratio of the Coulomb and
Josephson energies is (for review see~\citet{Zaikin}). The
resistive shunt in this approach is included via the
Caldeira-Leggett approach~\cite{Caldeira81,Caldeira83} that allows
to take into account the dissipative processes on the quantum
mechanical level. Similar approach based on the inclusion of the
phenomenological dissipative term that models the resistive
coupling of the grains was applied to the granular system in order
to resolve the disagreement with experiment~\cite {Chakravarty86,
Simanek86, Fisher86}. It was found that in the limit $E_{C}/J>1,$
in agreement with experiment, the boundary between the
superconducting and insulating states is controlled by the
resistance of the shunts only rather than by the ratio of
$E_{C}/J.$

However, the presence of the phenomenological resistive coupling
between the grains is rather difficult to justify, especially in
the limit of low temperatures where all quasiparticles are frozen
out. This problem was finally resolved by~\citet{Chakravarty87} on
the basis of the model that took into account the direct electron
tunneling processes between the grains in addition to the
Josephson couplings. The phase diagram obtained in this approach
turned out to be similar to the one obtained within the
"dissipative" model. At the same time, the considered model did
not rely on any phenomenological assumptions and the virtual
intergrain tunneling processes were well described in terms of the
original tunneling Hamiltonian.

\begin{figure}[t]
\hspace{-0.4cm}
\includegraphics[width=1.8in]{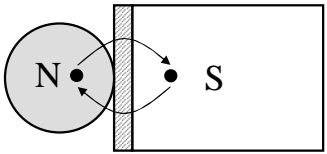}
\caption{ Illustration of the renormalization of the Coulomb
energy of a normal grain placed in a contact with a bulk
superconductor: Due to the virtual electron tunnelling processes
the Coulomb energy of a normal grain $E_c$ is renormalized down to
the values $\tilde E_c =\Delta/2g.$
 The conductance of the contact is supposed
to be large $g \gg 1 $ and the initial Coulomb energy is assumed
to be larger than the superconducting gap $\Delta.$ }
\label{Coulomb_Renormalization}
\end{figure}

According to the approach of~\citet{Chakravarty87}, the originally
strong Coulomb interaction is reduced by the electron tunneling to
other grains. This renormalization of the Coulomb energy can be
well understood on example of a simplified model of a normal grain
placed in a contact with the bulk superconductor shown in
Fig.~\ref{Coulomb_Renormalization}. It is clear that the Coulomb
charging energy is reduced due to the possibility for an electron
to be present virtually in the superconductor. Actually, this
renormalization is nothing but the screening of the Coulomb
interaction by free charges, which is usual for metals.

\begin{figure}[t]
\hspace{-0.4cm}
\includegraphics[width=2.2in]{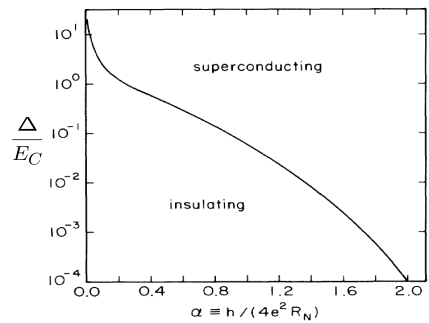}
\caption{ Zero temperature phase diagram of the granular
superconducting array according to~\citet{Chakravarty87}.
Dimensionless parameter $\alpha$ is related to the conductance $g$
as $\alpha= \pi g.$ } \label{Phase_Diagram_Halperin}
\end{figure}

The value of the screened charging energy in the case of a strong
coupling between the grains, $g\gg 1$, and strong Coulomb
interaction, $E_{c}\gg \Delta $ turns out to be $\sim \Delta /g;$
and what is remarkable, is independent on the original energy
$E_{c}$. Similar result holds for the superconducting grain as
well. We see that in the limit $g\gg 1$ the Josephson energy $J
\sim g\Delta $ is always larger than the effective Coulomb energy
$\Delta /g$ meaning that a sample with $g\gg 1$ should always be a
superconductor at sufficiently low
temperature~\cite{Larkin83,Eckern84}. Another, important
conclusion is that the superconductor-insulator transition occurs
at $g\sim 1,$ which is in agreement with the experiments. The
phase diagram obtained by~\citet{Chakravarty87} is presented in
Fig.~\ref{Phase_Diagram_Halperin}.

The fact that an array consisting of strongly coupled
superconducting grains ($g\gg 1$) has to be in a superconducting
state at sufficiently low temperatures can be understood from a
somewhat more general consideration. In the strong coupling
regime, the granular system resembles the homogeneously disordered
system. According to Anderson~\cite{Anderson59}, within the
accuracy of the mean field BCS theory, the thermodynamics of such
a system is essentially the same as that of a pure bulk electron
system. This means that the granular sample turns to the
superconducting state immediately after the superconducting gap
appears locally in each grain.

Thus, the insulating state where the superconductivity exists
locally, while being absent in the whole sample, may be obtained
only beyond the BCS approximation. Corrections to the BCS theory
may come from the Coulomb interaction and superconducting
fluctuations but they are expected to be small in the limit $g\gg
1.$ The fact that the corrections due to the Coulomb interaction
are small in normal metals in the limit $g\gg 1$ has been
demonstrated in Sec.~\ref{normal}. This result allows to come to
the conclusion that at $g\gg 1$ the BCS theory is a good starting
point for description of the granular system.

This in particular means that the critical temperature of a good
conducting granular sample ($g\gg 1$) is approximately given by
the single grain BCS critical temperature and that the corrections
to $T_{c}$ can be studied by means of the perturbation theory in
$1/g.$ In the case of homogeneously disordered films such a
perturbative approach to calculation of $T_{c}$ is well
known~\cite{Ovchinnikov73,Maekawa83, Finkelstein87}. Analogously,
one can consider the corrections to the critical temperature in
granular systems using a perturbation theory in the inverse
conductance. We discuss this approach and results obtained in
Sec.~\ref{suppression}.

In the opposite limit, $g\ll 1$, the critical temperature can be
found from an effective model that includes the Josephson and
Coulomb interaction only. As the probability of the tunneling from
grain to grain is small, the effect of the screening due to the
electron tunneling can be neglected. In particular, in the
simplest case of a very small charging energy, one can estimate
the critical temperature $T_{c}$ by comparing the Josephson energy
$J\sim \Delta g$ with the temperature, which leads to the estimate
$T_{c}\sim \Delta g.$ We will consider this question within a mean
field approximation in more detail in the next subsection.

\subsubsection{Granular superconductor in a magnetic field}

Now we present new features in the behavior of a granular array that
manifest in an applied magnetic field. It turns out that magnetic properties
of granular metals are very different from those of the corresponding bulk
systems even in the limit of large tunneling conductances, $g\gg 1.$ In
particular, there appears a new characteristic conductance value
\begin{equation}
g^{\ast }\approx 0.16 \, (a\delta )^{-1}\sqrt{\Delta _{0}D},
\label{e7}
\end{equation}
where $\Delta _{0}$ is the superconducting order parameter for a single
grain at zero magnetic field and zero temperature and $D$ is the diffusion
coefficient in the grain.

The critical magnetic field $H_{c}$ of samples with $g<g^{\ast }$
is close to the critical field $H_{c}^{\mathrm{gr}}$ of a single
grain, while for samples with $g>g^{\ast }$, it is close to the
one in a homogeneously disordered metals with the diffusion
coefficient $D_{eff}=g\delta a^{2}$ (c.f. Eqs.~(\ref{Gamma},
\ref{Deffective}). The latter is the effective diffusion
coefficient of the granular array. We note that the conductance $
g^{\ast }$ corresponds to the well conducting samples. Indeed,
expressing $ g^{\ast }$ in terms of the Thouless energy
$E_{Th}\sim D/a^{2}$ we obtain
\begin{equation}
g^{\ast }\sim \left( \frac{E_{Th}\Delta _{0}}{\delta ^{2}}\right)
^{1/2}\gg 1.
\end{equation}

The dependence of the critical magnetic field on the sample
conductance and temperature will be considered in detail in
Sec.~\ref{phase_gs}. Here we will only show how the new
conductance scale~(\ref{e7}) appears: The critical magnetic field
$H_{c}^{\mathrm{gr}}$ destroying the superconductivity in a single
grain of a spherical form due to the orbital mechanism is given
by~\cite{Larkin65} (the derivation is presented in
Sec.~\ref{phase_gs})
\begin{equation}
H_{c}^{\mathrm{gr}}={\frac{c}{e}}\sqrt{\frac{{\ 5\Delta
_{0}}}{{2R^{2}D}}}, \label{H_single_grain}
\end{equation}
where the grain radius $R$ is related to to the period of a cubic
granular array as $R=a/2.$ At sufficiently strong coupling the
critical field of the sample, on the contrary, has to be given by
the bulk value expressed via the effective diffusion coefficient:
\begin{equation}
H_{c}^{\mathrm{bulk}}=\frac{\Phi _{0}}{2\pi }\,\frac{\Delta _{0}}{D_{eff}},
\label{Hbulk}
\end{equation}
where $\Phi _{0} = h c/ 2e$ is the flux quantum. Comparing
Eqs.~(\ref{H_single_grain}) and (\ref{Hbulk}) one can see that the
two critical magnetic fields $H_{c}^{\mathrm{gr}}$ and
$H_{c}^{\mathrm{bulk}}$ become of the same order at tunneling
conductances $g\simeq g^{\ast }.$

The orbital mechanism of the destruction of the superconductivity
leading to Eqs.~(\ref{H_single_grain}) and (\ref{Hbulk}) is
dominant provided the grains are not very small. However, in the
limit of very small sizes of the grain, the Zeeman mechanism of
the destruction can become more
important~\cite{Tinkham96,Kee98,Beloborodov00} (everywhere in this
review we consider the $s $-wave singlet pairing).

If the Zeeman mechanism dominates the orbital one the critical magnetic
field $H_{c}^{z}$ is given by the Clogston value
\begin{equation}
H_{c}^{z}=\Delta _{0}/\sqrt{2}\mu _{B}  \label{s6}
\end{equation}
where $\mu _{B}=e\hbar /2mc$ is the Bohr magneton,~\cite{Clogston60}. We
note that the phase transition between the superconducting and normal states
is in this case of the first order.

In the opposite limit, when the superconductivity is destroyed
mainly by the orbital mechanism, the Zeeman splitting has no
dramatic effect on the phase diagram. Comparing the orbital and
Zeeman critical magnetic fields, Eqs.~(\ref{H_single_grain},
\ref{s6}) one can see that the latter can be neglected as long as
the grain is not too small $a>a_{c}$ with the critical grain size
$a_{c}$ given by
\begin{equation}
a_{c}\approx \sqrt{\frac{{\ 5}}{{\ D\Delta _{0}m^{2}}}}\sim
\frac{1}{p_{0}} \,\sqrt{\frac{E_{Th}}{\Delta _{0}}}.
\end{equation}
Throughout our review we assume that the grains are not too small such that
the phase transition within a singe grain remains of the second order. Of
course, we speak about the phase transition within the mean field
approximation. Fluctuations of the order parameter $\Delta $ smear the
transition in the isolated grain.

\subsubsection{Transport properties of granular superconductors}

Experimentally observed dependencies of the resistivity of the
granular superconductors on temperature and applied magnetic field
are very reach and all details are currently far from being well
theoretically understood. In our review we discuss only briefly
the general transport properties touching in detail only the
latest developments related to the low temperature
magnetoresistance in granular superconductors. A broader
discussion of theoretical advances in understanding the transport
properties of granular superconductors (networks of Josephson
junctions) can be found in Ref.~\cite{Fazio01}.

In the absence of the magnetic field, the dependence of the
resistivity of a granular superconductor on temperature is
typically non-monotonic. One of the examples is shown in
Fig.~\ref{Shapira_Deutscher}. Decreasing the temperature first
leads to a growth of the resistivity, which is due to the
enhancement of the Coulomb correlations that suppress the current.
However, the resistivity starts decreasing closer to the
superconducting transition temperature. In this region the
conductivity is enhanced due to the fluctuation contribution of
the Cooper pairs~\cite{Aslamazov68}.

Granular films with the normal state resistance close to the
critical value $R_{0}\sim 6.4\,k\Omega $ exhibit often a more
complicated reentrant behavior as it is seen in
Fig.~\ref{Figure_Jaeger89}. The resistance of such samples
increases again after a drop associated with the fluctuating
conductivity. This effect can be a consequence of the competition
between the conductance increase due to the fluctuating Cooper
pairs and the freezing of the excitations due to the opening of
the superconducting gap in the density of states.

The systematic theoretical description of this phenomenon is not
available yet. We can only note that the reentrant behavior of the
conductivity seems to be a consequence of the granularity.
Homogeneous films made from the same material having no (or less
pronounced) granular structure do not show such a
behavior~\cite{Liu93}.

Magnetic field applied to a granular system brings an extra control
parameter allowing to change the resistivity and to tune the state of the
granular system. In particular, sufficiently strong magnetic field can
always destroy the superconductivity even at low temperatures.

What is interesting, the resistance of a granular superconductor
has a non-monotonic dependence on the applied magnetic field at
low temperatures. As an example we present results of a
measurement on $Al$ granular from~\citet{Gerber97}. The samples
used by~\citet{Gerber97} were three dimensional and monodisperse
with a typical diameter of the grains $\approx 120$A. A surprising
feature of the experimental curves is that there is a region with
a negative magnetoresistance and a pronounced peak in the
resistivity at the magnetic field of several Tesla is seen. Only
at extremely strong fields, $H>6T$, the resistivity is almost
independent of the field.

\begin{figure}[t]
\hspace{-0.4cm} \includegraphics[width=2.2in]{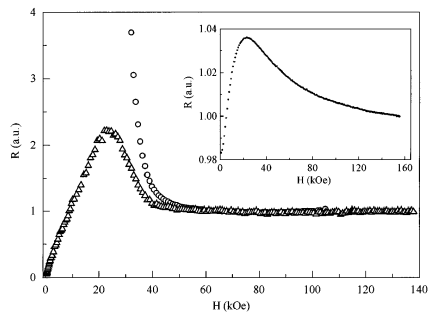}
\caption{ Low temperature resistance of $3d$ Al granular samples
as a function of magnetic field at low temperatures, $T \ll T_c$,
~\protect\cite{Gerber97}. The two curves are shown for two samples
with different normal temperature resistance. The insert shows
much less pronounce peak in the resistance at higher temperatures,
$T - T_c \ll T_c$. } \label{Gerber_Fig}
\end{figure}

The non-monotonic behavior of resistivity can be understood in
terms of the corrections to the conductivity due to the
superconducting fluctuations~\cite{Beloborodov99,Beloborodov00}.
In the vicinity of the phase transition the resistivity decreases
due to the opening of the additional transport channel via
fluctuating Cooper pairs. This is what is called the
Aslmazov-Larkin conductivity correction~\cite{Aslamazov68}.

We note that this correction gives always a positive contribution
to the conductivity and thus cannot explain the negative
magnetoresistance at higher fields. However, at stronger fields
another conductivity correction that appears as a result of the
reduction of the density of states due to fluctuating Cooper pairs
begins to dominate electron transport~\cite{Beloborodov99}. As a
result, in this regime the conductivity becomes lower than that in
the normal metal approaching the latter only in the limit $H\gg
H_{c},$ where all the superconducting fluctuations are completely
suppressed by the magnetic field. We discuss the magnetoresistance
of granular superconductors in more detail in
Sec.~\ref{magnetoresistance}.

Changing the magnetic field in the limit $T\rightarrow 0$ one can
cross the boundary between the superconducting and
metallic/insulating states and investigate the transport around
the zero temperature (quantum) phase transition. The corresponding
transition point at $T=0$ is referred as quantum critical
point~\cite{Sachdevbook,Sondhi97}.

Quantum phase transitions are analogous to the classical phase transitions
with the differences that at $T\rightarrow 0$ the system has been described
not only via the characteristic spatial (coherence length) $\xi $ but also
via the characteristic time scale $\tau .$ Both $\tau $ and $\xi ,$ as in
the case of classical transitions, are assumed to scale as powers of the
parameter $\delta $ that controls the closeness of the system to the quantum
critical point.

Scaling ideas were applied to the problem of the superconductor to
insulator transition in Refs.~\cite{Fisher90(1),Fisher90(2)} that
gave a substantial impact on the field. Scaling expressions for
the resistivity seem to work well for many samples. However, one
of the central prediction of Refs.~\cite{Fisher90(1),Fisher90(2)}
- the universal conductance at the superconductor to insulator
transition is not supported by all
experiments~\cite{Goldman98,Zant92,Chervenak00}.

Of course, the scaling approach to the resistivity behavior is not
restricted to the granular systems and, on the contrary,
homogeneously disordered films are much more suitable candidates
for experimental study of such phenomena. For this reason we will
not further touch this question referring the reader to
Refs.~\cite{Fazio01,Goldman98,Sondhi97}.

\subsection{ Phase diagram of granular superconductors}

\label{phase_gs}

In this section we discuss the phase diagram of granular
superconductors in the absence of the magnetic field. We will
mainly concentrate on the mean field approach and will not touch
the critical behavior in the very vicinity of the phase transition
and we will not discuss in detail the superconductor- insulator
transition. The latter phenomenon relates to a wide class of
disordered superconductors and represents the field of research on
its own. For reviews on this subject we refer the reader to
Refs.~\cite{Goldman98,Larkin99,Fazio01}.

The arrays of the superconducting grains can be conveniently
described using a phase functional analogous to the functional
$S\left[ \varphi \right] $ derived in Chapter~\ref{normal} for the
normal granular metals. In the next subsection we introduce such a
functional.

\subsubsection{Phase functional for the granular superconductors}

Description of a granular superconductor in terms of the phase
action is very similar to that used for the normal metals and the
derivation can rather simply be extended to include the
superconducting order parameter. In order to simplify the
discussion we consider the limit of low temperatures $T\ll
T_{c0}$, where $T_{c0}$ is the critical temperature in the bulk in
the BCS approximation. We assume that the superconducting gap is
still smaller than the Thouless energy, Eq.~(\ref{thouless}),
\begin{equation}
\Delta \ll E_{th},  \label{s07}
\end{equation}
which allows us to consider a single grain as zero-dimensional. Another way
to write this inequality is
\begin{equation}
\xi _{eff}\gg a,  \label{s007}
\end{equation}
where $\xi _{eff}=\sqrt{\xi _{0}l},$ $\xi _{0}=v_{F}/T_{c0}$, and $l$ is the
mean free path.

The derivation of the action $S\left[ \varphi \right] $ can be
carried out in the same way as in Chapter~\ref{normal} decoupling
the Coulomb interaction by an auxiliary field $V_{i}$ and the
making the gauge transformation. The phase $\varphi $ is
determined by Eqs.~(\ref{e29} - \ref{a5}). Expanding the action in
the tunneling amplitude $t_{ij}$ up to the second order we obtain
the action $S\left[ \varphi \right] $ in the form of a sum of
three terms
\begin{equation}
S\left[ \varphi \right] =S_{c}\left[ \varphi \right] +S_{ts}\left[
\varphi \right] +S_{J}\left[ \varphi \right]  \label{s7}
\end{equation}
The first term $S_{c}\left[ \varphi \right] $ stands for the
charging energy and is given by Eq.~(\ref{a10}). This means that
the charging energy of the superconducting grain is the same as
that of the normal one provided the same charge is placed onto the
grains.

The second term $S_{ts}\left[ \varphi \right] $ can be written in
the form of Eq.~(\ref{a11}) but with another function $\alpha
_{s}\left( \tau \right) $ instead of $\alpha \left( \tau \right)
,$ Eq.~(\ref{f7}). In the low temperature limit this function
takes the form~\cite{Eckern84}
\begin{equation}
\alpha _{s}(\tau )=\frac{1}{\pi ^{2}}\Delta ^{2}\,K_{1}^{2}(\Delta |\tau |),
\label{alpha_super}
\end{equation}
where $K_{1}(x)$ is the modified Bessel function. This difference
between the functions $\alpha \left( \tau \right) $,
Eq.~(\ref{f7}), and $\alpha _{s}\left( \tau \right) $ is very
important. The function $\alpha _{s}\left( \tau \right)
$~(\ref{alpha_super}) decays exponentially when increasing $\tau
$, while the functions $\alpha (\tau )$ decays only algebraically
at large $\tau $. The good convergence of the
kernel~(\ref{alpha_super}) allows us to expand the term $S_{st}$,
Eqs.~(\ref{a11}, \ref{alpha_super})~ in powers of $\tau -\tau
^{\prime }$ and to write the term $S_{ts}$ in a simplified form
\begin{equation}
S_{ts}= \frac{3\pi g}{ 64 \Delta }\,\sum_{<ij>}\int_{0}^{\beta
}\dot{\varphi}_{ij}^{2}(\tau )d\tau ,  \label{charge_renorm}
\end{equation}
which should be correct for $ \tau \gg \Delta ^{-1}.$ Only the
third term $S_{J}$ in Eq.~(\ref{s7}) is new as compared with
normal metals; it describes the Josephson coupling between the
grains and in the limit $ \tau \gg \Delta ^{-1}$ can be written as
\begin{equation}
S_{J}=-{1\over 2 }\sum\limits_{\langle ij\rangle }\int_{0}^{\beta
}J_{ij}\cos \left( 2\left[ \varphi _{i}\left( \tau \right)
-\varphi _{j}\left( \tau \right) \right] \right) d\tau .
\label{s9}
\end{equation}
The sum in Eq.~(\ref{s9}) should be performed over the nearest
neighbors $i$ , $j$ and $J=J_{ij}$ for these values of $i,j$ is
determined by Eq.~(\ref{s5a}).

Equations~(\ref{s7}, \ref{a10}, \ref{a11}, \ref{alpha_super},
\ref{s9}, \ref{s5a}) describe completely the granular
superconductors in the low temperature limit. They can easily be
extended to higher temperatures by writing proper expressions for
the functions $\alpha _{s}\left( \tau \right) $ and $J_{ij}$.
Trying to present the physical picture in the simplest way we do
not consider here these complications.

The action $S\left[ \varphi \right] $ contains only the phases
$\varphi ,$ whereas fluctuations of the modulus of the order
parameter are neglected. This is justified at low temperatures
provided the inequality (\ref{s1}) is fulfilled. A microscopic
derivation of the action $S_{c}+$ $S_{J}$ was originally done
using another approach in~\citet{Efetov80a}. It was also written
phenomenologically in~\citet{Simanek79} assuming however that the
variable $\varphi $ varied from $-\infty $ to $\infty .$ The
existence of the term $S_{ts}$ was realized later~\cite{Eckern84}
and used for a system of Josephson junctions
by~\citet{Chakravarty87}.

Assuming that the most important are large times $\tau \gg
\Delta^{-1}$ we use in the subsequent calculations
Eq.~(\ref{charge_renorm}) for the term $S_{ts}$. Then, denoting
\begin{equation}
S_{0}\left[ \varphi \right] =S_{c}\left[ \varphi \right]
+S_{ts}\left[ \varphi \right],  \label{s10}
\end{equation}
we write the quadratic form $S_{0}\left[ \varphi \right] $ as
\begin{equation}
S_{0}\left[ \varphi \right] =\frac{1}{4}\sum_{i,j}\int \left(
\tilde E_c^{-1}\right) _{ij}\dot{\varphi}_{i}\left( \tau \right)
\dot{\varphi}_{j}\left( \tau \right) d\tau,  \label{s11}
\end{equation}
where
\begin{equation}
\tilde E_c\left( \mathbf{q}\right) =\frac{1}{2}\left[ C\left(
\mathbf{q}\right) /e^{2}+(3\pi g/16\Delta ) \lambda _{\mathbf{q}}
\right] ^{-1}  \label{s12}
\end{equation}%
and $\lambda _{\mathbf{q}}$ is to be taken from Eq.~(\ref{e240}).
The function $E\left( \mathbf{q}\right) $ is the Fourier transform
of the function $E_{ij}$ and $\mathbf{q}$ is quasimomentum. When
integrating over the phases $\varphi $, one should remember about
the winding numbers. Looking at Eq.~(\ref{s12}) one can understand
that the term $S_{ts}$, Eqs.~(\ref{a11}, \ref{alpha_super}), leads
to a screening of the Coulomb interactions.

\subsubsection{Mean field approximation}

In spite of many simplifications, the action
\begin{equation}
S\left[ \varphi \right] =S_{0}\left[ \varphi \right] +S_{J}\left[
\varphi \right]  \label{s13}
\end{equation}
is still complicated and it is not possible to describe the system
without making approximations. A common way to understand, at
least qualitatively, the properties of a model is to develop a
mean field theory. Following this approach one should replace the
initial Hamiltonian (action) by a simplified one and determine
parameters of the effective model self-consistently.

Now we will consider the action $S\left[ \varphi \right] $,
Eq.~(\ref{s13}) in the mean field approximation.
Following~\citet{Efetov80a} we make in the action $S_{J}$,
Eq.~(\ref{s9}), the following replacement
\begin{equation}
\sum_{\left\langle i,j\right\rangle }J_{ij}\cos [ 2\left( \varphi
_{i}-\varphi _{j}\right) ] \rightarrow Jz\langle \cos 2\varphi
\rangle _{MF}\sum_{j}\cos 2\varphi _{j}.  \label{s14}
\end{equation}
The average $\langle \cos \varphi \rangle _{MF}$ in
Eq.~(\ref{s14}) should be calculated with the action
$S_{eff}\left[ \varphi \right] $ that can be obtained from
$S\left[ \varphi \right] $, Eq.~(\ref{s13}) by making the
replacement, Eq.~(\ref{s14}), in $S_{J}\left[ \varphi \right] $,
Eq.~(\ref{s9}). Actually, this average is the order parameter for
the macroscopic superconductivity.

Assuming that the transition between the superconductor and insulator is of
the second order we can find the critical point from the condition that at
this point the mean field $\langle \cos 2\varphi \rangle _{MF}$ vanishes.
The corresponding equation determining the boundary of the superconducting
state takes the form
\begin{equation}
1=\frac{zJ}{2}\int\limits_{0}^{\beta }\Pi _{s}\left( \tau \right) d\tau ,
\label{mean_field}
\end{equation}
where $z$ is the coordination number of the lattice and $\Pi
_{s}\left( \tau \right) $ is determined by Eq.~(\ref{s3}) assuming
however that the average $\langle \ldots \rangle $ in that
equation should be calculated with the action $S_{0}\left[ \varphi
\right] $, Eqs.~(\ref{s11}) and (\ref{s12}).

In order to simplify calculations further we will neglect the
off-diagonal terms in the matrix $E_{ij}$, Eq.~(\ref{s11}),
keeping the value $E_{0}\left( g\right) =\tilde E_c^{ii}$ only.
This value can be written as
\begin{equation}
E_{0}\left( g\right) =a^d \int \tilde
E_c(\mathbf{q})\frac{d^{d}\mathbf{q}}{(2\pi )^{d}}. \label{s15}
\end{equation}%
where $\tilde E_c\left( \mathbf{q}\right) $ is determined by
Eq.~(\ref{s12}). Then for $\Pi _{s}\left( \tau \right) $ instead
of Eq.~(\ref{s4}) we obtain the following expression
\begin{equation}
\Pi _{s}\left( \tau \right) =\exp \left( -4E_{0}\tau \right).
\label{s16}
\end{equation}
Solving Eq.~(\ref{mean_field}) at $T\rightarrow 0$ one obtains the
relation between the Josephson and Coulomb energies at the phase
transition
\begin{equation}
zJ=8E_{0}\left( g\right). \label{AAcondition}
\end{equation}
Equation~(\ref{AAcondition}) agrees at small $g\ll 1$ (up to the
coefficient) with the Anderson-Abeles criterion~\cite{Abeles77}.
In this limit we obtain using Eq.~(\ref{s5a}) the critical value
for the conductance in the form
\begin{equation}
g_{c}^{\ast }= 16 E_{0}\left( 0\right) /\pi z\Delta .  \label{gc2}
\end{equation}%
Samples with the tunneling conductances $g<g_{c}^{\ast }$ are insulators,
while those with $g>g_{c}^{\ast }$ are superconductors. We note that this
result suggests that even samples with weakly coupled granules $g_{c}^{\ast
}<g\ll 1$ are still superconductors in spite of the fact that the
corresponding array of normal grains with the same Coulomb interaction would
be an insulator in this regime.

\begin{figure}[t]
\includegraphics[width=3.0in]{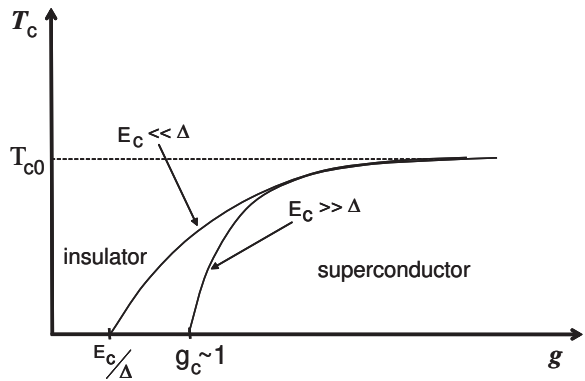} \vspace{-0.3cm}
\caption{ The sketch of the phase diagram of a granular
superconductor in coordinates critical temperature vs. tunnelling
conductance for two cases $ E_c \ll \Delta_0$ and $E_c \gg
\Delta_0$. At large tunnelling conductances $g \gg 1$ the Coulomb
interaction is screened due to electron intergrain tunnelling and
the transition temperature is approximately given by the single
grain BCS value. In the opposite case $g \ll 1$ the boundary
between the insulating and superconducting states is obtained
comparing the Josephson $E_J = g\Delta$ and Coulomb $E_c$
energies. } \label{tcfigure}
\end{figure}

It is clear from Eq.~(\ref{mean_field}) that the zero temperature
result (\ref{gc2}) holds also at finite temperatures so long as
$T\ll E_{c}.$ In the opposite limit $T\gg E_{c}$ (but still $T\ll
T_{c0}$), one obtains $\Pi _{s}\left( \tau \right) =1,$ which
leads to the critical temperature
\begin{equation}
T_{c}=\frac{1}{2}zJ=\frac{1}{4}z\pi g\Delta .  \label{highTc}
\end{equation}%
The results~(\ref{gc2}) and~(\ref{highTc}) correspond to the low coupling
part (small $g$) of the phase diagram shown in Fig.~(\ref{tcfigure}).

As we have mentioned previously, the Anderson-Abeles criterion fails to
produce the correct phase boundary between the insulating and
superconducting states for arrays with a stronger coupling $g\geq 1$ .
According to the experiments \cite{Orr86,Jaeger89} the
superconductor-insulator transition is controlled by the resistance of the
array only and the critical resistance $R_{c}$ (in the case of granular
films) is close to the resistance quantum $R_{0}\approx 6.4k\Omega $. The
Coulomb interaction drops out from the result.

This interesting observation can be understood using
Eqs.~(\ref{AAcondition} , \ref{s15}, \ref{s12}, \ref{s5a}). It is
important to notice that the energy $E_{0}\left( g\right) $
depends on the dimensionality. Using Eqs.~(\ref{s12}, \ref{s15})
we see that in $3d$ arrays of the grains the function $E_{0}\left(
g\right)$ equals
\begin{equation}
E_{0}\left( g\right) =c\Delta /g,  \label{s17}
\end{equation}
where $c$ is a number of the order unity, while in $2d$ we obtain with the
logarithmic accuracy
\begin{equation}
E_{0}\left( g\right) ={\frac{{ 2  \Delta }}{3 \pi^2 g}}\ln
(gE_{c}/\Delta ). \label{s18}
\end{equation}
One can see from Eqs.~(\ref{s17}) and (\ref{s18}) that the
effective charging energy is independent on the originally strong
Coulomb interaction in $3d$ case and is almost independent on it
in the $2d$ case. Substituting Eqs.~(\ref{s17}) and (\ref{s18})
into Eqs.~(\ref{AAcondition}), we come to the conclusion that in
the limit $E_{C}\gg \Delta $ the superconductor-insulator
transition occurs at $g_{c}\sim 1$, which is in agreement the with
experiments.

Considerations presented in the original work~\cite
{Chakravarty87} lead to results qualitatively similar to those
presented here. In particular, the phase boundary at $T\rightarrow
0$ in two dimensional case was found to be
\begin{equation}
\alpha _{c}^{2}={ \frac{4}{{3\pi }}}\ln \left( 1+{\frac{{3\pi
}}{{16}}}\alpha _{c}{\frac{{E_{c} }}{{\Delta }}}\right),
\label{s19}
\end{equation}
where $\alpha_c$ is the critical value of the dimensionless
conductance related to our conductance $g$ as $\alpha=\pi g.$ The
phase boundary following from the calculations
of~\citet{Chakravarty87} is represented in
Fig.~\ref{Phase_Diagram_Halperin}.

Summarizing our discussion we present in Fig.~\ref{tcfigure} the phase
diagram for a granular superconductor in the temperature vs. conductance
coordinates for two cases of $E_{c}\ll \Delta $ and $E_{c}\gg \Delta $.

The strong coupling regime, $g>1$, can also be analyzed using the
diagrammatic technique. This approach is advantageous in many respects since
it allows one to find the corrections to the critical temperature in a
straightforward and rigorous way not relying on the mean field approximation
used in the treatment of the effective functional. This approach is
considered in detail in Sec.~\ref{suppression}.

We note that while the mean field theory gives qualitatively
correct results, it certainly cannot be viewed as a complete
description of the model, Eq.~(\ref{s7}) because fluctuations near
the mean field solutions can be important. In particular,
Eq.~(\ref{mean_field}) assumes that the system is not affected by
the Josephson couplings so long as the mean value of $\varphi $ is
zero. However, the Josephson coupling is known to be important
even in the insulating state. In the simplest case of the
insulating state with the low Josephson coupling the effect of the
later can be included via the perturbation theory. As a result,
the charging energy is reduced, $ E_{c}\rightarrow
\tilde{E}_{c}=E_{c}+\delta E_{c}$ with the correction $ \delta
E_{c}$ given by~\cite{Sachdevbook}
\begin{equation}
\delta E_{c}=-zJ/2.
\end{equation}
Closer to the superconductor - insulator transition the charging
energy gets renormalized even more strongly and in the vicinity of
the transition it is expected to have the scaling
form,~\cite{Sachdevbook}
\begin{equation}
E_{0}(g)\sim |g_{c}^{\ast }-g|^{\gamma },  \label{EcScaling}
\end{equation}%
where $\gamma >0$ is the critical index. One may interpret the
energy $E_{0}\left( g\right) $ as the increase of charging energy
when adding a Cooper pair to a neutral grain. Exactly at the phase
transition this energy vanishes according to Eq.~(\ref{EcScaling})
and this ensures the continuity of the Coulomb gap at the phase
transition, since in the superconducting state this gap has to be
zero.

The Coulomb gap is an important characteristic of the insulating state. In
particular, it controls the low temperature conductivity that has an
activation form
\begin{equation}
\sigma \sim e^{-4E_{0}\left( g\right) /T}.  \label{s20}
\end{equation}%
Experimentally measured conductivity of granular samples is,
however, usually more complicated than that given by
Eq.~(\ref{s20}) and typically has a variable-range hopping type of
behavior. Such a behavior can be obtained taking into account the
electrostatic disorder neglected in the model, Eq.~(\ref{s7}).

\subsection{Upper critical field of a granular superconductor}

\label{phase_gs}

As we already mentioned, the granular superconductors have rather
unusual magnetic properties. In particular, at low temperatures,
depending on the tunneling conductance $g,$ the critical magnetic
field is determined by either bulk formula~(\ref{Hbulk}) at
$g>g^{\ast }$ or by the critical magnetic field of a single
grain~(\ref{H_single_grain}) at $g<g^{\ast }$, where the
characteristic conductance $g^{\ast }$ is defined by Eq.~(\ref{e7}
). In this section we will discuss in detail the dependence of the
upper critical field $H_{c_{2}}$ of granular superconductors on
the tunneling conductance $g$ at arbitrary temperatures. We will
consider granular samples that are relatively good metals in their
normal state, such that $g\gg 1$. This will allow us to neglect
the effect of the suppression of the critical temperature by the
Coulomb interaction and fluctuations.

\subsubsection{Critical field of a single grain}

\label{Hsingle}

We begin with a discussion of the critical magnetic field of a
single grain. This problem was first considered long ago
by~\citet{Larkin65} by means of averaging the Gor'kov equations
over disorder. Another way to consider this problem is based on
using a later technique, namely, a semiclassical Usadel
equation,~\cite{Usadel70}. This way is more convenient for our
purposes since it can easily be generalized to the case of a
granular array. Assuming that the transition between the
superconducting and normal states is of the second order we define
the critical field as a field at which the self-consistency
equation~\cite{Abrikosov65}
\begin{equation}
\Delta (\mathbf{r})=\lambda _{0}\pi T\sum_{\omega }f_{\omega
}(\mathbf{r}), \label{BCS}
\end{equation}
where $\omega =\pi T\left( 2n+1\right) ,$ acquires a nontrivial
solution under decrease of the magnetic field. The function
$f_{\omega }\left( \mathbf{r}\right) $ is a quasiclassical Green
function that can be obtained from the anomalous Gor'kov Green
function $F$ by integration over $\xi $, Eqs.~(\ref{s0a}) and
(\ref{s01a}). The Green function $f_{\omega }(\mathbf{r})$ can be
determined from the Usadel equation that can be linearized close
to the phase transition, see e.g.~\cite{Kopninbook}
\begin{equation}
\Big(|\omega |+D(-i\nabla -2e\mathbf{A}/c)^{2}/2\Big)f_{\omega
}(\mathbf{r} )=\Delta (\mathbf{r),}  \label{Usadel}
\end{equation}
where $D$ is the classical diffusion coefficient in the
superconducting grain and $\mathbf{A}$ is the vector potential. We
choose the vector potential corresponding to a homogeneous
magnetic field $\mathbf{H}$ in the form $\mathbf{A}\left(
\mathbf{r}\right) =[\mathbf{H}\times \mathbf{r}]/2$, which allows
us to reduce Eq.~(\ref{Usadel}) to the form
\begin{equation}
\Big(|\omega |-D\nabla ^{2}/2 +
2D(e\mathbf{A}/c)^{2}\Big)f_{\omega }(\mathbf{r })=\Delta
(\mathbf{r).}  \label{Usade2}
\end{equation}
Equation~(\ref{Usade2}) can be solved via perturbative expansion
in $\mathbf{A}^{2}.$ In the limit determined by Eq.~(\ref{s007}),
one can neglect dependence of the functions $f_{\omega
}(\mathbf{r)}$ and $\Delta (\mathbf{r)}$ on coordinates. In the
main approximation we can simply replace the term with
$\mathbf{A}^{2}\left( \mathbf{r}\right) $ in Eq.~(\ref{Usade2}) by
its average over the volume of the grain. Thus, we obtain the
simple algebraic equation relating constant components of $f$ and
$\Delta $
\begin{equation}
f_{\omega }\,(|\omega |+\alpha )=\Delta.  \label{f_function}
\end{equation}
where the depairing parameter $\alpha $ is the case of a spherical
grain with the radius $R$ is
\begin{equation}
\alpha =R^{2}D(eH/c)^{2}/5.  \label{s21}
\end{equation}
Substituting the solution of Eq.~(\ref{f_function}) into
Eq.~(\ref{BCS}) we come to the standard
equation~\cite{Abrikosov61} relating the critical temperature
$T_{c}$ to the depairing parameter $\alpha $, Eq.~(\ref{s21}),
\begin{equation}
\ln (T_{c}/T_{c0})=\psi (1/2)-\psi (1/2+\alpha /2\pi T_{c}).
\label{Tc_vs_H}
\end{equation}
Here $\psi (x)$ is the digamma function and $T_{c0}\equiv
T_{c}(H=0)$ is the single grain superconducting critical
temperature in the absence of the magnetic field. One can see that
at zero temperature the superconductivity is destroyed at
\begin{equation}
\alpha _{c0}=\Delta _{0}/2,
\end{equation}
where $\Delta _{0}$ is the superconducting order parameter at
$T,H=0$ that is related in the standard way to the critical
temperature $T_{c0}$ in the absence of the magnetic field as
$\Delta _{0}=\pi T_{c0}/\gamma $ with $\ln \gamma \approx 0.577$
being the Euler constant. For a single grain of a spherical shape
we obtain Eq.~(\ref{H_single_grain}) for the critical field $
H_{c}^{\mathrm{gr}}$ at zero temperature.

The above approach can be easily generalized to take into account
the Zeeman splitting by making the shift $|\omega |\rightarrow
|\omega |+h\, \mathrm{sign}\,\omega ,$ where $h$ is the Zeeman
energy $h=\mu _{B}H.$ Then, Eq.~(\ref{Tc_vs_H}) is modified to
\begin{equation}
\ln (T_{c}/T_{c0})=\psi (1/2)-\mathrm{Re}\left[ \;\psi
(1/2+(\alpha +ih)/2\pi T_{c})\right] .  \label{Tc_vs_H_}
\end{equation}
We note that Eq.~(\ref{Tc_vs_H_}) is valid as long as the phase
transition is of the second order. However, in the limit where the
Zeeman effect dominates the orbital one~\citet{Clogston60} the
phase transition is known to be first order. Analysis of the first
order phase transition at finite temperatures is rather involved
and here we present only the corresponding phase diagram at zero
temperature (Fig.~\ref{Phase_diagram single grain}) that
illustrates how the second order phase transition turns into the
first order one~\cite{Beloborodov06}. To simplify the discussion
we assume that the Clogston limit is not reached and
Eq.~(\ref{Tc_vs_H_}) describing the second order phase transition
is applicable.

\begin{figure}[t]
\centerline{\includegraphics[width=65mm]{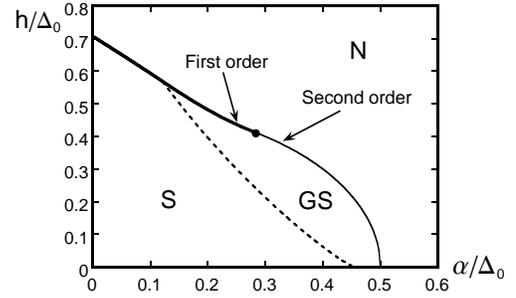}}
\caption{Phase diagram of a superconducting grain at $T=0$ in
coordinates Zeeman energy $h$ vs. pairbreaking parameter
$\protect\alpha $. The dashed line separates the gapless (GS) and
gapful (S) superconducting states. First and second order phase
transition lines are shown by thick and thin solid lines,
respectively~\cite{Beloborodov06}.} \label{Phase_diagram single
grain}
\end{figure}

At zero temperature, Eq.~(\ref{Tc_vs_H_}) reduces
to~\cite{Beloborodov00}
\begin{equation}
\alpha ^{2}+h^{2}=\Delta _{0}^{2}/4,  \label{T=0}
\end{equation}
such that Eq.~(\ref{H_single_grain}) for the critical field
$H_{c}^{\mathrm{ gr}}$ in the presence of the Zeeman coupling can
simply be generalized via the substitution
\begin{equation}
\Delta _{0}\rightarrow \tilde{\Delta}_{0}\equiv \sqrt{\Delta
_{0}^{2}-(2h)^{2}}.  \label{tilde_delta}
\end{equation}

\subsubsection{Critical magnetic field of a granular sample}

According to subsection~\ref{Hsingle} the effect of the magnetic
field on a single grain reduces to the renormalization of the
Matsubara frequency
\begin{equation}
\omega \rightarrow \tilde{\omega}=\omega + ih + \alpha.
\label{renorm}
\end{equation}
So, considering the critical temperature of the whole granular
sample one can take into account the local single grain effects
via the frequency renormalization,~(\ref{renorm}). The critical
temperature of the granular array can be again found from the
linearized Usadel equation that can be written for the granular
sample in the Fourier representation as
\begin{equation}
\Big(|\omega |+(g\delta /2)\,\lambda
(\mathbf{q}-2e\mathbf{A}/c)\Big)f( \mathbf{q})=\Delta
(\mathbf{q}),  \label{s22}
\end{equation}
where $\lambda
(\mathbf{q}-2e\mathbf{A}/c)=2\sum_{\mathbf{a}}(1-\cos [(
\mathbf{q}-2e\mathbf{A}/c)\mathbf{a}])$ and $\mathbf{q}$ is the
quasimomentum of the periodic lattice. In the regime under
consideration, Eq.~(\ref{s007}), the magnetic flux per a unite
cell of the lattice is always small. This allows us to find the
lowest eigenvalue of the operator $\lambda (\mathbf{q}-2e
\mathbf{A}/c)$ as the lowest Landau level of the operator for the
corresponding continuous system
$a^{2}(\hat{\mathbf{q}}-2e\mathbf{A}/c)^{2} $, which gives
$2eHa^{2}/c$.

Thus, the evaluation of the orbital effect that originates from
flux accumulation on scales involving many grains is very similar
to the analogous effect in the homogeneously disordered samples,
which is a direct consequence of the smallness of the magnetic
flux per unite cell. Now one can see that the orbital effect that
comes from large distances simply gives an additional contribution
$g\delta a^{2}eH/c$ to the depairing parameter $\alpha $. Using
Eq.~(\ref{Deffective}) for the effective diffusive coefficient
$D_{eff}$ we obtain for the total depairing parameter including
both single grain and bulk orbital effects of the magnetic field
\begin{equation}
\alpha =R^{2}D(eH/c)^{2}/5+D_{eff}eH/c.  \label{alpha_total}
\end{equation}
We note that although Eq.~(\ref{alpha_total}) has been obtained
within the periodic cubic lattice array model it is in fact more
general. Indeed, the first term in the right hand side of
Eq.~(\ref{alpha_total}) represents the single grain effect and is
sensitive to the grain size only. The second term, being expressed
in terms of the effective diffusive coefficient $D_{eff}$, is not
sensitive to the structure of the array at all.

Comparing the two contributions to the depairing parameter $\alpha
$ in Eq.~(\ref{alpha_total}) we obtain immediately the value
$H^{\ast }$ characterizing the crossover from the single grain to
the bulk orbital effects
\begin{equation}
H^{\ast }={\frac{{\ 5\,\phi _{0}}}{{\pi
R^{2}}}}{\frac{{D_{eff}}}{{\ D}}}, \label{s23}
\end{equation}
Equations~(\ref{alpha_total}) and (\ref{Tc_vs_H_}) determine the
superconducting transition line in the coordinates $T$ vs. $H$
plane.

In the limit of zero temperature $T\rightarrow 0,$
Eq.~(\ref{Tc_vs_H_}) reduces to Eq.~(\ref{T=0}), which allows us
to find a simple expression for the critical filed at $T=0$
resolving Eq.~(\ref{alpha_total}) with respect to $H:$
\begin{equation}
H_{c}(T=0)={\frac{{\ 5\phi _{0}}}{{2D\pi R^{2}}}}\left(
\sqrt{D_{eff}^{2}+{\frac{{\
2\tilde{\Delta}_{0}R^{2}D}}{{5}}}}-D_{eff}\right) ,
\end{equation}
where $\tilde{\Delta}_{0}$ defined by Eq.~(\ref{tilde_delta})
takes into account the Zeeman effect.

\subsection{Suppression of the superconducting critical temperature}

\label{suppression}

The BCS theory gives a very accurate description of
superconductors in bulk well conducting metals. However, in
strongly disordered or granular metals there can be considerable
deviations from this theory. This concerns, in particular, the
superconducting transition temperature.

In this subsection we consider suppression of the
superconductivity in granular metals at large tunneling
conductances between the grains, $g\gg 1$. In this limit, all
properties of the system should be close to those following from
the BCS theory. The mean field transition temperature $T_{c}$ of
the granular superconductors has to be close to that of the
corresponding disordered bulk superconductor. A difference between
the granular and disordered bulk superconductors may appear only
in corrections to the BCS theory. These corrections will be
discussed in the present subsection.

The main mechanisms of the suppression of the superconductivity
are the Coulomb repulsion and fluctuations of the superconducting
order parameter. For example, disorder shifts significantly the
superconducting transition temperature in the $2d$ thin
films~\cite{Ovchinnikov73,Fukuyama81,Maekawa83,Finkelstein87,Finkelstein94,Ishida98,Larkin99}.
The physical reason for the suppression of the critical
temperature is that in thin films the interaction amplitude in the
superconducting channel decreases due to peculiar disorder-induced
interference effects that enhance the effective Coulomb
interaction. We refer to this mechanism of the superconductivity
suppression as to the \textit{fermionic} mechanism.

The superconducting transition temperature can also be reduced by
the fluctuations of the order parameter, the effect being
especially strong in low dimensions. The corresponding mechanism
of the superconductivity suppression is called the
\textit{bosonic} mechanism. In particular, the bosonic mechanism
can lead to the appearance of the insulating state at zero
temperature~\cite{Efetov80a,Fisher90(2),Simanekbook}.

In this subsection we discuss the suppression of the
superconductivity in terms of the perturbation theory. Although
this method is restrictive and cannot be used to study the
superconductor -insulator transition, it is useful in a sense that
both the relevant mechanisms of the reduction of the critical
temperature can be included systematically within the same
framework. The usefulness of the perturbative calculations has
been demonstrated in the study of properties of normal granular
metals in Sec.~\ref{very.low} where the new important energy scale
$\Gamma $, Eq.~(\ref{Gamma}) was introduced. In view of these
findings one may expect that the correction to the superconducting
transition temperature can be different depending on whether the
temperature is larger or smaller than the energy scale $\Gamma $.

We start with the following Hamiltonian $\hat{H}_{s}$ describing
the granular system
\begin{equation}
\hat{H}_{s}=\hat{H}+\hat{H}_{e-ph},  \label{s24}
\end{equation}
where the Hamiltonian $\hat{H}$ for the normal granular metals is
given by Eq.~(\ref{a02}) and $\hat{H}_{e-ph}$ is an additional
electron-phonon interaction on each grain
\begin{equation}
\hat{H}_{e-ph}=-\lambda \sum\limits_{i,k,k^{\prime
}}a_{i,k}^{+}a_{i,-k}^{+}a_{i,-k^{\prime }}a_{i,k^{\prime }},
\label{e5}
\end{equation}%
where $i$ labels the grains, $k\equiv (\mathbf{k},\uparrow )$,
$-k\equiv (- \mathbf{k},\downarrow )$; $\lambda >0$ is the
interaction constant; $ a_{i,k}^{+}(a_{i,k})$ are the creation
(annihilation) operators for an electron in the state $k$ of the
$i$-th grain.

The superconducting transition temperature can be found by
considering the anomalous Green function $F$ in the presence of an
infinitesimal source of pairs $\Delta $,~\cite{Ovchinnikov73}.
Neglecting fluctuations and interaction effects the anomalous
Green function $F$ is given by the expression~\cite{Abrikosov65}
\begin{equation}
F(\xi ,\omega )=\Delta /(\omega ^{2}+\xi ^{2}),
\end{equation}
where $\xi =\mathbf{p}^{2}/2m-\mu ,$ and $\omega =2\pi T(n+1/2)$
is the fermionic Matsubara frequency. The suppression of the
transition temperature $T_{c}$ is determined by the correction to
the function $F(\xi ,\omega )$
\begin{equation}
\frac{\Delta T_{c}}{T_{c}}={\frac{T}{\Delta }}\,\int d\xi
\,\sum\limits_{\omega }\delta F(\xi ,\omega ),  \label{ovchin}
\end{equation}%
where the function $\delta F(\xi ,\omega )$ represents the leading
order corrections to the anomalous Green function $F(\xi ,\omega
)$ due to fluctuations of the order parameter and Coulomb
interaction.
\begin{figure}[t]
\resizebox{.43\textwidth}{!}{
\includegraphics{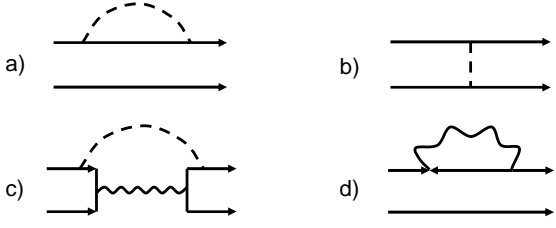}} \vspace{0.6cm}
\caption{Diagrams a) - c) describe the correction to the
superconducting transition temperature due to Coulomb repulsion.
The diagram d) describes correction to the transition temperature
due to superconducting fluctuations. All diagrams are shown before
averaging over the impurities. The solid lines denote the electron
propagators, the dashed lines describe screened Coulomb
interaction and the wavy lines describe the propagator of
superconducting fluctuations~\cite{Beloborodov05super}.}
\label{fluctuations}
\end{figure}
All diagrams (before impurity averaging) contributing to the
suppression of the transition temperature in Eq.~(\ref{ovchin})
are shown in Fig.~\ref{fluctuations}. One can see that there exist
two qualitatively different classes of diagrams. First, the
diagrams a) - c) describe corrections to the transition
temperature due to Coulomb repulsion and correspond to the so
called fermionic mechanism of the suppression of
superconductivity. The second type, diagram d), describes a
correction to the transition temperature due to the
superconducting fluctuations and represents the bosonic mechanism.
The details of calculations are presented in
Ref.~\cite{Beloborodov05super}. They are rather straightforward
and therefore we present and discuss below only the final results
for the suppression of $ T_{c}$.

It is convenient to separate corrections due to the bosonic and
fermionic mechanisms from each other and write the result for the
suppression $\Delta T_{c}$ of the superconductor transition
temperature as
\begin{equation}
{\frac{{\Delta T_{c}}}{{T_{c}}}}=\left( {\frac{{\Delta
T_{c}}}{{T_{c}}}}\right) _{b}+\left( {\frac{{\Delta
T_{c}}}{{T_{c}}}}\right) _{f}, \label{sum}
\end{equation}
where the two terms in the right hand side correspond to the
bosonic and fermionic mechanisms, respectively. The critical
temperature $T_{c}$ in Eq.~(\ref{sum}) is the BCS critical
temperature.

It is shown by~\citet{Beloborodov05super} that at high transition
temperatures, $T_{c}>\Gamma $, the fermionic correction to the
superconducting transition temperature does not depend on the
dimensionality of the sample and takes the form
\begin{equation}
\left( {\frac{{\Delta T_{c}}}{{T_{c}}}}\right)
_{f}=-c_{1}\,\frac{\delta }{ T_{c}},\hspace{0.8cm}d=2,3.
\label{totalTc}
\end{equation}
where $c_{1}=7\zeta (3)/2\pi ^{2}-(\ln 2)/4$ is the numerical
coefficient and $d$ is the dimensionality of the array of the
grains.

In contrast, in the low temperature regime, $T_{c}<\Gamma $, the
fermionic mechanism correction to the superconducting transition
temperature depends on the dimensionality of the sample and is
given by~\cite{Beloborodov05super}
\begin{equation}
\left( \frac{\Delta T_{c}}{T_{c}}\right) _{f}=-\left\{
\begin{array}{lr}
\frac{A}{2\pi \,g}\ln ^{2}\frac{\Gamma }{T_{c}},\hspace{1.6cm}d=3 &  \\
\frac{1}{24\,\pi ^{2}g}\ln ^{3}\frac{\Gamma
}{T_{c}},\hspace{1.3cm}d=2 &
\end{array}
\right. ,  \label{mainresult_fermion}
\end{equation}
where $A=a^{3}\int d^{3}q/(2\pi )^{3}\lambda
_{\mathbf{q}}^{-1}\approx 0.253$ is the dimensionless constant,
$a$ is the size of the grain and $\lambda _{\mathbf{q}}$ is given
by Eq.~(\ref{e240}). Note that in the low temperature regime,
$T_{c}<\Gamma ,$ the correction to the critical temperature in two
dimensions agrees with the one obtained for homogeneously
disordered superconducting films provided the substitution $\Gamma
\rightarrow \tau ^{-1}$ is done.

The correction to the transition temperature due to the bosonic
mechanism in Eq.~(\ref{sum}) is qualitatively different from the
fermionic one. It remains the same in both the regimes $T<\Gamma $
and $T>\Gamma $ and is given by~\cite{Beloborodov05super}
\begin{equation}
\left( \frac{\Delta T_{c}}{T_{c}}\right) _{b}=-\left\{
\begin{array}{lr}
\frac{14A\zeta (3)}{\pi ^{3}}\;{\frac{1}{{g}}},\hspace{1.75cm}d=3 &  \\
{\frac{{\ 7\zeta (3)}}{{2\pi ^{4}g}}}\;\ln \frac{g^{2}\delta
}{T_{c}}, \hspace{1.4cm}d=2 &
\end{array}
\right. ,  \label{mainresult_boson}
\end{equation}
where $\zeta (x)$ is the zeta-function and the dimensionless
constant $A$ is written below Eq.~(\ref{mainresult_fermion}).

Note that the energy scale $\Gamma $ does not appear in this
bosonic part of the suppression of superconducting temperature in
Eq.~(\ref{mainresult_boson}). This stems from the fact that the
characteristic length scale for the bosonic mechanism is the
coherence length $\xi _{eff}$ that is assumed to be much larger
than the size of the single grain. The result for the two
dimensional case in Eq.~(\ref{mainresult_boson}) is written with a
logarithmic accuracy assuming that $\ln (g^{2}\delta /T_{c})\gg
1$.

We see from Eqs.~(\ref{mainresult_fermion}) and
(\ref{mainresult_boson}) that the suppression of the
superconducting transition temperature becomes stronger with
diminishing the coupling $g$ between the grains. Of course, the
calculations may be justified only when the correction to the
transition temperature is much smaller than the temperature
itself.

The above expression for the correction to the transition
temperature due to the bosonic mechanism was obtained in the
lowest order in the propagator of superconducting fluctuations and
holds therefore as long as the value for the critical temperature
shift given by Eq.~(\ref{mainresult_boson}) is larger than the
Ginzburg region $(\Delta T)_{\scriptscriptstyle G}$
\begin{equation}
(\Delta T)_{\scriptscriptstyle G}\sim \left\{
\begin{array}{lr}
\frac{1}{g^{2}}\frac{T_{c}^{2}}{g\delta }\hspace{1.4cm}d=3, &  \\
\frac{T_{c}}{g}\hspace{1.8cm}d=2. &
\end{array}
\right.  \label{G_region}
\end{equation}
Comparing the correction to the transition temperature $T_{c}$
given by Eq.~(\ref{mainresult_boson}) with the width of the
Ginzburg region, Eq.~(\ref{G_region}), one concludes that for $3d$
granular metals the perturbative result~(\ref{mainresult_boson})
holds if
\begin{equation}
T_{c}\ll g^{2}\delta .  \label{intervalt}
\end{equation}
In two dimensions the correction to the transition temperature in
Eq.~(\ref{mainresult_boson}) is only logarithmically larger than
$(\Delta T)_{G}$ in Eq.~(\ref{G_region}). The two dimensional
result (\ref{mainresult_boson}) holds with the logarithmic
accuracy in the same temperature interval (\ref{intervalt}) as for
the three dimensional samples.

Note that inside the Ginzburg region higher order fluctuation
corrections become important. Moreover, the non perturbative
contributions that appear, in particular, due to superconducting
vortices should be taken into account as well. These effects
destroy the superconducting long range order and lead to
Berezinskii-Kosterlitz-Thouless transition in $2d$ systems. The
suppression of the transition temperature in the limit
$T>g^{2}\delta $ in $3d$ should be studied considering strong
critical fluctuations within the Ginzburg-Landau free energy
functional.

It follows from Eq.~(\ref{totalTc}) that in the limit $T\gg \Gamma
$ the fermionic mechanism of the suppression of the
superconductivity is no longer efficient. This can be seen rather
easily in another way using the phase approach presented in
Sec.~\ref{phaseaction}. After the decoupling of Coulomb
term~(\ref{e26}) by the integration over the auxiliary field
$\bar{V} $ and gauge transformation~(\ref{a4}) the phase enters
the tunneling term, Eq.~(\ref{f4}) only. However, this term is not
important in the limit $T\gg \Gamma $ and we conclude that the
long range part of the Coulomb interaction leading to charging of
the grains is completely removed in this limit. Therefore, the
effect of the Coulomb interaction on the superconducting
transition temperature must be small and this is seen from
Eq.~(\ref{totalTc}). This conclusion matches well the fact that
the upper limit in the logarithms in
Eq.~(\ref{mainresult_fermion}) is just $\Gamma $ and, at
temperatures exceeding this energy, the logarithms should
disappear.

At lower temperatures $T<\Gamma $ the phase description is not
applicable and, as a consequence, we obtain the non-trivial
result, Eq.~(\ref{mainresult_fermion}). This result is of the pure
quantum origin, and the interference effects are very important
for its derivation (one should consider a contribution of the
diffusion modes). In the high temperature limit $T\gg \Gamma $ the
interference effects are suppressed and this is the reason why the
fermionic mechanism of the suppression of the superconductivity is
no longer efficient.

The results obtained, Eqs.~(\ref{totalTc}) - (\ref{intervalt}),
suggest an experimental method of extracting the information about
the morphology of the samples from the $T_{c}$ data. Indeed, one
can study the dependence of the superconducting transition
temperature $T_{c}$ of the granular metals as a function of the
tunneling conductance $g$ by comparing several granular samples
with different tunneling conductances (different oxidation
coating). The experimental curves for the suppression of $T_{c}$
should have a different slope at high $T_{c}>\Gamma $ and low
$T_{c}<\Gamma $ critical temperatures due to the fact that the
suppression of the superconductivity is given by the two different
mechanisms. The information on the morphology of the sample,
\textit{i.e.} whether the samples are homogeneously disordered or
granular is then obtained from the dependencies of the critical
temperatures on the tunneling conductance $T_{c}(g)$. (We remind
the reader that the scale $\Gamma $ exists in the granular samples
only).

Another consequence of Eqs.~(\ref{totalTc}) - (\ref{intervalt}) is
the following: since at low critical temperatures $T_{c}\ll \Gamma
$ the suppression of the superconductivity in granular metals is
provided by the fermionic mechanism and coincides, upon the
substitution $\Gamma \rightarrow \tau ^{-1}$, with the proper
result for homogeneously disordered samples,~\cite{Ovchinnikov73},
one can generalize the renormalization group result
by~\citet{Finkelstein87} for the $T_{c}$ suppression. The latter
result obtained for homogeneously disordered films can be directly
applied to the granular superconductors by making the proper
substitution for the diffusion coefficient $D=\Gamma a^{2}$, where
$a$ is the size of the single grain.

\subsection{Magnetoresistance of granular superconductors}

\label{magnetoresistance}

In this subsection we discuss magnetoresistance of a granular
superconductor in a strong magnetic field. We start our
consideration with a discussion of the experiment~\cite{Gerber97}
where transport properties a system of $Al$ superconducting grains
in a strong magnetic field were studied. The samples were three
dimensional and quite homogeneous with a typical grain diameter
$120\pm 20\mathring{A}.$ Bulk superconductivity could be destroyed
by application the strong magnetic field leading to the appearance
of a finite resistivity. Magnetic field above $17T$ was enough to
destroy even the superconducting gap within each grain.

The dependence of the resistivity on the magnetic field observed
in~\citet{Gerber97} was not simple. Although at extremely strong
fields the resistivity was almost independent of the field, it
\textit{increased } when decreasing the magnetic field. Only at
sufficiently weak magnetic fields the resistivity started to
decrease and finally the samples displayed superconducting
properties. A similar behavior had been reported in a number of
publications~\cite{Gantmakher96,Parthasarathy04}.

A negative magnetoresistance due to weak localization effects is
not unusual in disordered metals~\cite{Altshuler80,Lee_review}.
However, the characteristic magnetic field in
experiment~\cite{Gerber97} was several tesla such that all weak
localization effects had to be strongly suppressed.

We want to present an explanation for this unusual behavior based
on consideration of superconducting fluctuations. At first glance
this idea looks counterintuitive, because, naively,
superconducting fluctuations are expected to increase the
conductivity. Nevertheless, it turns out that the
magnetoresistance of a good granulated metal ($g\gg 1$) in a
strong magnetic field, $H>H_{c}$, and at low temperature, $T\ll
T_{c}$, \textit{is} \textit{negative.} In our model, the
superconducting gap in each granule is assumed to be suppressed by
the strong magnetic field. All the interesting behavior considered
below originates from the superconducting fluctuations that lead
to a suppression of the density of states (DOS) and decrease the
conductivity.

Theory of superconducting fluctuations near the transition into
the superconducting state has been developed long ago~\cite
{Aslamazov68,Maki68,Maki68b,Thompson70,Abrahams70} (for a review
see~\citet{Larkinbook}). Above the transition temperature $T_{c}$,
non-equilibrium Cooper pairs are formed and a new channel of
charge transfer opens (Aslamazov-Larkin
contribution)~\cite{Aslamazov68}. This correction always gives
positive contribution to the conductivity. Another fluctuation
contribution comes from a coherent scattering of the electrons
forming a Cooper pair on impurities (Maki-Thompson
contribution)~\cite{Maki68,Maki68b,Thompson70}. This correction
also results in a positive contribution to conductivity, though,
in principle, the sign of it is not prescribed.

Formation of the non-equilibrium Cooper pairs results also in a
fluctuational gap in the one-electron spectrum~\cite{Abrahams70}
but in conventional (non granular) superconductors the first two
mechanisms are more important near $T_{c}$ and the conductivity
increases when approaching the transition.

The total conductivity for a bulk sample above the transition
temperature $T_{c}$ can be written in the following form
\begin{equation}
\sigma =\sigma _{Drude}+\delta \sigma ,  \label{Drud}
\end{equation}%
where $\sigma _{Drude}=(e^{2}\tau n)/m$ is the conductivity of a
normal metal without electron-electron interaction, $\tau $ is the
elastic mean free time, $m$ and $n$ are the effective mass and the
density of electrons, respectively. In Eq.~(\ref{Drud}), $\delta
\sigma $ is a correction to the conductivity due to the Cooper
pair fluctuations
\begin{equation}
\delta \sigma =\delta \sigma _{DOS}+\delta \sigma _{AL}+\delta
\sigma _{MT}, \label{All_corrections}
\end{equation}%
where $\delta \sigma _{DOS}$ is the correction to the conductivity
due to the reduction of the DOS and $\delta \sigma _{AL}$ and
$\delta \sigma _{MT}$ stand for the Aslamazov-Larkin (AL) and
Maki-Thompson (MT) contributions to the conductivity. Close to the
critical temperature $T_{c}$ the AL correction is more important
than both the MT and DOS corrections and its contribution can be
written as follows~\cite{Aslamazov68}
\begin{equation}
\frac{\delta \sigma _{AL}}{\sigma _{Drude}}=\gamma \left(
\frac{T_{c}}{T-T_{c}}\right) ^{\beta },  \label{AL1}
\end{equation}%
where $\gamma $ is a small dimensionless positive parameter,
$\gamma \ll 1$, depending on dimensionality and $\beta =1/2$ for
the three-dimensional case ($3d$), $1$ for $2d$ and $3/2$ for
quasi-$1d$. Equation~(\ref{AL1}) is valid for $\delta \sigma
_{AL}/\sigma _{Drude}\ll 1$.

For quite a long time the superconducting fluctuations have been
studied near the critical temperature $T_{c}$ in a zero or a weak
magnetic field. Only recently the limit of very low temperature,
$T\ll T_{c}$, and a strong magnetic field, $H>H_{c}$ was
considered~\cite{Beloborodov99,Beloborodov00} and some
qualitatively new effects were predicted. In this subsection we
concentrate on this limit.

As in conventional bulk superconductors, we can write corrections
to the classical conductivity $\sigma _{0}$,
Eq.~(\ref{conductivity0}), as a sum of corrections to the DOS,
Aslamazov-Larkin (AL)\thinspace , $\sigma _{AL}$, and
Maki-Thompson (MT), $\sigma _{MT}$, corrections. Diagrams
describing these contributions are represented in
Fig.~\ref{fluctuations}. They have the conventional form but the
wavy lines standing for the superconducting fluctuations should be
written taking into account the granular structure.
\begin{figure}[tbp]
\epsfysize =4cm \centerline{\epsfbox{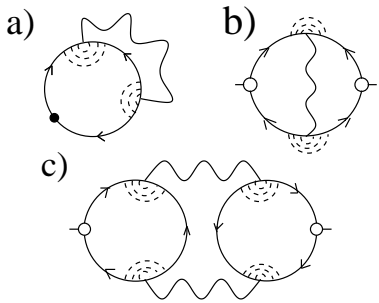}}
\caption{Diagram a) describes correction to DOS, diagrams b) and
c) describe corrections to conductivity due to superconducting
fluctuations. The wavy lines denote the propagator of the
fluctuations, the dashed lines stand for the impurity
scattering~\cite{Beloborodov99}.} \label{fluctuations}
\end{figure}
It turns out that in the regime $T \ll T_c$ and $H > H_c$ the DOS
corrections plays a very important role: This correction reaches
its maximum at $H\rightarrow H_{c}$, where $H_{c}$ is the field
destroying the superconducting gap in the single grain. At zero
temperature and close to the critical field $H_{c}$ such that
$\Gamma /\Delta _{0}\gg h$ , where $h=(H-H_{c})/H_{c}$, the
maximum value $\delta \sigma _{DOS}$ for $ 3d $ granular
superconductors is~\cite{Beloborodov00}
\begin{equation}
\frac{\delta \sigma _{DOS}}{\sigma
_{0}}=-\frac{1}{3g}\,\frac{\Gamma }{\Delta _{0}}\,\ln \left(
\frac{\Delta _{0}}{\Gamma }\right). \label{estDOS0}
\end{equation}
We see from Eq.~(\ref{estDOS0}) that the correction to the
conductivity $\delta \sigma _{DOS}$ i) is negative and its
absolute value decreases when the magnetic field increases, ii) it
is smaller than unity and it becomes comparable with $\sigma _{0}$
when the tunneling conductance $g\sim 1$. However, such values of
$g$ mean that we would be in this case not far from the
metal-insulator transition. Then, we would have to take into
account all localization effects, which is not done here.

Even in the limit of strong magnetic fields, $H\gg H_{c}$, the
correction to $\sigma _{0}$ can still be
noticeable~\cite{Beloborodov00}
\begin{equation}
\frac{\delta \sigma _{DOS}}{\sigma
_{0}}=-\frac{1}{3g}\,\frac{\Gamma }{E_{0}(H)}\,\ln ^{-1}\left[
\frac{E_{0}(H)}{\Delta _{0}}\right] , \label{highfield}
\end{equation}%
where, for spherical grain, $E_{0}\left( H\right)
=\frac{2}{5}\left( \frac{eHR}{c}\right) ^{2}D$ with $D$ being the
diffusion coefficient in the single grain.
Equation~(\ref{highfield}) shows that in the region $H\gg H_{c}$
the correction to the conductivity behaves essentially as $\delta
\sigma _{DOS}\sim H^{-2}$, which is a rather slow decay.

We emphasize that the correction to the conductivity coming from
the DOS, Eqs.~(\ref{estDOS0}) and (\ref{highfield}) remains finite
in the limit $ T\rightarrow 0$, thus indicating the existence of
the virtual Cooper pairs even at $T=0$.

In order to calculate the entire conductivity we must investigate
the AL and MT contributions (Figs.~\ref{fluctuations}c and
\ref{fluctuations}b). Near $T_{c}$ these contributions are most
important leading to an increase of the conductivity. At low
temperatures $T\ll T_{c}$ and strong magnetic fields $H>H_{c}$ the
situation is completely different. It turns out that both the AL
and MT contributions \textit{vanish} in the limit $T\rightarrow 0$
at all $H>H_{c}$ and thus, the correction to the conductivity
comes from the DOS only.

So, at low temperatures, estimating the total correction to the
classical conductivity $\sigma _{0}$, Eq.~(\ref{a4}), one may use
the formulae~(\ref{estDOS0}) and (\ref{highfield}). We present the
final result for AL correction shown in Fig.~\ref{fluctuations}c
at $T\ll T_{c}$ and $h\ll \Gamma /\Delta _{0}$ for $3d$ granular
superconductors~\cite{Beloborodov00}
\begin{equation}
\frac{\delta \sigma _{AL}}{\sigma _{0}}\sim
\frac{1}{g}\,\frac{T^{2}}{\Delta _{0}^{3/2}\Gamma ^{1/2}}\,\left(
\frac{H_{c}}{H-H_{c}}\right) ^{3/2}. \label{estALT0}
\end{equation}
It follows from Eq.~(\ref{estALT0}) that at low temperatures the
AL correction to the conductivity is proportional to the square of
the temperature $\delta \sigma _{AL}\sim T^{2}$ and vanishes in
the limit $T\rightarrow 0$.

For MT corrections shown in Fig.~\ref{fluctuations}b one obtains
in the limit $T\ll T_{c}$ and $h\ll \Gamma /\Delta _{0}$ for $3d$
granular superconductors~\cite{Beloborodov00}
\begin{equation}
\frac{\sigma _{MT}}{\sigma _{0}}\sim \frac{1}{g}\,\frac{\Gamma
}{\Delta _{0}} \,\frac{T^{2}}{\Delta _{0}^{2}}.  \label{estMT0}
\end{equation}

The temperature and magnetic field dependence of $\delta \sigma
_{AL}$ and $ \delta \sigma _{MT}$ is rather complicated but they
are definitely positive. The competition between these corrections
and $\sigma _{DOS}$ determines the sign of the magnetoresistance.
One can see from Eqs.~(\ref{estALT0}) and (\ref{estMT0}) that both
the AL and MT contributions are proportional at low temperatures
to $T^{2}$. Therefore the $\sigma _{DOS}$ in this limit is larger
and the magnetoresistance is negative for all $H_{c}$. In
contrast, at $T\sim T_{c}$ and close to $H_{c}$, the AL and MT
corrections can become larger than $\sigma _{DOS}$ resulting in a
positive magnetoresistance in this region. Far from $H_{c}$ the
magnetoresistance is negative again.

In the above considerations we did not take into account
interaction between the magnetic field and spins of the electrons.
This approximation is justified if the size of the grains is not
very small. Then, the orbital magnetic critical field
$H_{c}^{or\text{ }}$ destroying the superconducting gap is smaller
than the paramagnetic limit $\mu H_{c}^{Z} \approx \Delta _{0}$
and the orbital mechanism dominates the magnetic field effect on
the superconductivity. However, the Zeeman splitting leading to
the destruction of the superconducting pairs can become important
if one further decreases the size of the grains. The results for
conductivity corrections including Zeeman splitting can be found
in Ref.~\cite{Beloborodov00}.

The results presented in this subsection show that, even if the
superconducting gap in each granule is destroyed by the magnetic
field, the virtual Cooper pairs can persist up to extremely strong
magnetic fields. However, the contribution of the Cooper pairs to
transport is proportional at low temperatures to $T^{2}$ and
vanishes in the limit $T\rightarrow 0$. In contrast, they reduce
the one-particle density of states in the grains even at $T=0$,
thus diminishing the macroscopic conductivity. The conductivity
can reach its classical value only in extremely strong magnetic
fields when all the virtual Cooper pairs do not exist anymore.
This leads to the negative magnetoresistance.

\begin{figure}[t]
\epsfysize =4.0cm \centerline{\epsfbox{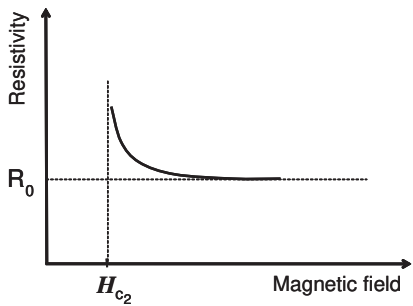}}
\caption{Schematic picture for resistivity behavior of the
granulated superconductors as a function of the magnetic field at
low temperatures $ T\ll T_{c}$. The resistivity at low
temperatures grows monotonously when decreasing the magnetic
field. It reaches asymptotically the value of the classical
resistivity $R_{0}$ only at extremely strong magnetic fields. The
transition into the superconducting state occurs at a field
$H_{c_{2}}$~\cite{Beloborodov00}.} \label{magnetoresistance_fig}
\end{figure}

Qualitatively, the results for resistivity behavior of granular
superconductors at low temperatures $T\ll T_{c}$ and strong
magnetic fields $ H>H_{c}$ are summarized in
Fig.~\ref{magnetoresistance_fig}. The resistivity $R$ at low
temperatures grows monotonously when decreasing the magnetic field
and it reaches asymptotically the value of the classical
resistivity $ R_{0}$ only at extremely strong magnetic fields. The
transition into the superconducting state occurs at a field
$H_{c_{2}}$ that may be lower than $ H_{c}$

\section{Discussion of the results}

\label{discussions}

\subsection{Quantitative comparison with experiments}

The models of the normal and superconducting granular electronic
systems considered in this review are relevant to many experiments
on different materials and we have presented quite a few of them.
This is not accidental because we selected for the presentation
such theoretical results that could be relevant for existing
experiments.

On the one hand, the review is not as complete as it could be
because we did not include many interesting theoretical works
where such physical quantities, as tunneling density of
states~\cite{BeloborodovDOS}, thermal
conductivity~\cite{Beloborodovthermal,Loh05,Biagini05}, etc. were
considered. The only reason for this omission is that at present
we are not aware of clear measurements of these quantities in the
granular materials.

On the other hand, this gave us the opportunity to concentrate on
experimentally well studied phenomena. Some of them, like the
logarithmic temperature dependence of the conductivity of well
conducting granular metals or the hopping like temperature
dependence of the insulating very regular granular systems
remained unexplained for several decades. As we have seen, these
dependencies and many interesting features of the experiments can
be explained using rather simple models of the metallic grains.
The simplicity of the models and a small number of parameters
characterizing the system suggests that there should be not only a
good qualitative but also a quantitative agreement between the
theory and experiment.

So, we present now a quantitative comparison of several
theoretical results with experimental data. Of course, we cannot
discuss here all existing measurements and therefore only three
different experiments are considered.

\subsubsection{Logarithmic temperature dependence of the conductivity}

\label{logarithm}

In this subsection we will  compare with experiments the
logarithmic temperature dependence of the conductivity,
Eq.~(\ref{cond_general}). This formulae should be applicable at
not very low temperature that simplifies the experimental check.
Specially prepared granular metals like those considered in
Refs.~\cite{Gerber97,Abeles75,Simon} should be the first objects
of the application of the theory developed.

As an example, we compare now Eq.~(\ref{cond_general}) with the
experimental results of~\citet{Gerber97} on films made of $Al$
grains embedded in an amorphous $Ge$ matrix. At low temperatures,
superconductivity in $Al$ grains was destroyed by a strong
magnetic field. Depending on the coupling between the grains
(extracted from the conductivity at room temperatures) the samples
of the experiment~\cite{Gerber97} were macroscopically either in
an insulating state with the temperature dependence of the
resistivity $R\sim \exp \left( a/T^{1/2}\right) $ or in a
\textquotedblleft metallic\textquotedblright\ one. However, the
resistivity of the metallic state depended on temperature and the
authors suggested a power law, $R \sim T^{\alpha}$, with an
exponent $\alpha \ll 1.$  As the exponent $\alpha $ for the
\textquotedblleft metallic\textquotedblright\ sample was small we
can argue that Eq.~(\ref{cond_general}) should not be worse for
fitting the experimental data. Then, we can estimate the exponent
$\alpha $ without using fitting parameters.

The sample of the experiment~\cite{Gerber97} had the room
temperature resistivity $R_{0}=7.3\times 10^3 \Omega $cm. The
diameter of the grains was $120\pm 20$\AA $,$ which allows, using
the value $\hbar /e^{2}=4.1\times 10^{-3}\Omega $, and the
relation between the conductivity and conductance for 3d samples
$\sigma = 2 e^2 g /\hbar\, a,$  to estimate the dimensionless
tunneling conductance as $g=0.34.$ Taking $d=3$ and $(z=6)$ in
Eq.~(\ref{cond_general}) we obtain $\alpha =0.15$, which agrees
fairly well with the experimental value $\alpha =0.117$. Keeping
in mind that the theoretical dependence~(\ref{cond_general}) was
obtained for a periodic cubic array while in realistic samples the
coordination number can be only approximately close to 6 one can
hardly hope for a better agreement.

A logarithmic dependence of the resistivity on temperature has
been observed in other granular materials. In Ref.~\cite{Simon} a
granular cermet consisting of $Nb$ grains in a boron nitride
insulating matrix was studied. Again, at small coupling between
the grains the temperature dependence of the resistivity $\exp
\left( a/T^{1/2}\right) $ was observed in a very broad interval of
temperatures. The resistivity of samples with a strong coupling
between the grains was very well described by the law
\begin{equation}
R=R_{0}\ln \left( T_{0}/T\right),  \label{k1}
\end{equation}
which is close to Eq.~(\ref{cond_general}) provided the
temperature interval is not very large, such that the variation of
the resistivity is small. However, Eq.~(\ref{k1}) gave a good
description for the temperature dependence of the resistivity in a
very broad region and the changing of the resistivity was not
small. The reason for the applicability of Eq.~(\ref{k1}) in a so
broad interval of temperatures is not clear because according to
the results of the renormalization group analysis of
Sec.~\ref{not.low} not the resistivity but the conductivity should
obey Eq.~(\ref{k1}). A more careful experimental study might
clarify this question.

The unusual logarithmic behavior of the type, Eq.~(\ref{k1}), has
been observed not only in \textquotedblleft
standard\textquotedblright\ granular systems but also in
high-$T_{c}$ cuprates at very strong magnetic fields. The first
observation of this dependence was done on underdoped $
La_{2-x}Sr_{x}CuO_{4}$ crystals~\cite{boeb}. The superconductivity
in this experiment was suppressed with pulsed magnetic fields of
$61T$. It was found that both the in-plane resistivity $\rho
_{ab}$ and out-of-plane resistivity $\rho _{c}$ diverged
logarithmically with decreasing the temperature. This means that a
$3d$ effect was observed in a very strong magnetic field and
traditional explanations for a logarithmic behavior like a weak
localization or the Kondo effect could not clarify the situation.

In a subsequent publication~\cite{boeb1} a metal-insulator
crossover was observed in the same material at a $Sr$
concentration near optimum doping ($x\simeq 0.16$). In underdoped
samples both $\rho _{ab}$ and $\rho _{c}$ showed no evidence of
saturation at low temperatures and diverged as the logarithm of
the temperature. The authors called this state \textquotedblleft
insulator\textquotedblright\ in contrast to the state at high
doping where the resistivity did not have a pronounced dependence
on the temperature. It was conjectured by~\citet{boeb1} that the
logarithmic behavior they observed might be related to the one
seen in the experiment~\cite{Simon} on granular $Nb$. This would
demand a phase segregation throughout the underdoped regime of
$LSCO$.

We hope that our results for the model of the granular materials
may be applicable to the experiments on the
$La_{2-x}Sr_{x}CuO_{4}$ crystals~\cite{boeb,boeb1}, which would
mean that the underdoped crystals have a granular structure and
the logarithmic behavior is due to the Coulomb interaction. The
transition to the metallic state of Refs.~\cite{boeb,boeb1} would
mean that at higher doping the granularity disappears. A
quantitative comparison of Eq.~(\ref{cond_general}) with the data
of Refs.~\cite{boeb,boeb1} was done in~\citet{Beloborodov03} and a
good agreement was found.

What is interesting, a microscopic granularity was directly
experimentally observed in the superconducting state of
$Bi_{2}Sr_{2}CaCu_{2}O_{8+\delta }$ by the STM probe~\cite{Lang}.
As this cuprate is rather similar in its structure to those
studied in Refs.~\cite{boeb,boeb1} the assumption that the
materials studied there were granular does not look groundless.

The logarithmic dependence of the resistivity on temperature has
also been observed in many other experiments. For example,
in~\citet{gerber90} this dependence was observed in granular $Pb$
films. It was also observed in phase compounds of
$Nd_{2-x}Ce_{x}CuO_{4-y}$~\cite{radha}.

\subsubsection{Hopping conductivity}

As we have mentioned in the previous subsection, the logarithmic
temperature dependence is usually seen in samples with a good
coupling between the grains. If the coupling is weak, one observes
usually the Efros-Shklovskii law, Eq.~(\ref{e6}).

According to the theory developed in Sec.~\ref{hopping} the
characteristic temperature $T_{0}$ entering Eq.~(\ref{e6}) is
determined by Eqs.~(\ref{T0}) and (\ref{p6}) for the elastic and
inelastic co-tunneling, respectively.

Now we want to compare our results for the VRH conductivity of
granular metals obtained in Sec.~\ref{hopping} with the most
recent experimental data. In Ref.~\cite{Tran05} the zero-bias
conductivity was studied for bilayers, trilayers, tetralayers and
thick films of $Au$ nanoparticles with the particle diameters
around $5.5$ \textrm{nm} and their dispersion less than $5\%$.

Figure~\ref{Tran05_picture} demonstrates that the zero-bias
conductivity follows $\sigma(T)\sim \exp [-(T_{0}/T)^{1/2}]$ over
the range $30-90K$ for the multilayers and $30-150K$ for the thick
films. The fits indicated by the lines give $T_{0}=(4.00\pm
0.02)\times 10^{3}K$ and $(3.00\pm 0.01)\times 10^{3}K$ for the
multilayers and thick films, respectively.

At temperatures slightly exceeding $100K$, the conductivity for
the multilayers starts deviating from the low-temperature behavior
(dashed lines) and crosses over to Arrhenius behavior $\sigma(T)
\sim \exp [-U/k_{B}T]$ (Fig.~\ref{Tran05_picture}, inset) with the
activation energies $U/k_{B}\approx 320\pm 8K$.

Associating this high-temperature behavior with the
nearest-neighbor tunneling between the particles we can use
$U\approx 0.2E_{c}$, in analogy with the
monolayers~\cite{Parthasarathy04}. This gives us an estimate $
E_{c}\approx 1600K,$ where $E_{c}=e^{2}/\tilde{\kappa}a$ is the
Coulomb charging energy of an individual grain, expressed in terms
of the grain radius $a$, the electron charge $e$, and the
dielectric constant of the surrounding medium $\tilde{\kappa}$.
The charging energy $E_{c}\approx 1600K$ for this system leads to
$\tilde{\kappa}\approx 4$.

The energy scale $T_{0}$ is related to the localization length
$\xi $ by Eqs.~(\ref{T0}, \ref{p6}). Using the above values for
$\tilde{\kappa}$ and $T_{0}$, we find $3nm<\xi <4nm$ for both the
multilayers and thick films, which corresponds to the localization
within one grain. This is an important check because larger values
would imply strongly coupled clusters of the grains.

The typical hopping distance $r^{\ast }\left( T\right) $ is given
by Eq.~( \ref{e101}). The number of grains $N^{\ast }$ involved in
a typical hop is $ N^{\ast }=r^{\ast }/d_{0}$, with a
center-to-center distance $d_{0}\approx 8nm$ between neighboring
grains. At $T=10K$ this leads to $N^{\ast }=4$ for multilayers and
$N^{\ast }=4-5$ for the thick films.

As temperature increases, $N^{\ast }$ decreases down to $N^{\ast
}\sim 1$ and this is the point of the crossover to a standard
activation transport, Eq.~(\ref{activation1}). For the
multilayers, this happens at $T\approx 90-95K$ but for the thick
films only above $T\approx 130K$. Both the estimates are in
excellent agreement with the data in Fig.~\ref{Tran05_picture}.

\subsubsection{Negative magnetoresistance}

\label{sec7}

Let us compare the theoretical predictions on the negative
magnetoresistance presented in Sec.~\ref{magnetoresistance} with
the available experimental results of~\citet{Gerber97}. In that
work three samples were studied. We will concentrate our attention
on the sample $2$, Fig.~4, of~\citet{Gerber97} that has a metallic
conductivity behavior above the critical temperature.

We analyze the case of low temperatures $T\ll T_{c}$ and magnetic
fields $ H>H_{c}$, where $T_{c}\approx 1.6K$ is the critical
temperature for $Al$ grains studied in the experiment and $H_{c}$
is the critical magnetic field
that suppresses the superconductivity in a single grain, $%
E_{0}(H_{c})=\Delta _{0}$, where $E_{0}(H)$ was defined below
Eq.~(\ref{highfield}). At temperature $T\simeq 0.3K$ and magnetic
field $H\simeq 4T$ this sample shows a large negative
magnetoresistance. The resistivity has the maximum at $H=2.5T$ and
the value of this peak is more than twice as large as the
resistivity in the normal state (that is, at $H\gg H_{c}$, when
all superconducting fluctuations are completely suppressed). A
negative magnetoresistance due to weak localization (WL) is
typical for disordered metals and in order to describe the
experimental data, its value should be estimated along with the
effects of the superconducting fluctuations discussed in the
previous subsection.

The total conductivity of the granular metal under consideration
including effects of WL and superconducting fluctuations can be
written in the form:
\begin{equation}
\sigma =\sigma _{0}+\delta \sigma _{DOS}+\delta \sigma
_{AL}+\delta \sigma _{MT}+\delta \sigma _{WL}.  \label{totalcond}
\end{equation}%
At low temperatures, $T\ll T_{c}$, the contribution $\delta \sigma
_{DOS}$ originating from the reduction of DOS due to the formation
of the virtual Cooper pairs is larger than the contributions
$\delta \sigma _{AL}$ and $ \delta \sigma _{MT}$ since the latter
vanish in the limit $T\rightarrow 0$. So, let us concentrate on
estimating the contributions $\delta \sigma _{DOS}$ and $\delta
\sigma _{WL}$.

It is not difficult to show that in the case under consideration
the weak localization corrections originating from a contribution
of Cooperons are strongly suppressed by the magnetic field. Using
approximations developed by~\citet{Beloborodov99} one can easily
obtain the following expression for a $3d$ cubic lattice of
metallic grains at tunneling conductances $g\ll E_{0}(H)/\delta $
\begin{equation}
\frac{\delta \sigma _{WL}}{\sigma _{0}}\sim -\frac{1}{g}\left[
\frac{\Gamma }{E_{0}\left( H\right) }\right] ^{2} . \label{exp3}
\end{equation}
Equation~(\ref{exp3}) shows that the weak localization correction
in the strong magnetic fields $H>H_{c}$ considered here is always
small. Comparing Eq.~(\ref{exp3}) with Eq.~(\ref{estDOS0}) for DOS
correction one can see that in the limit $g\ll \Delta _{0}/\delta
$ the contribution from the weak localization correction can be
neglected.

Now, let us estimate the corrections $\delta \sigma _{DOS}$ and
$\delta \sigma _{WL}$ using the parameters of the
experiment~\cite{Gerber97}. For the typical diameter $120\pm
20\mathring{A}$ of $Al$ grains studied by~\citet{Gerber97}, the
mean level spacing $\delta $ is approximately $\delta \approx 1K$.
Using the critical temperature $T_{c}\simeq 1.6K$ for $Al$ we
obtain for the BCS gap in a single grain the following result
$\Delta _{0}\approx 1.8T_{c}\approx 3K$. Substituting the
extracted values of the parameters into Eq.~(\ref{estDOS0}) we can
estimate the maximal increase of the resistivity. As a result, we
obtain $\left( \delta \rho /\rho _{0}\right) _{\max }\approx 0.4$,
which is somewhat smaller but not far from the value $\left(
\delta \rho /\rho \right) _{\exp }\approx 1$ observed
experimentally.

Although Eq.~(\ref{estDOS0}) gives smaller values of $\left(
\delta \rho /\rho _{0}\right) _{\max }$ than the experimental
ones, the discrepancy cannot be attributed to the weak
localization effects. Using the experimental values of the
conductance $g$, mean level spacing $\delta $ and the
superconducting gap $\Delta _{0}$ we find from Eq.~(\ref{exp3})
that $\delta \sigma _{WL}$ is $10$ times smaller than $\delta
\sigma _{DOS}$. The value of the correction $\delta \sigma
_{WL}/\sigma _{0}$, Eq.~(\ref{exp3}), near $H_{c}$ equals
$2.8\times 10^{-2}$.

At the same time, we should not take this disagreement too
seriously because all the results were obtained under the
assumption of a large conductance $ g\gg 1$, while experimentally
this parameter is not large. As the experimental value of $\delta
\sigma /\sigma _{0}$ is not small, we come again to the conclusion
that the parameters of the system are such that the
Eq.~(\ref{estDOS0}) is no longer applicable and one should develop
a more sophisticated theory to describe this region more
accurately.

\subsection{Outlook}

We discussed in this review a rather simple general model that allowed us to
understand many properties of granular materials. This is a model of
disordered or chaotic grains coupled to each other by the electron
tunneling. A very important ingredient of the model is the long range part
of the Coulomb interaction taken in the form of the charging energy of the
grains. For the description of the superconducting grains we included the
superconducting BCS gap into the consideration.

We considered the limit of temperatures $T$ exceeding the mean level spacing
$\delta $ in the grains and neglected quantum size effects. Considering the
superconducting grains we assumed that the superconducting gap $\Delta $ was
larger than $\delta $. These regimes are most easily achieved experimentally
and this was the main reason for choosing them.

In spite of its simplicity, a variety of interesting phenomena has
been derived from the model involved. Such phenomena as the
logarithmic dependence of the conductivity on temperature or
Efros-Shklovskii law for a regular system of grains with an almost
perfect shape remained unexplained for several decades and have
been clarified within the model only recently. Another interesting
effect concerns the negative magnetoresistance due to
superconducting fluctuations. For these effects we have obtained
not only qualitative but a good quantitative description.

The granularity is a rather general phenomenon and it may be
encountered rather unexpectedly. Thin metallic films are often
rather granular than homogeneously
disordered,~\cite{Gantmakher96,Goldman98}. Another unexpected
conclusion about the granularity of some underdoped high $T_{c}$
cuprates has been arrived at using STM technique,~\cite{Lang}.

Studying the granular systems may help to understand some
properties of homogeneously disordered metals and superconductors.
As all scales involved in the model we studied are much larger
than the electron wavelength, it might in some cases simplify
explicit calculations. This concerns, in particular, study of the
regime of $g\sim 1$. We presented arguments that the
metal-insulator transition is possible at least in three
dimensions. The superconductor-insulator transition is also
possible. Studying these transitions is the most challenging
extension of the present study but this may be simpler than
investigation of strongly disordered systems with interaction.

From the experimental side, an evident extension of the works presented here
is fabricating grains made of ferromagnets and studying properties of such
systems. One can couple these grains directly or put them into normal metals
or superconductors. One can make superconducting grains and embed them into
normal metals or not very strong ferromagnets, etc. All these systems
promise very unusual properties.

One of the examples of an unexpected behavior in the system of
ferromagnetic grains imbedded in a superconductor is inducing a
magnetic moment in the superconductor over large distances of the
order of the size of the Cooper pairs. The direction of this
magnetic moment is opposite to that of the ferromagnet and one
comes to a phenomenon of the \textquotedblleft spin
screening\textquotedblright\,~\cite{bergeret,bergeretrmp}.
Different directions of the magnetic moments of the grains may
lead to the phenomenon of an \textquotedblleft odd
triplet\textquotedblright\
superconductivity,~\cite{bergeretodd,bergeretrmp}.

The list of new theoretical and experimental possibilities that are
anticipated in granular systems can be continued. Taking into account the
growing number of existing and potential industrial applications of the
granular materials we are confident that all this is only the beginning of
an exciting development.

\section*{Acknowledgements}

We thank Igor Aleiner, Anton Andreev, Mikhail Feigel'man,
Alexander Finkel'stein, Frank Hekking, Alexei Koshelev for useful
discussions. We especially grateful to Anatoly Ivanovich Larkin
for invaluable discussions. We thank Thu Tran, Heinrich Jaeger,
Xiao-Min Lin for continues communication of relevant experimental
data and stimulating discussions. K.~E. thanks SFB Transregio 12
of DFG and DIP for a financial support.

\appendix

\section{Calculation of the tunneling probability $P_{el}$ in the elastic
regime, Eq.~(\protect\ref{P_el_result})}

\label{elastic_hopping}

In this appendix we derive the probability of elastic electron
tunneling $ P_{el}$ between two distant grains through a chain of
other grains, Eq.~(\ref{P_el_result}). Such a probability can be
found comparatively easily only for the model with the diagonal
Coulomb interaction $E_{ij}^{c}=E_{i}\delta _{ij}.$ For this
reason we consider first such a simplified model and then discuss
the generalization of the results obtained for the case of
realistic capacitance matrices $E_{ij}^{c}$ with nonzero
off-diagonal elements.

In the model with the diagonal Coulomb interaction, the
electron-hole excitation energies $E_{i}^{\pm }$ that will enter
the final results for the hopping probability are given by
\begin{equation}
E_{i}^{\pm }=E_{c}^{i}-V_{i},
\end{equation}%
where $V_{i}$ is the local potential that models the presence of
the electrostatic disorder. These energies have to be strictly
positive and larger than the energies of the initial and final
states otherwise the tunneling path could have been cut into two
or more independent parts. This allows us to assume that the
temperature is less than all $E_{i}^{\pm }.$ For this reason we
will consider the probability of the elastic process in the limit
$T\rightarrow 0.$

The tunneling probability $P$ between the sites $i_{0}$ and
$i_{N+1}$ is proportional to the square of the absolute value of
the amplitude $A_{i_{0},i_{N+1}}$ of the corresponding tunneling
process
\begin{equation}
A_{i_{0},i_{N+1}}=<0|\;\hat{c}_{i_{N+1}}\hat{S}\;\hat{c}_{i_{0}}^{+}\,|0>.
\end{equation}%
Here $\hat{S}$ is the evolution operator written in the
interaction representation that takes into account only the
tunneling part of the Hamiltonian~(\ref{a02}). The conjugate
amplitude $A^{\ast }$ can be written as the probability of the
inverse process of the tunneling between $i_{N+1}-$st and
$i_{0}-$th grains
\begin{equation}
A_{i_{0},i_{N+1}}^{\ast
}=A_{i_{N+1},i_{0}}=<0|\;\hat{c}_{i_{0}}\,\hat{S}\;
\hat{c}_{i_{N+1}}^{+}\,|0>.
\end{equation}

The probability of the tunneling process $P=A^{\ast }A$  can be
found by perturbative expansion of the amplitude  in the tunneling
matrix elements $t_{ij}$ and further averaging over the
realizations of $t_{ij}.$ At the same time the Coulomb interaction
cannot be considered perturbatively and should be taken into
account exactly. In order to construct a proper perturbative
expansion we use the gauge transformation described in
Sec.~\ref{gauge} that allows us to transfer the strong Coulomb
interaction into the phase factors that accompany the tunneling
matrix elements (see Eqs. (\ref{a4}, \ref{tilde_t}).

Fluctuations of the phases $\varphi _{i}\left( \tau \right) $ are
governed by the Coulomb action~(\ref{a10}) and we obtain the
following expression for the correlation function $\Pi \left( \tau
\right) $ that plays a very important role for a description of
the insulating state (c.f. Eq. (\ref{s4} ))
\begin{eqnarray}
\Pi (\tau _{1}-\tau _{2}) &\equiv &\langle \exp \left( i\phi
_{i}(\tau
_{1})-i\phi _{i}(\tau _{2})\right) \rangle  \label{phase_correlator} \\
&=&\exp \left( -E_{c}^{i}|\tau _{1}-\tau _{2}|-V_{i}(\tau
_{1}-\tau _{2})\right) .  \nonumber
\end{eqnarray}%
Since we are interested in the optimal tunneling path we will
consider only trajectories with no return points (no loops). This
simplifies the consideration substantially because the phases on
different sites are not correlated for the diagonal model under
consideration.

\begin{figure}[tbp]
\includegraphics[width=3.0in]{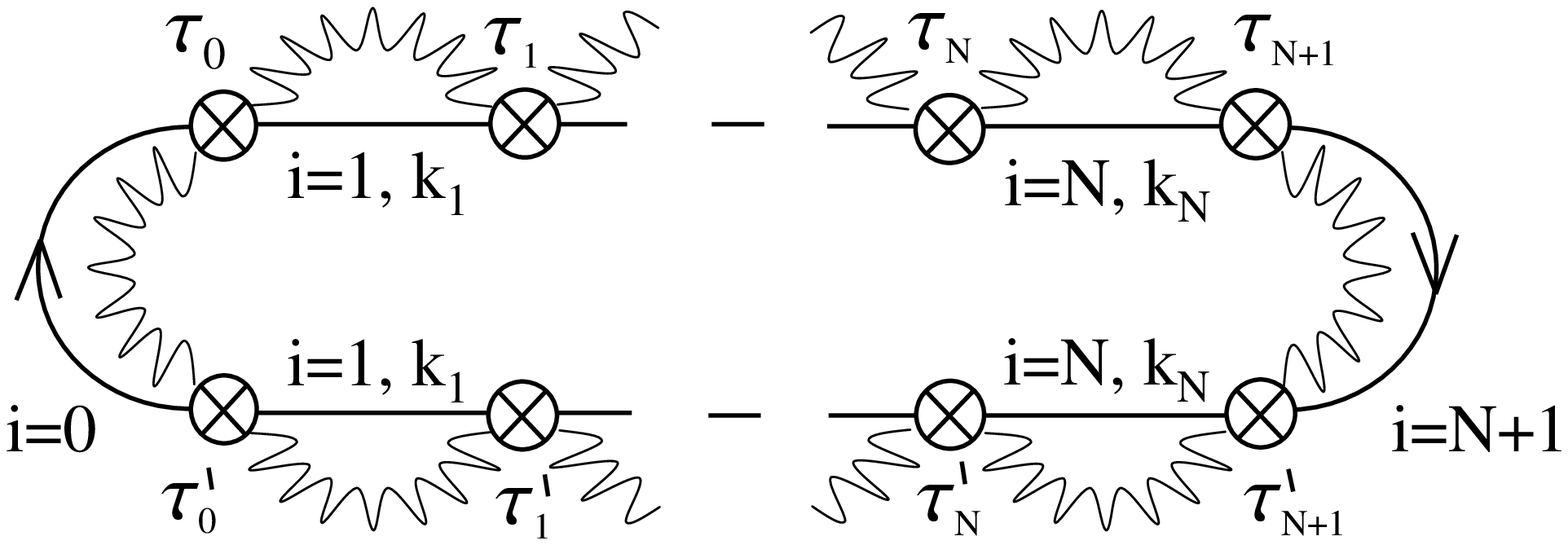} \vspace{-0.3cm}
\caption{ This diagram represents the tunnelling probability via
elastic co-tunneling processes. The crossed circles represent the
tunnelling matrix elements $t^{ij}_{k,k^\prime}\,
e^{i\protect\phi_i(\protect\tau)} \, e^{-i
\protect\phi_j(\protect\tau)}$ where phase factors appear from the
gauge transformation. Wavy lines represent the average of the
phase factors $ \langle e^{i\protect\phi(\protect\tau_1)} \,
e^{-i\protect\phi(\protect\tau_2)} \rangle$ with respect to the
Coulomb action. } \label{Elastic_Cotunneling}
\end{figure}

The gauge transformation approach allows us to represent the
tunneling probability $P$ by the diagram shown in
Fig.~\ref{Elastic_Cotunneling}. In order to simplify the
derivation we work now using the basis of the \textit{exact}
eigenstates of the single particle Green functions that
automatically takes into account the presence of the disorder
within each grain. An alternative procedure based on the momentum
representation leading to the same results would require the
averaging over disorder in each grain and inclusion of the
diffusion propagators. The Green function lines on the initial
$i_{0}$ and final $i_{N+1}$ grains describe the processes for a
particle located on these sites, while the intermediate Green
function lines represent the tunneling amplitudes $A$ (upper part)
and $A^{\ast }$ (lower part). The wavy lines denote the
correlation function~(\ref{phase_correlator}) that takes into
account the Coulomb correlations.

It may seem surprising that the two wavy lines that belong to the
same $i-$th site are consider independently, i.e. that four
exponent correlation function is factorized in a certain way into
the two second order ones. One can check however that this
factorization is justified indeed as long as the time intervals
$(\tau _{i-1},\tau _{i})$ and $(\tau _{i-1}^{\prime },\tau
_{i}^{\prime })$ do not overlap.

This factorization can most easily be understood by presenting the
correlation function of $N$ phase exponents as the exponent of the
interaction energy of $N$ charges interacting via a
\textit{linear} one dimensional Coulomb potential. The two nearby
charges with the opposite signs create an "electric field" only in
the region between the two. Thus, the energy of the dipoles can be
considered independently as long as these dipoles do not overlap
geometrically. For a detailed discussion of the Coulomb analogy we
refer the reader to~\citet{Beloborodov05}.

It follows from this discussion that all the electron Green
functions in the diagram Fig.~\ref{Elastic_Cotunneling} are
accompanied by the correlation function $\Pi (\tau ).$ For this
reason, it is convenient to introduce a modified Green function
$\tilde{G}(\tau )$ in the following way
\begin{equation}
\tilde{G}(\tau )=G(\tau )\Pi (\tau ).  \label{ap1}
\end{equation}%
The Green function $\tilde{G}(\tau )$ has a very simple form in
the time representation: For positive arguments the Green function
describes the electron excitation~\cite{Abrikosov65}
\begin{equation}
\tilde{G}_{\xi _{i}}(\tau >0)=[n(\xi _{i})-1]\;e^{-\xi _{i}\tau
-E_{i}^{+}\tau },
\end{equation}%
where $\xi _{i}$ is the bare single particle electron energy
counted with respect to the Fermi energy, Eq.~(\ref{s0a}) and
$n(\xi _{i})$ is the Fermi distribution function. One can see that
the Coulomb part of the electron excitation energy $E_{i}^{+}$
appears naturally in addition to the single particle energy $\xi
_{i}.$ For negative time arguments the Green function $\tilde{G}$
describes the hole excitation and is given by
\begin{equation}
\tilde{G}_{\xi _{i}}(\tau <0)=n(\xi _{i})\;e^{-\xi _{i}\tau
+E_{i}^{-}\tau },
\end{equation}%
where $\xi _{i}<0$ is the bare electron energy counted with
respect to the Fermi energy and $E_{i}^{-}$ is the Coulomb part of
the hole excitation energy.

Calculating the diagram in Fig.~\ref{Elastic_Cotunneling} we
further note that the intermediate time intervals are of the order
of the inverse Coulomb energy: $|\tau _{i+1}-\tau _{i}|,|\tau
_{i+1}^{\prime }-\tau _{i}^{\prime }|\sim E_{c}^{-1}.$ In the
limit of strong Coulomb interaction ($E_{c}\gg \delta $)
considered here, we can integrate over the intermediate times
independently. Each intermediate block gives rise to the factor
\begin{equation}
P_{k}={{ \Gamma_k } \over {2\pi } } \int_{-\infty }^{\infty }d\xi
\int \,d\tau _{1}\,d\tau _{2}\tilde{G}_{\xi }(\tau
_{1})\,\tilde{G}_{\xi }(\tau _{2})={\frac{ \Gamma_k }{{\pi
\tilde{E}_{k}}}},
\end{equation}%
where the energy $\tilde{E}_{k}$ is the combination of the
energies $E_{k}^{+}$ and $E_{k}^{-}$ defined by
Eq.~(\ref{tilde_E}) and $\Gamma_k = g_k \delta_k,$ with $g_k$
being the conductance between $k-$th and $k+1-$st grains.  We note
that in the insulating system under consideration the energy scale
$\Gamma$ cannot be any more  interpreted as the escape rate,
nevertheless its introduction is convenient even in this regime.

In order to determine the time dependence of the tunneling process
we notice that the time intervals $\tau _{0}^{\prime }-\tau _{0}$
and $\tau _{N+1}^{\prime }-\tau _{N+1}$ coincide within the
accuracy of the inverse charging energy. In the limit of the
strong Coulomb interaction one can take both the intervals to be
equal to the "instanton" time $\tau $ that electron spends out of
its original place at $i=0.$ Thus, the time dependence of the
tunneling process is simply given by
\begin{equation}
e^{-(\varepsilon _{N+1}-\varepsilon _{0})\tau },
\label{time_dependence}
\end{equation}
where $\varepsilon _{N+1}$ and $\varepsilon _{0}$ are the electron
and hole excitation energies on the cites $N+1$-st and $0-$th
respectively. Making an analytical continuation to real times
$\tau =it$ and taking the last integral over $t$ we obtain the
delta function
\begin{equation}
2\pi \delta (\varepsilon _{N+1}-\varepsilon _{0}),
\end{equation}
that indeed shows that the process is elastic, i.e. electron can
tunnel only to the state with exactly the same energy.

Finally, we obtain for the tunneling probability of the elastic
process
\begin{equation}
P_{el}=w\,\delta (\varepsilon _{N+1}-\varepsilon
_{0})\,g_{0}\,\prod_{k=1}^{N}P_{k},  \label{P}
\end{equation}%
where the factor $w=n(\xi _{0})[1-n(\xi _{N+1})]$ takes into
account the occupation numbers of the initial and final states.

We see that the total probability $P_{el}$ contains the product of
the ratios $\Gamma_i /\tilde{E}_{i}$. For this reason it is
convenient to introduce the geometrical averages of theses
quantities along the tunneling path that were introduced in
Sec.~\ref{sum_el_cot}. Writing the total probability $P_{el}$ in
terms of the quantities $\bar{\Gamma},\bar{E}$ we obtain the
result~(\ref{P_el_result}).

The factorization of the probability $P_{el}$, Eq.~(\ref{P}), into
the product of the probabilities $P_{k}$ means that the hops from
grain to grain are independent from each other, which is a
consequence of the diagonal form of the Coulomb interaction.

If the Coulomb interaction is so long ranged that it involves many
granules, the situation is more complicated since the integrals
over the time variables cannot be taken on each site
independently. Nevertheless, one can generalize the obtained
results to the case of the long range Coulomb interaction as
follows: The most important effect of the off diagonal part of the
interaction $E_{c}^{ij}$ is the renormalization of the excitation
energies. Since the final result contains the single particle
excitations only, this effect may be included by the proper
definition of the electron-hole excitation energies
\begin{equation}
E_{i}^{\pm }=E_{ii}-\mu _{i},  \label{A_substitution}
\end{equation}
where $\mu _{i}$ represents the local potential formed by both the
external potential $V_{i}$ and all charges surrounding the grain
$i.$ This procedure, however, does not include all the effects of
the presence of the off diagonal part of $E_{c}^{ij}.$ The problem
is that the virtual process represented by the diagram in
Fig.~\ref{Elastic_Cotunneling} does not correspond to a causal
classical process. For this reason the virtual field created by an
electron $i$ at a time $t_{i}$ affects in general the same
electron when it is present on the other site at a different time.

However, such an interaction decays at least as $1/r^{2}$ with the
distance. This assumes that the long range effects of the Coulomb
interaction are not important, since the corresponding integral
along the chain converges. Thus, all the effects that are not
taken into account by the substitution~(\ref{A_substitution}) are
short range and may only lead to the change of a constant under
the logarithm of the effective localization length
(\ref{localization}). Using the analogy with classical Coulomb
problem \cite{Beloborodov05} one can estimate the boundaries for
the factor under the logarithm as $0.5\lesssim c<1.0.$

\section{Calculation of the tunneling probability $P_{in}$ in the inelastic
regime, Eq.~(\protect\ref{result_in1})}

\label{inelastic_hopping}

In this appendix we derive the probability of the inelastic
cotunneling $P_{in}$ through a chain of grains. As in the case of
the elastic processes considered in the previous section, the
probability of the inelastic processes can easily be found for the
model with the diagonal Coulomb interaction. For this model, we
follow the same steps as previously, making the gauge
transformation described in Sec.~\ref{gauge} and expanding the
tunneling probability in the tunneling matrix elements.

\begin{figure}[tbp]
\includegraphics[width=2.8in]{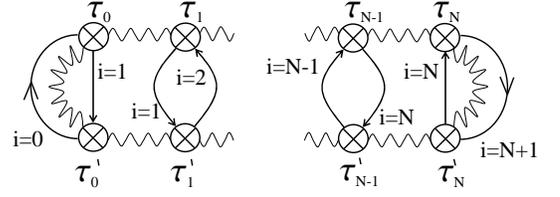} \vspace{-0.3cm}
\caption{ This diagram represents the tunnelling probability via
inelastic cotunneling processes. The crossed circles represent the
tunnelling matrix elements $t^{ij}_{k,k^\prime}\,
e^{i\protect\phi_i(\protect\tau)} \, e^{-i
\protect\phi_j(\protect\tau)}$ where phase factors appear from the
gauge transformation. Wavy lines represent the average of the
phase factors $ \langle e^{i\protect\phi(\protect\tau_1)} \,
e^{-i\protect\phi(\protect\tau_2)} \rangle$ with respect to the
Coulomb action. } \label{Inelastic_Cotunneling}
\end{figure}

The diagram describing the inelastic process is shown in
Fig.~\ref{Inelastic_Cotunneling}. We see that it consists only of
short electron loops, which means that the charge on each step is
transferred by different electrons as it should be in the
inelastic process.

All intermediate Green functions with the grain index $i=0,...N$
enter the diagram in Fig.~\ref{Inelastic_Cotunneling} being
integrated over the internal states of the grain, such that each
Green function is
\begin{equation}
G(\tau )\equiv \sum_{k}G_{k}(\tau )={\frac{{\pi T}}{{\sin (\pi
T\tau )}}}.
\end{equation}

The processes for the electrons located on the initial and final
sites ($i_{0}$ and $i_{N}$) are given by the Green functions that
are accompanied by the phase correlations $\Pi (\tau )$ exactly in
the same way as in the case of the elastic cotunneling and, thus,
can be described in terms of the Green functions $\tilde{G}$.

The intervals $\tau _{0}-\tau _{0}^{\prime }$ and $\tau _{N}-\tau
_{N}^{\prime }$ coincide with each other within the accuracy of
the inverse Coulomb energy and we just put $\tau _{0}-\tau
_{0}^{\prime }\approx \tau _{N}-\tau _{N}^{\prime }\equiv \tau .$
In the same limit, the integrals over the intermediate times
intervals $\Delta \tau _{k}=\tau _{k}-\tau _{k-1}$ (and over $\tau
_{k}^{\prime }-\tau _{k-1}^{\prime }$) can be calculated
independently resulting in the contribution
\begin{equation}
\int_{-\infty }^{\infty }d\left( \Delta \tau _{k}\right)
\;e^{-E_{k}^{c}|\Delta \tau _{k}|+\mu _{k}\Delta \tau
_{k}}=2/\tilde{E}_{k}, \label{Coulomb_in}
\end{equation}%
where the energy $\tilde{E}_{k}$ is defined in
Eq.~(\ref{tilde_E}). Collecting all the terms for the probability
$P_{in}$ one obtains
\begin{equation}
P_{in}(\tau )={\frac{{\ \bar{g}^{N+1}}}{{2\pi ^{N+1}}}}\left(
{\frac{{\ 2\pi T}}{{\ {\bar{E}}\sin (\pi T\tau )}}}\right)
^{2N}e^{-\tau \Delta \varepsilon },  \label{analit}
\end{equation}%
where $\Delta \varepsilon =\varepsilon _{N+1}-\varepsilon _{0}$ is
the difference between the energies of the initial and final
states and $\bar{E}$ and $\bar{g}$ are geometrical averages along
a the tunneling path (see Eqs. (\ref{geoave})).

Finally, in order to find the tunneling probability $P_{in}$ for
the inelastic cotunneling one has to make the analytical
continuation in Eq.~(\ref{analit}) to the real times $\tau
_{w}=it_{w}$ and integrate over $t_{w}$ arriving at
\begin{equation}
P_{in}={\frac{{\ w\,g^{N+1}}}{{2\pi ^{N+1}}}}\int\limits_{-\infty
}^{+\infty }dt\left[ {\frac{{\ 2\pi iT}}{{\ {\bar{E}}\sinh (\pi
Tt)}}}\right] ^{2N}e^{-it\Delta \varepsilon }.  \label{P_in}
\end{equation}
Here the singularity of the function $\sinh ^{-2N}(\pi Tt)$ is
assumed to be in the upper half of the complex plain. At zero
temperature one can easily calculate the integral over the
variable $t$ in Eq.~(\ref{P_in}) by shifting the contour of
integration in the complex plain either to $-i\infty $ for
positive $\Delta \varepsilon $ or to $+i\infty $ for negative
$\Delta \varepsilon .$

In the first case we obtain $P_{in}=0$. This reflects the fact
that the real tunneling process with an increase of the energy of
the electron is forbidden at $T=0$. In the latter case, $\Delta
\varepsilon <0$, the zero temperature probability is determined by
the pole of the function $\sinh ^{-2N}(\pi Tt)$ that results in
Eq.~(\ref{inel}).

At finite temperatures the integral in Eq.~(\ref{P_in}) can be
expressed in terms of the Euler Gamma functions that leads to the
general expression (\ref{result_in1}).

Generalization of the result to include the long-range interaction
consists, as in the case of the elastic co-tunneling, of a proper
redefinition (\ref{A_substitution}) of the electron/hole
excitation energies. Considerations analogous to those presented
in the previous appendix allow us to conclude that inclusion of
the off diagonal terms in the capacitance matrix results only in
the appearance of the  coefficient under the
logarithm~\cite{Beloborodov05} in the expression for the
localization length~(\ref{inelastic_loc_len}).


\begin{thebibliography}{165}
\expandafter\ifx\csname
natexlab\endcsname\relax\def\natexlab#1{#1}\fi
\expandafter\ifx\csname bibnamefont\endcsname\relax
  \def\bibnamefont#1{#1}\fi
\expandafter\ifx\csname bibfnamefont\endcsname\relax
  \def\bibfnamefont#1{#1}\fi
\expandafter\ifx\csname citenamefont\endcsname\relax
  \def\citenamefont#1{#1}\fi
\expandafter\ifx\csname url\endcsname\relax
  \def\url#1{\texttt{#1}}\fi
\expandafter\ifx\csname
urlprefix\endcsname\relax\def\urlprefix{URL }\fi
\providecommand{\bibinfo}[2]{#2}
\providecommand{\eprint}[2][]{\url{#2}}

\bibitem[{\citenamefont{Abeles}(1977)}]{Abeles77}
\bibinfo{author}{\bibnamefont{Abeles}, \bibfnamefont{B.}},
  \bibinfo{year}{1977}, \bibinfo{journal}{Phys. Rev. B}
  \textbf{\bibinfo{volume}{15}}, \bibinfo{pages}{2828}.

\bibitem[{\citenamefont{Abeles} \emph{et~al.}(1975)\citenamefont{Abeles, Sheng,
  Coutts, and Arie}}]{Abeles75}
\bibinfo{author}{\bibnamefont{Abeles}, \bibfnamefont{B.}},
  \bibinfo{author}{\bibfnamefont{P.}~\bibnamefont{Sheng}},
  \bibinfo{author}{\bibfnamefont{M.}~\bibnamefont{Coutts}}, and
  \bibinfo{author}{\bibfnamefont{Y.}~\bibnamefont{Arie}}, \bibinfo{year}{1975},
  \bibinfo{journal}{Adv. Phys.} \textbf{\bibinfo{volume}{24}},
  \bibinfo{pages}{407}.

\bibitem[{\citenamefont{Abrahams} \emph{et~al.}(1979)\citenamefont{Abrahams,
  Anderson, Licciardello, and Ramakrishnan}}]{Abrahams79}
\bibinfo{author}{\bibnamefont{Abrahams}, \bibfnamefont{E.}},
  \bibinfo{author}{\bibfnamefont{P.}~\bibnamefont{Anderson}},
  \bibinfo{author}{\bibfnamefont{D.}~\bibnamefont{Licciardello}}, and
  \bibinfo{author}{\bibfnamefont{T.}~\bibnamefont{Ramakrishnan}},
  \bibinfo{year}{1979}, \bibinfo{journal}{Phys. Rev. Lett.}
  \textbf{\bibinfo{volume}{42}}, \bibinfo{pages}{673}.

\bibitem[{\citenamefont{Abrahams} \emph{et~al.}(2001)\citenamefont{Abrahams,
  Kravchenko, and Sarachik}}]{Abrahams_review}
\bibinfo{author}{\bibnamefont{Abrahams}, \bibfnamefont{E.}},
  \bibinfo{author}{\bibfnamefont{S.~V.} \bibnamefont{Kravchenko}}, and
  \bibinfo{author}{\bibfnamefont{M.~P.} \bibnamefont{Sarachik}},
  \bibinfo{year}{2001}, \bibinfo{journal}{Rev. Mod. Phys.}
  \textbf{\bibinfo{volume}{73}}, \bibinfo{pages}{251}.

\bibitem[{\citenamefont{Abrahams} \emph{et~al.}(1970)\citenamefont{Abrahams,
  Redi, and Woo}}]{Abrahams70}
\bibinfo{author}{\bibnamefont{Abrahams}, \bibfnamefont{E.}},
  \bibinfo{author}{\bibfnamefont{M.}~\bibnamefont{Redi}}, and
  \bibinfo{author}{\bibfnamefont{J.}~\bibnamefont{Woo}}, \bibinfo{year}{1970},
  \bibinfo{journal}{Phys. Rev. B} \textbf{\bibinfo{volume}{1}},
  \bibinfo{pages}{208}.

\bibitem[{\citenamefont{Abrikosov}(1988)}]{Abrikosovbook}
\bibinfo{author}{\bibnamefont{Abrikosov}, \bibfnamefont{A.}},
  \bibinfo{year}{1988}, \emph{\bibinfo{title}{Fundamentals of the Theory of
  Metals}} (\bibinfo{publisher}{North Holland, Amsterdam}).

\bibitem[{\citenamefont{Abrikosov and Gor'kov}(1961)}]{Abrikosov61}
\bibinfo{author}{\bibnamefont{Abrikosov}, \bibfnamefont{A.~A.}}, and
  \bibinfo{author}{\bibfnamefont{L.~P.} \bibnamefont{Gor'kov}},
  \bibinfo{year}{1961}, \bibinfo{journal}{Sov. Phys. JETP}
  \textbf{\bibinfo{volume}{12}}, \bibinfo{pages}{1243}.

\bibitem[{\citenamefont{Abrikosov} \emph{et~al.}(1965)\citenamefont{Abrikosov,
  Gor'kov, and Dzyaloshinski}}]{Abrikosov65}
\bibinfo{author}{\bibnamefont{Abrikosov}, \bibfnamefont{A.~A.}},
  \bibinfo{author}{\bibfnamefont{L.~P.} \bibnamefont{Gor'kov}}, and
  \bibinfo{author}{\bibfnamefont{I.~E.} \bibnamefont{Dzyaloshinski}},
  \bibinfo{year}{1965}, \emph{\bibinfo{title}{Methods of quantum field theory
  in statistical physics}} (\bibinfo{publisher}{Dover Publications, Inc. New
  York}).

\bibitem[{\citenamefont{Aleiner} \emph{et~al.}(2002)\citenamefont{Aleiner,
  Brouwer, and Glazman}}]{Aleiner02}
\bibinfo{author}{\bibnamefont{Aleiner}, \bibfnamefont{I.~I.}},
  \bibinfo{author}{\bibfnamefont{P.~W.} \bibnamefont{Brouwer}}, and
  \bibinfo{author}{\bibfnamefont{L.~I.} \bibnamefont{Glazman}},
  \bibinfo{year}{2002}, \bibinfo{journal}{Phys. Rep.}
  \textbf{\bibinfo{volume}{358}}, \bibinfo{pages}{309}.

\bibitem[{\citenamefont{Alhassid}(2000)}]{Alhassid00}
\bibinfo{author}{\bibnamefont{Alhassid}, \bibfnamefont{Y.}},
  \bibinfo{year}{2000}, \bibinfo{journal}{Rev. Mod. Phys.}
  \textbf{\bibinfo{volume}{72}}, \bibinfo{pages}{895}.

\bibitem[{\citenamefont{Altland} \emph{et~al.}(2004)\citenamefont{Altland,
  Glazman, and Kamenev}}]{kamenev}
\bibinfo{author}{\bibnamefont{Altland}, \bibfnamefont{A.}},
  \bibinfo{author}{\bibfnamefont{L.}~\bibnamefont{Glazman}}, and
  \bibinfo{author}{\bibfnamefont{A.}~\bibnamefont{Kamenev}},
  \bibinfo{year}{2004}, \bibinfo{journal}{Phys. Rev. Lett.}
  \textbf{\bibinfo{volume}{92}}, \bibinfo{pages}{026801}.

\bibitem[{\citenamefont{Altland} \emph{et~al.}(2005)\citenamefont{Altland,
  Glazman, Kamenev, and Meyer}}]{Altland05}
\bibinfo{author}{\bibnamefont{Altland}, \bibfnamefont{A.}},
  \bibinfo{author}{\bibfnamefont{L.~I.} \bibnamefont{Glazman}},
  \bibinfo{author}{\bibfnamefont{A.}~\bibnamefont{Kamenev}}, and
  \bibinfo{author}{\bibfnamefont{J.~S.} \bibnamefont{Meyer}},
  \bibinfo{year}{2005}, \bibinfo{journal}{cond-mat/0507695} .

\bibitem[{\citenamefont{Altshuler and Aronov}(1985)}]{Altshuler85}
\bibinfo{author}{\bibnamefont{Altshuler}, \bibfnamefont{B.}}, and
  \bibinfo{author}{\bibfnamefont{A.}~\bibnamefont{Aronov}},
  \bibinfo{year}{1985}, \emph{\bibinfo{title}{Electron-Electron Interaction in
  Disordered Conductors}} (\bibinfo{publisher}{North-Holland, Amsterdam}),
  volume~\bibinfo{volume}{10} of \emph{\bibinfo{series}{Modern Problems in
  Condensed Matter Science}}, p.~\bibinfo{pages}{1}.

\bibitem[{\citenamefont{Altshuler} \emph{et~al.}(1982)\citenamefont{Altshuler,
  Aronov, and Khmel'nitskii}}]{Altshuler82}
\bibinfo{author}{\bibnamefont{Altshuler}, \bibfnamefont{B.~L.}},
  \bibinfo{author}{\bibfnamefont{A.~G.} \bibnamefont{Aronov}}, and
  \bibinfo{author}{\bibfnamefont{D.~E.} \bibnamefont{Khmel'nitskii}},
  \bibinfo{year}{1982}, \bibinfo{journal}{J. Phys. C}
  \textbf{\bibinfo{volume}{15}}, \bibinfo{pages}{7367}.

\bibitem[{\citenamefont{Altshuler} \emph{et~al.}(1980)\citenamefont{Altshuler,
  Khmelnitskii, Larkin, and Lee}}]{Altshuler80}
\bibinfo{author}{\bibnamefont{Altshuler}, \bibfnamefont{B.~L.}},
  \bibinfo{author}{\bibfnamefont{D.~E.} \bibnamefont{Khmelnitskii}},
  \bibinfo{author}{\bibfnamefont{A.~I.} \bibnamefont{Larkin}}, and
  \bibinfo{author}{\bibfnamefont{P.~A.} \bibnamefont{Lee}},
  \bibinfo{year}{1980}, \bibinfo{journal}{Phys. Rev. B}
  \textbf{\bibinfo{volume}{22}}.

\bibitem[{\citenamefont{Ambegaokar and Baratoff}(1963)}]{ambegaokar63}
\bibinfo{author}{\bibnamefont{Ambegaokar}, \bibfnamefont{V.}}, and
  \bibinfo{author}{\bibfnamefont{A.}~\bibnamefont{Baratoff}},
  \bibinfo{year}{1963}, \bibinfo{journal}{Phys. Rev. Lett.}
  \textbf{\bibinfo{volume}{10}}, \bibinfo{pages}{486}.

\bibitem[{\citenamefont{Ambegaokar}
  \emph{et~al.}(1982)\citenamefont{Ambegaokar, Eckern, and Sch\"{o}n}}]{AES}
\bibinfo{author}{\bibnamefont{Ambegaokar}, \bibfnamefont{V.}},
  \bibinfo{author}{\bibfnamefont{U.}~\bibnamefont{Eckern}}, and
  \bibinfo{author}{\bibfnamefont{G.}~\bibnamefont{Sch\"{o}n}},
  \bibinfo{year}{1982}, \bibinfo{journal}{Phys. Rev. Lett.}
  \textbf{\bibinfo{volume}{48}}, \bibinfo{pages}{1745}.

\bibitem[{\citenamefont{Anderson}(1958)}]{Anderson58}
\bibinfo{author}{\bibnamefont{Anderson}, \bibfnamefont{P.~W.}},
  \bibinfo{year}{1958}, \bibinfo{journal}{Phys. Rev.}
  \textbf{\bibinfo{volume}{109}}, \bibinfo{pages}{1492}.

\bibitem[{\citenamefont{Anderson}(1959)}]{Anderson59}
\bibinfo{author}{\bibnamefont{Anderson}, \bibfnamefont{P.~W.}},
  \bibinfo{year}{1959}, \bibinfo{journal}{J. Phys. Chem. Solid}
  \textbf{\bibinfo{volume}{1}}, \bibinfo{pages}{26}.

\bibitem[{\citenamefont{Anderson}(1964)}]{Anderson64}
\bibinfo{author}{\bibnamefont{Anderson}, \bibfnamefont{P.~W.}},
  \bibinfo{year}{1964}, \emph{\bibinfo{title}{in Lectures on the Mony-Body
  Problems}}, volume~\bibinfo{volume}{2} (\bibinfo{publisher}{Academic, New
  York}).

\bibitem[{\citenamefont{Ando} \emph{et~al.}(1995)\citenamefont{Ando, Boebinger,
  Passner, Kimura, and Kishio}}]{boeb}
\bibinfo{author}{\bibnamefont{Ando}, \bibfnamefont{Y.}},
  \bibinfo{author}{\bibfnamefont{G.}~\bibnamefont{Boebinger}},
  \bibinfo{author}{\bibfnamefont{A.}~\bibnamefont{Passner}},
  \bibinfo{author}{\bibfnamefont{T.}~\bibnamefont{Kimura}}, and
  \bibinfo{author}{\bibfnamefont{K.}~\bibnamefont{Kishio}},
  \bibinfo{year}{1995}, \bibinfo{journal}{Phys. Rev. Lett.}
  \textbf{\bibinfo{volume}{75}}, \bibinfo{pages}{4662}.

\bibitem[{\citenamefont{Andreev and Beloborodov}(2004)}]{Andreev04}
\bibinfo{author}{\bibnamefont{Andreev}, \bibfnamefont{A.~V.}}, and
  \bibinfo{author}{\bibfnamefont{I.~S.} \bibnamefont{Beloborodov}},
  \bibinfo{year}{2004}, \bibinfo{journal}{Phys. Rev. B}
  \textbf{\bibinfo{volume}{69}}, \bibinfo{pages}{081406(R)}.

\bibitem[{\citenamefont{Aslamazov and Larkin}(1968)}]{Aslamazov68}
\bibinfo{author}{\bibnamefont{Aslamazov}, \bibfnamefont{L.~G.}}, and
  \bibinfo{author}{\bibfnamefont{A.~I.} \bibnamefont{Larkin}},
  \bibinfo{year}{1968}, \bibinfo{journal}{Fiz. Tverd. Tela~Leningrad} ,
  \bibinfo{pages}{1104}\bibinfo{note}{[Sov. Phys. Solid State {\bf 10}, 875
  (1968)]}.

\bibitem[{\citenamefont{Averin and Nazarov}(1992)}]{Averin92}
\bibinfo{author}{\bibnamefont{Averin}, \bibfnamefont{D.}}, and
  \bibinfo{author}{\bibfnamefont{Y.}~\bibnamefont{Nazarov}},
  \bibinfo{year}{1992}, \emph{\bibinfo{title}{in Single Charge Tunneling}}
  (\bibinfo{publisher}{Plenum, New York}), volume \bibinfo{volume}{294}.

\bibitem[{\citenamefont{Averin and Nazarov}(1990)}]{Averin}
\bibinfo{author}{\bibnamefont{Averin}, \bibfnamefont{D.~A.}}, and
  \bibinfo{author}{\bibfnamefont{Y.~V.} \bibnamefont{Nazarov}},
  \bibinfo{year}{1990}, \bibinfo{journal}{Phys. Rev. Lett.}
  \textbf{\bibinfo{volume}{65}}, \bibinfo{pages}{2446}.

\bibitem[{\citenamefont{Averin and Likharev}(1991)}]{Averin91}
\bibinfo{author}{\bibnamefont{Averin}, \bibfnamefont{D.~V.}}, and
  \bibinfo{author}{\bibfnamefont{K.~K.} \bibnamefont{Likharev}},
  \bibinfo{year}{1991}, \emph{\bibinfo{title}{in Mesoscopic Phenomena in
  Solids}} (\bibinfo{publisher}{Elsevier, North-Holland, Amsterdam}), p.
  \bibinfo{pages}{173}.

\bibitem[{\citenamefont{Basko} \emph{et~al.}(2005)\citenamefont{Basko, Aleiner,
  and Altshuler}}]{basko}
\bibinfo{author}{\bibnamefont{Basko}, \bibfnamefont{D.}},
  \bibinfo{author}{\bibfnamefont{I.}~\bibnamefont{Aleiner}}, and
  \bibinfo{author}{\bibfnamefont{B.}~\bibnamefont{Altshuler}},
  \bibinfo{year}{2005}, \bibinfo{journal}{cond-mat/0506617} .

\bibitem[{\citenamefont{Beenakker}(1997)}]{Beenakker97}
\bibinfo{author}{\bibnamefont{Beenakker}, \bibfnamefont{C.~W.~J.}},
  \bibinfo{year}{1997}, \bibinfo{journal}{Rev. Mod. Phys.}
  \textbf{\bibinfo{volume}{69}}, \bibinfo{pages}{731}.

\bibitem[{\citenamefont{Belitz and Kirkpatrick}(1994)}]{Belitz94}
\bibinfo{author}{\bibnamefont{Belitz}, \bibfnamefont{D.}}, and
  \bibinfo{author}{\bibfnamefont{T.~R.} \bibnamefont{Kirkpatrick}},
  \bibinfo{year}{1994}, \bibinfo{journal}{Rev. Mod. Phys.}
  \textbf{\bibinfo{volume}{66}}, \bibinfo{pages}{261}.

\bibitem[{\citenamefont{Beloborodov and Efetov}(1999)}]{Beloborodov99}
\bibinfo{author}{\bibnamefont{Beloborodov}, \bibfnamefont{I.~S.}}, and
  \bibinfo{author}{\bibfnamefont{K.~B.} \bibnamefont{Efetov}},
  \bibinfo{year}{1999}, \bibinfo{journal}{Phys. Rev. Lett.}
  \textbf{\bibinfo{volume}{82}}, \bibinfo{pages}{3332}.

\bibitem[{\citenamefont{Beloborodov}
  \emph{et~al.}(2001)\citenamefont{Beloborodov, Efetov, Altland, and
  Hekking}}]{Beloborodov01}
\bibinfo{author}{\bibnamefont{Beloborodov}, \bibfnamefont{I.~S.}},
  \bibinfo{author}{\bibfnamefont{K.~B.} \bibnamefont{Efetov}},
  \bibinfo{author}{\bibfnamefont{A.}~\bibnamefont{Altland}}, and
  \bibinfo{author}{\bibfnamefont{F.~W.~J.} \bibnamefont{Hekking}},
  \bibinfo{year}{2001}, \bibinfo{journal}{Phys. Rev. B}
  \textbf{\bibinfo{volume}{63}}, \bibinfo{pages}{115109}.

\bibitem[{\citenamefont{Beloborodov}
  \emph{et~al.}(2000)\citenamefont{Beloborodov, Efetov, and
  Larkin}}]{Beloborodov00}
\bibinfo{author}{\bibnamefont{Beloborodov}, \bibfnamefont{I.~S.}},
  \bibinfo{author}{\bibfnamefont{K.~B.} \bibnamefont{Efetov}}, and
  \bibinfo{author}{\bibfnamefont{A.~I.} \bibnamefont{Larkin}},
  \bibinfo{year}{2000}, \bibinfo{journal}{Phys. Rev. B.}
  \textbf{\bibinfo{volume}{61}}, \bibinfo{pages}{9145}.

\bibitem[{\citenamefont{Beloborodov}
  \emph{et~al.}(2003)\citenamefont{Beloborodov, Efetov, Lopatin, and
  Vinokur}}]{Beloborodov03}
\bibinfo{author}{\bibnamefont{Beloborodov}, \bibfnamefont{I.~S.}},
  \bibinfo{author}{\bibfnamefont{K.~B.} \bibnamefont{Efetov}},
  \bibinfo{author}{\bibfnamefont{A.~V.} \bibnamefont{Lopatin}}, and
  \bibinfo{author}{\bibfnamefont{V.~M.} \bibnamefont{Vinokur}},
  \bibinfo{year}{2003}, \bibinfo{journal}{Phys. Rev. Lett}
  \textbf{\bibinfo{volume}{91}}, \bibinfo{pages}{246801}.

\bibitem[{\citenamefont{Beloborodov}
  \emph{et~al.}(2005{\natexlab{a}})\citenamefont{Beloborodov, Efetov, Lopatin,
  and Vinokur}}]{Beloborodov05super}
\bibinfo{author}{\bibnamefont{Beloborodov}, \bibfnamefont{I.~S.}},
  \bibinfo{author}{\bibfnamefont{K.~B.} \bibnamefont{Efetov}},
  \bibinfo{author}{\bibfnamefont{A.~V.} \bibnamefont{Lopatin}}, and
  \bibinfo{author}{\bibfnamefont{V.~M.} \bibnamefont{Vinokur}},
  \bibinfo{year}{2005}{\natexlab{a}}, \bibinfo{journal}{Phys.~Rev.~B}
  \textbf{\bibinfo{volume}{71}}, \bibinfo{pages}{184501}.

\bibitem[{\citenamefont{Beloborodov}
  \emph{et~al.}(2006)\citenamefont{Beloborodov, Fominov, Lopatin, and
  Vinokur}}]{Beloborodov06}
\bibinfo{author}{\bibnamefont{Beloborodov}, \bibfnamefont{I.~S.}},
  \bibinfo{author}{\bibfnamefont{Y.}~\bibnamefont{Fominov}},
  \bibinfo{author}{\bibfnamefont{A.~V.} \bibnamefont{Lopatin}}, and
  \bibinfo{author}{\bibfnamefont{V.~M.} \bibnamefont{Vinokur}},
  \bibinfo{year}{2006}, \bibinfo{journal}{cond-mat/0509386} .

\bibitem[{\citenamefont{Beloborodov}
  \emph{et~al.}(2005{\natexlab{b}})\citenamefont{Beloborodov, Lopatin, Hekking,
  Fazio, and Vinokur}}]{Beloborodovthermal}
\bibinfo{author}{\bibnamefont{Beloborodov}, \bibfnamefont{I.~S.}},
  \bibinfo{author}{\bibfnamefont{A.~V.} \bibnamefont{Lopatin}},
  \bibinfo{author}{\bibfnamefont{F.~J.~W.} \bibnamefont{Hekking}},
  \bibinfo{author}{\bibfnamefont{R.}~\bibnamefont{Fazio}}, and
  \bibinfo{author}{\bibfnamefont{V.~M.} \bibnamefont{Vinokur}},
  \bibinfo{year}{2005}{\natexlab{b}}, \bibinfo{journal}{Europhys. Lett.}
  \textbf{\bibinfo{volume}{69}}, \bibinfo{pages}{435}.

\bibitem[{\citenamefont{Beloborodov}
  \emph{et~al.}(2004{\natexlab{a}})\citenamefont{Beloborodov, Lopatin,
  Schwiete, and Vinokur}}]{BeloborodovDOS}
\bibinfo{author}{\bibnamefont{Beloborodov}, \bibfnamefont{I.~S.}},
  \bibinfo{author}{\bibfnamefont{A.~V.} \bibnamefont{Lopatin}},
  \bibinfo{author}{\bibfnamefont{G.}~\bibnamefont{Schwiete}}, and
  \bibinfo{author}{\bibfnamefont{V.~M.} \bibnamefont{Vinokur}},
  \bibinfo{year}{2004}{\natexlab{a}}, \bibinfo{journal}{Phys Rev B}
  \textbf{\bibinfo{volume}{70}}, \bibinfo{pages}{073404}.

\bibitem[{\citenamefont{Beloborodov}
  \emph{et~al.}(2004{\natexlab{b}})\citenamefont{Beloborodov, Lopatin, and
  Vinokur}}]{Lopatin04}
\bibinfo{author}{\bibnamefont{Beloborodov}, \bibfnamefont{I.~S.}},
  \bibinfo{author}{\bibfnamefont{A.~V.} \bibnamefont{Lopatin}}, and
  \bibinfo{author}{\bibfnamefont{V.~M.} \bibnamefont{Vinokur}},
  \bibinfo{year}{2004}{\natexlab{b}}, \bibinfo{journal}{Phys.~Rev.~B}
  \textbf{\bibinfo{volume}{70}}, \bibinfo{pages}{205120}.

\bibitem[{\citenamefont{Beloborodov}
  \emph{et~al.}(2004{\natexlab{c}})\citenamefont{Beloborodov, Lopatin, and
  Vinokur}}]{BLV_2004}
\bibinfo{author}{\bibnamefont{Beloborodov}, \bibfnamefont{I.~S.}},
  \bibinfo{author}{\bibfnamefont{A.~V.} \bibnamefont{Lopatin}}, and
  \bibinfo{author}{\bibfnamefont{V.~M.} \bibnamefont{Vinokur}},
  \bibinfo{year}{2004}{\natexlab{c}}, \bibinfo{journal}{Phys. Rev. Lett.}
  \textbf{\bibinfo{volume}{92}}, \bibinfo{pages}{207002}.

\bibitem[{\citenamefont{Beloborodov}
  \emph{et~al.}(2005{\natexlab{c}})\citenamefont{Beloborodov, Lopatin, and
  Vinokur}}]{Beloborodov05}
\bibinfo{author}{\bibnamefont{Beloborodov}, \bibfnamefont{I.~S.}},
  \bibinfo{author}{\bibfnamefont{A.~V.} \bibnamefont{Lopatin}}, and
  \bibinfo{author}{\bibfnamefont{V.~M.} \bibnamefont{Vinokur}},
  \bibinfo{year}{2005}{\natexlab{c}}, \bibinfo{journal}{Phys.~Rev.~B}
  \textbf{\bibinfo{volume}{72}}, \bibinfo{pages}{125121}.

\bibitem[{\citenamefont{Bergeret} \emph{et~al.}(2001)\citenamefont{Bergeret,
  Volkov, and Efetov}}]{bergeretodd}
\bibinfo{author}{\bibnamefont{Bergeret}, \bibfnamefont{F.~S.}},
  \bibinfo{author}{\bibfnamefont{A.~F.} \bibnamefont{Volkov}}, and
  \bibinfo{author}{\bibfnamefont{K.~B.} \bibnamefont{Efetov}},
  \bibinfo{year}{2001}, \bibinfo{journal}{Phys. Rev. Lett.}
  \textbf{\bibinfo{volume}{86}}, \bibinfo{pages}{4096}.

\bibitem[{\citenamefont{Bergeret} \emph{et~al.}(2004)\citenamefont{Bergeret,
  Volkov, and Efetov}}]{bergeret}
\bibinfo{author}{\bibnamefont{Bergeret}, \bibfnamefont{F.~S.}},
  \bibinfo{author}{\bibfnamefont{A.~F.} \bibnamefont{Volkov}}, and
  \bibinfo{author}{\bibfnamefont{K.~B.} \bibnamefont{Efetov}},
  \bibinfo{year}{2004}, \bibinfo{journal}{Europhys. Lett.}
  \textbf{\bibinfo{volume}{66}}, \bibinfo{pages}{111}.

\bibitem[{\citenamefont{Bergeret} \emph{et~al.}(2005)\citenamefont{Bergeret,
  Volkov, and Efetov}}]{bergeretrmp}
\bibinfo{author}{\bibnamefont{Bergeret}, \bibfnamefont{F.~S.}},
  \bibinfo{author}{\bibfnamefont{A.~F.} \bibnamefont{Volkov}}, and
  \bibinfo{author}{\bibfnamefont{K.~B.} \bibnamefont{Efetov}},
  \bibinfo{year}{2005}, \bibinfo{journal}{Rev. Mod. Phys.}
  \textbf{\bibinfo{volume}{77}}, \bibinfo{pages}{1321}.

\bibitem[{\citenamefont{Biagini}
  \emph{et~al.}(2005{\natexlab{a}})\citenamefont{Biagini, Caneva, Tognetti, and
  Varlamov}}]{Varlamov05}
\bibinfo{author}{\bibnamefont{Biagini}, \bibfnamefont{C.}},
  \bibinfo{author}{\bibfnamefont{T.}~\bibnamefont{Caneva}},
  \bibinfo{author}{\bibfnamefont{V.}~\bibnamefont{Tognetti}}, and
  \bibinfo{author}{\bibfnamefont{A.~A.} \bibnamefont{Varlamov}},
  \bibinfo{year}{2005}{\natexlab{a}}, \bibinfo{journal}{Phys. Rev. B}
  \textbf{\bibinfo{volume}{72}}, \bibinfo{pages}{041102}.

\bibitem[{\citenamefont{Biagini}
  \emph{et~al.}(2005{\natexlab{b}})\citenamefont{Biagini, Ferone, Fazio,
  Hekking, and Tognetti}}]{Biagini05}
\bibinfo{author}{\bibnamefont{Biagini}, \bibfnamefont{C.}},
  \bibinfo{author}{\bibfnamefont{R.}~\bibnamefont{Ferone}},
  \bibinfo{author}{\bibfnamefont{R.}~\bibnamefont{Fazio}},
  \bibinfo{author}{\bibfnamefont{F.~W.~J.} \bibnamefont{Hekking}}, and
  \bibinfo{author}{\bibfnamefont{V.}~\bibnamefont{Tognetti}},
  \bibinfo{year}{2005}{\natexlab{b}}, \bibinfo{journal}{Phys. Rev. B}
  \textbf{\bibinfo{volume}{72}}, \bibinfo{pages}{134510}.

\bibitem[{\citenamefont{Black} \emph{et~al.}(1996)\citenamefont{Black, Ralph,
  and Tinkham}}]{Black96}
\bibinfo{author}{\bibnamefont{Black}, \bibfnamefont{C.}},
  \bibinfo{author}{\bibfnamefont{D.}~\bibnamefont{Ralph}}, and
  \bibinfo{author}{\bibfnamefont{M.}~\bibnamefont{Tinkham}},
  \bibinfo{year}{1996}, \bibinfo{journal}{Phys. Rev. Lett.}
  \textbf{\bibinfo{volume}{76}}, \bibinfo{pages}{688}.

\bibitem[{\citenamefont{Boebinger} \emph{et~al.}(1996)\citenamefont{Boebinger,
  Ando, Passner, Kimura, Okuya, Shimoyama, Kishio, Tamasaku, Ichikawa, and
  Uchida}}]{boeb1}
\bibinfo{author}{\bibnamefont{Boebinger}, \bibfnamefont{G.}},
  \bibinfo{author}{\bibfnamefont{Y.}~\bibnamefont{Ando}},
  \bibinfo{author}{\bibfnamefont{A.}~\bibnamefont{Passner}},
  \bibinfo{author}{\bibfnamefont{T.}~\bibnamefont{Kimura}},
  \bibinfo{author}{\bibfnamefont{M.}~\bibnamefont{Okuya}},
  \bibinfo{author}{\bibfnamefont{J.}~\bibnamefont{Shimoyama}},
  \bibinfo{author}{\bibfnamefont{K.}~\bibnamefont{Kishio}},
  \bibinfo{author}{\bibfnamefont{K.}~\bibnamefont{Tamasaku}},
  \bibinfo{author}{\bibfnamefont{N.}~\bibnamefont{Ichikawa}}, and
  \bibinfo{author}{\bibfnamefont{S.}~\bibnamefont{Uchida}},
  \bibinfo{year}{1996}, \bibinfo{journal}{Phys. Rev. Lett.}
  \textbf{\bibinfo{volume}{77}}, \bibinfo{pages}{5417}.

\bibitem[{\citenamefont{Bulgadaev}(1984)}]{bulgadaev85}
\bibinfo{author}{\bibnamefont{Bulgadaev}, \bibfnamefont{S.}},
  \bibinfo{year}{1984}, \bibinfo{journal}{Pis'ma Zh. Eksp. Teor. Phys.}
  \textbf{\bibinfo{volume}{39}}, \bibinfo{pages}{264}, \bibinfo{note}{[Sov.
  Phys. JETP Lett. {\bf 39}, 314 (1985)]}.

\bibitem[{\citenamefont{Caldeira and Leggett}(1981)}]{Caldeira81}
\bibinfo{author}{\bibnamefont{Caldeira}, \bibfnamefont{A.~O.}}, and
  \bibinfo{author}{\bibfnamefont{A.~J.} \bibnamefont{Leggett}},
  \bibinfo{year}{1981}, \bibinfo{journal}{Phys. Rev. Lett.}
  \textbf{\bibinfo{volume}{46}}, \bibinfo{pages}{211}.

\bibitem[{\citenamefont{Caldeira and Leggett}(1983)}]{Caldeira83}
\bibinfo{author}{\bibnamefont{Caldeira}, \bibfnamefont{A.~O.}}, and
  \bibinfo{author}{\bibfnamefont{A.~J.} \bibnamefont{Leggett}},
  \bibinfo{year}{1983}, \bibinfo{journal}{Ann. Phys.}
  \textbf{\bibinfo{volume}{149}}, \bibinfo{pages}{374}.

\bibitem[{\citenamefont{Chakravarty}
  \emph{et~al.}(1986)\citenamefont{Chakravarty, Ingold, and
  S.~Kivelson}}]{Chakravarty86}
\bibinfo{author}{\bibnamefont{Chakravarty}, \bibfnamefont{S.}},
  \bibinfo{author}{\bibfnamefont{G.~L.} \bibnamefont{Ingold}}, and
  \bibinfo{author}{\bibfnamefont{A.~L.} \bibnamefont{S.~Kivelson}},
  \bibinfo{year}{1986}, \bibinfo{journal}{Phys Rev Lett}
  \textbf{\bibinfo{volume}{56}}, \bibinfo{pages}{2303}.

\bibitem[{\citenamefont{Chakravarty}
  \emph{et~al.}(1987)\citenamefont{Chakravarty, Kivelson, Zimanyi, and
  Halperin}}]{Chakravarty87}
\bibinfo{author}{\bibnamefont{Chakravarty}, \bibfnamefont{S.}},
  \bibinfo{author}{\bibfnamefont{S.}~\bibnamefont{Kivelson}},
  \bibinfo{author}{\bibfnamefont{G.}~\bibnamefont{Zimanyi}}, and
  \bibinfo{author}{\bibfnamefont{B.}~\bibnamefont{Halperin}},
  \bibinfo{year}{1987}, \bibinfo{journal}{Phys Rev B}
  \textbf{\bibinfo{volume}{35}}, \bibinfo{pages}{7256}.

\bibitem[{\citenamefont{Chervenak and J.~M.~Valles}(2000)}]{Chervenak00}
\bibinfo{author}{\bibnamefont{Chervenak}, \bibfnamefont{J.~A.}}, and
  \bibinfo{author}{\bibfnamefont{J.}~\bibnamefont{J.~M.~Valles}},
  \bibinfo{year}{2000}, \bibinfo{journal}{Phys Rev B}
  \textbf{\bibinfo{volume}{61}}, \bibinfo{pages}{R9245}.

\bibitem[{\citenamefont{Clogston}(1960)}]{Clogston60}
\bibinfo{author}{\bibnamefont{Clogston}, \bibfnamefont{A.~M.}},
  \bibinfo{year}{1960}, \bibinfo{journal}{Phys. Rev. Lett.}
  \textbf{\bibinfo{volume}{5}}, \bibinfo{pages}{464}.

\bibitem[{\citenamefont{Cohen} \emph{et~al.}(1962)\citenamefont{Cohen, Falicov,
  and Phillips}}]{cohen}
\bibinfo{author}{\bibnamefont{Cohen}, \bibfnamefont{M.}},
  \bibinfo{author}{\bibfnamefont{L.}~\bibnamefont{Falicov}}, and
  \bibinfo{author}{\bibfnamefont{J.}~\bibnamefont{Phillips}},
  \bibinfo{year}{1962}, \bibinfo{journal}{Phys. Rev. Lett.}
  \textbf{\bibinfo{volume}{8}}, \bibinfo{pages}{316}.

\bibitem[{\citenamefont{Collier} \emph{et~al.}(1998)\citenamefont{Collier,
  Vossmeyer, and Heath}}]{Collier98}
\bibinfo{author}{\bibnamefont{Collier}, \bibfnamefont{C.}},
  \bibinfo{author}{\bibfnamefont{T.}~\bibnamefont{Vossmeyer}}, and
  \bibinfo{author}{\bibfnamefont{J.}~\bibnamefont{Heath}},
  \bibinfo{year}{1998}, \bibinfo{journal}{Annual review of Physical Chemistry}
  \textbf{\bibinfo{volume}{49}}, \bibinfo{pages}{371}.

\bibitem[{\citenamefont{Collier} \emph{et~al.}(1997)\citenamefont{Collier,
  Saykally, Shiang, Henrichs, and Health}}]{Collier97}
\bibinfo{author}{\bibnamefont{Collier}, \bibfnamefont{C.~P.}},
  \bibinfo{author}{\bibfnamefont{R.~J.} \bibnamefont{Saykally}},
  \bibinfo{author}{\bibfnamefont{J.~J.} \bibnamefont{Shiang}},
  \bibinfo{author}{\bibfnamefont{S.~E.} \bibnamefont{Henrichs}}, and
  \bibinfo{author}{\bibfnamefont{J.~R.} \bibnamefont{Health}},
  \bibinfo{year}{1997}, \bibinfo{journal}{Science}
  \textbf{\bibinfo{volume}{277}}, \bibinfo{pages}{1978}.

\bibitem[{\citenamefont{Davidovi\'c and Tinkham}(1999)}]{Davidovic99}
\bibinfo{author}{\bibnamefont{Davidovi\'c}, \bibfnamefont{D.}}, and
  \bibinfo{author}{\bibfnamefont{M.}~\bibnamefont{Tinkham}},
  \bibinfo{year}{1999}, \bibinfo{journal}{Phys. Rev. Lett.}
  \textbf{\bibinfo{volume}{83}}, \bibinfo{pages}{688}.

\bibitem[{\citenamefont{von Delft} \emph{et~al.}(1996)\citenamefont{von Delft,
  Zaikin, Golubev, and Tichy}}]{VonDelft96}
\bibinfo{author}{\bibnamefont{von Delft}, \bibfnamefont{J.}},
  \bibinfo{author}{\bibfnamefont{A.~D.} \bibnamefont{Zaikin}},
  \bibinfo{author}{\bibfnamefont{D.~S.} \bibnamefont{Golubev}}, and
  \bibinfo{author}{\bibfnamefont{W.}~\bibnamefont{Tichy}},
  \bibinfo{year}{1996}, \bibinfo{journal}{Phys. Rev. Lett.}
  \textbf{\bibinfo{volume}{77}}, \bibinfo{pages}{3189}.

\bibitem[{\citenamefont{Du} \emph{et~al.}(2002)\citenamefont{Du, Chen,
  Krishnan, Krauss, Harbold, Wise, Thomas, and Silcox}}]{Du02}
\bibinfo{author}{\bibnamefont{Du}, \bibfnamefont{H.}},
  \bibinfo{author}{\bibfnamefont{C.~L.} \bibnamefont{Chen}},
  \bibinfo{author}{\bibfnamefont{R.}~\bibnamefont{Krishnan}},
  \bibinfo{author}{\bibfnamefont{T.~D.} \bibnamefont{Krauss}},
  \bibinfo{author}{\bibfnamefont{J.~M.} \bibnamefont{Harbold}},
  \bibinfo{author}{\bibfnamefont{F.~W.} \bibnamefont{Wise}},
  \bibinfo{author}{\bibfnamefont{M.~G.} \bibnamefont{Thomas}}, and
  \bibinfo{author}{\bibfnamefont{J.}~\bibnamefont{Silcox}},
  \bibinfo{year}{2002}, \bibinfo{journal}{Nano Lett}
  \textbf{\bibinfo{volume}{2}}, \bibinfo{pages}{1321}.

\bibitem[{\citenamefont{Eckern} \emph{et~al.}(1984)\citenamefont{Eckern,
  Sch\"{o}n, and Ambegaokar}}]{Eckern84}
\bibinfo{author}{\bibnamefont{Eckern}, \bibfnamefont{U.}},
  \bibinfo{author}{\bibfnamefont{G.}~\bibnamefont{Sch\"{o}n}}, and
  \bibinfo{author}{\bibfnamefont{V.}~\bibnamefont{Ambegaokar}},
  \bibinfo{year}{1984}, \bibinfo{journal}{Phys. Rev. B}
  \textbf{\bibinfo{volume}{30}}, \bibinfo{pages}{6419}.

\bibitem[{\citenamefont{Efetov}(1983)}]{efetov83}
\bibinfo{author}{\bibnamefont{Efetov}, \bibfnamefont{K.}},
  \bibinfo{year}{1983}, \bibinfo{journal}{Adv. Phys,}
  \textbf{\bibinfo{volume}{32}}, \bibinfo{pages}{53}.

\bibitem[{\citenamefont{Efetov}(1980)}]{Efetov80a}
\bibinfo{author}{\bibnamefont{Efetov}, \bibfnamefont{K.~B.}},
  \bibinfo{year}{1980}, \bibinfo{journal}{Zh.~Eksp.~Teor.~Fiz.}
  \textbf{\bibinfo{volume}{78}}, \bibinfo{pages}{2017}, \bibinfo{note}{[Sov.
  Phys. JETP}.

\bibitem[{\citenamefont{Efetov}(1997)}]{efetov}
\bibinfo{author}{\bibnamefont{Efetov}, \bibfnamefont{K.~B.}},
  \bibinfo{year}{1997}, \emph{\bibinfo{title}{Supersymmetry in Disorder and
  Chaos}} (\bibinfo{publisher}{Cambridge University Press, New York}).

\bibitem[{\citenamefont{Efetov} \emph{et~al.}(1980)\citenamefont{Efetov,
  Larkin, and Khmel'nitskii}}]{Efetov80}
\bibinfo{author}{\bibnamefont{Efetov}, \bibfnamefont{K.~B.}},
  \bibinfo{author}{\bibfnamefont{A.~I.} \bibnamefont{Larkin}}, and
  \bibinfo{author}{\bibfnamefont{D.~E.} \bibnamefont{Khmel'nitskii}},
  \bibinfo{year}{1980}, \bibinfo{journal}{Zh. Eksp. Teor. Fiz.}
  \textbf{\bibinfo{volume}{79}}, \bibinfo{pages}{1120}, \bibinfo{note}{[Sov.
  Phys. JETP {\bf 52}, 568 (1980)]}.

\bibitem[{\citenamefont{Efetov and Tschersich}(2002)}]{Efetov02}
\bibinfo{author}{\bibnamefont{Efetov}, \bibfnamefont{K.~B.}}, and
  \bibinfo{author}{\bibfnamefont{A.}~\bibnamefont{Tschersich}},
  \bibinfo{year}{2002}, \bibinfo{journal}{Europhys. Lett.}
  \textbf{\bibinfo{volume}{59}}, \bibinfo{pages}{114}.

\bibitem[{\citenamefont{Efetov and Tschersich}(2003)}]{Efetov02b}
\bibinfo{author}{\bibnamefont{Efetov}, \bibfnamefont{K.~B.}}, and
  \bibinfo{author}{\bibfnamefont{A.}~\bibnamefont{Tschersich}},
  \bibinfo{year}{2003}, \bibinfo{journal}{Phys. Rev. B}
  \textbf{\bibinfo{volume}{67}}, \bibinfo{pages}{174205}.

\bibitem[{\citenamefont{Efros and Shklovskii}(1975)}]{Efros}
\bibinfo{author}{\bibnamefont{Efros}, \bibfnamefont{A.~L.}}, and
  \bibinfo{author}{\bibfnamefont{B.~I.} \bibnamefont{Shklovskii}},
  \bibinfo{year}{1975}, \bibinfo{journal}{J.~Phys.~C}
  \textbf{\bibinfo{volume}{8}}, \bibinfo{pages}{L49}.

\bibitem[{\citenamefont{Falci} \emph{et~al.}(1995)\citenamefont{Falci,
  Sch\"{o}n, and Zimanyi}}]{falci}
\bibinfo{author}{\bibnamefont{Falci}, \bibfnamefont{G.}},
  \bibinfo{author}{\bibfnamefont{G.}~\bibnamefont{Sch\"{o}n}}, and
  \bibinfo{author}{\bibfnamefont{G.}~\bibnamefont{Zimanyi}},
  \bibinfo{year}{1995}, \bibinfo{journal}{Phys. Rev. Lett.}
  \textbf{\bibinfo{volume}{74}}, \bibinfo{pages}{3257}.

\bibitem[{\citenamefont{Fazio and van~der Zant}(2001)}]{Fazio01}
\bibinfo{author}{\bibnamefont{Fazio}, \bibfnamefont{R.}}, and
  \bibinfo{author}{\bibfnamefont{H.}~\bibnamefont{van~der Zant}},
  \bibinfo{year}{2001}, \bibinfo{journal}{Phys. Rep.}
  \textbf{\bibinfo{volume}{355}}.

\bibitem[{\citenamefont{Feigel'man and Ioselevich}(2005)}]{Feigelman05}
\bibinfo{author}{\bibnamefont{Feigel'man}, \bibfnamefont{M.~V.}}, and
  \bibinfo{author}{\bibfnamefont{A.~S.} \bibnamefont{Ioselevich}},
  \bibinfo{year}{2005}, \bibinfo{journal}{Pis'ma Zh. Eksp. Teor. Fiz.}
  \textbf{\bibinfo{volume}{81}}, \bibinfo{pages}{341}, \bibinfo{note}{[Sov.
  Phys. JETP Lett. {\bf 81}, 227 (2005)]}.

\bibitem[{\citenamefont{Finkelstein}(1987)}]{Finkelstein87}
\bibinfo{author}{\bibnamefont{Finkelstein}, \bibfnamefont{A.~M.}},
  \bibinfo{year}{1987}, \bibinfo{journal}{Pis'ma Zh. Eksp. Teor. Fiz.}
  \textbf{\bibinfo{volume}{45}}, \bibinfo{pages}{37}, \bibinfo{note}{[Sov.
  Phys. JETP Lett. {\bf 45}, 46 (1987)]}.

\bibitem[{\citenamefont{Finkelstein}(1990)}]{Finkelstain_review}
\bibinfo{author}{\bibnamefont{Finkelstein}, \bibfnamefont{A.~M.}},
  \bibinfo{year}{1990}, \emph{\bibinfo{title}{Electron liquid in Disordered
  Conductors}} (\bibinfo{publisher}{Soviet Scientific Reviews, Harwood,
  London}), volume~\bibinfo{volume}{14}.

\bibitem[{\citenamefont{Finkelstein}(1994)}]{Finkelstein94}
\bibinfo{author}{\bibnamefont{Finkelstein}, \bibfnamefont{A.~M.}},
  \bibinfo{year}{1994}, \bibinfo{journal}{Physica B}
  \textbf{\bibinfo{volume}{197}}, \bibinfo{pages}{636}.

\bibitem[{\citenamefont{Fisher}(1986)}]{Fisher86}
\bibinfo{author}{\bibnamefont{Fisher}, \bibfnamefont{M.~P.~A.}},
  \bibinfo{year}{1986}, \bibinfo{journal}{Phys. Rev. Lett.}
  \textbf{\bibinfo{volume}{57}}, \bibinfo{pages}{885}.

\bibitem[{\citenamefont{Fisher}(1990)}]{Fisher90(2)}
\bibinfo{author}{\bibnamefont{Fisher}, \bibfnamefont{M.~P.~A.}},
  \bibinfo{year}{1990}, \bibinfo{journal}{Phys. Rev. Lett.}
  \textbf{\bibinfo{volume}{65}}, \bibinfo{pages}{923}.

\bibitem[{\citenamefont{Fisher} \emph{et~al.}(1990)\citenamefont{Fisher,
  Grinstein, and Girvin}}]{Fisher90(1)}
\bibinfo{author}{\bibnamefont{Fisher}, \bibfnamefont{M.~P.~A.}},
  \bibinfo{author}{\bibfnamefont{G.}~\bibnamefont{Grinstein}}, and
  \bibinfo{author}{\bibfnamefont{S.~M.} \bibnamefont{Girvin}},
  \bibinfo{year}{1990}, \bibinfo{journal}{Phys. Rev. Lett.}
  \textbf{\bibinfo{volume}{64}}, \bibinfo{pages}{587}.

\bibitem[{\citenamefont{Frydman} \emph{et~al.}(2002)\citenamefont{Frydman,
  Naaman, and Dynes}}]{Frydman02}
\bibinfo{author}{\bibnamefont{Frydman}, \bibfnamefont{A.}},
  \bibinfo{author}{\bibfnamefont{O.}~\bibnamefont{Naaman}}, and
  \bibinfo{author}{\bibfnamefont{R.~C.} \bibnamefont{Dynes}},
  \bibinfo{year}{2002}, \bibinfo{journal}{Phys. Rev. B}
  \textbf{\bibinfo{volume}{66}}, \bibinfo{pages}{052509}.

\bibitem[{\citenamefont{Fujimori} \emph{et~al.}(1994)\citenamefont{Fujimori,
  Mitani, Ohnuma, Ikeda, Shima, and Matsumoto}}]{fujimori}
\bibinfo{author}{\bibnamefont{Fujimori}, \bibfnamefont{H.}},
  \bibinfo{author}{\bibfnamefont{S.}~\bibnamefont{Mitani}},
  \bibinfo{author}{\bibfnamefont{S.}~\bibnamefont{Ohnuma}},
  \bibinfo{author}{\bibfnamefont{T.}~\bibnamefont{Ikeda}},
  \bibinfo{author}{\bibfnamefont{T.}~\bibnamefont{Shima}}, and
  \bibinfo{author}{\bibfnamefont{T.}~\bibnamefont{Matsumoto}},
  \bibinfo{year}{1994}, \bibinfo{journal}{Mater. Sci. Eng. A}
  \textbf{\bibinfo{volume}{181}}, \bibinfo{pages}{897}.

\bibitem[{\citenamefont{Gantmakher}
  \emph{et~al.}(1996)\citenamefont{Gantmakher, Golubkov, Lok, and
  Geim}}]{Gantmakher96}
\bibinfo{author}{\bibnamefont{Gantmakher}, \bibfnamefont{V.~F.}},
  \bibinfo{author}{\bibfnamefont{M.}~\bibnamefont{Golubkov}},
  \bibinfo{author}{\bibfnamefont{J.~G.~S.} \bibnamefont{Lok}}, and
  \bibinfo{author}{\bibfnamefont{A.}~\bibnamefont{Geim}}, \bibinfo{year}{1996},
  \bibinfo{journal}{Zh. Eksp. Teor. Fiz.} \textbf{\bibinfo{volume}{82}},
  \bibinfo{pages}{958}, \bibinfo{note}{[JETP {\bf 82}, 951 (1996)]}.

\bibitem[{\citenamefont{Gaponenko}(1998)}]{Gaponenko98}
\bibinfo{author}{\bibnamefont{Gaponenko}, \bibfnamefont{S.}},
  \bibinfo{year}{1998}, \emph{\bibinfo{title}{Optical Properties of
  Semiconductor Nanocrystals}} (\bibinfo{publisher}{Cambridge, University
  Press}).

\bibitem[{\citenamefont{Gerber}(1990)}]{gerber90}
\bibinfo{author}{\bibnamefont{Gerber}, \bibfnamefont{A.}},
  \bibinfo{year}{1990}, \bibinfo{journal}{J. Phys.: Cond. Matter}
  \textbf{\bibinfo{volume}{2}}, \bibinfo{pages}{8161}.

\bibitem[{\citenamefont{Gerber} \emph{et~al.}(1997)\citenamefont{Gerber,
  Milner, Deutscher, Karpovsky, and Gladkikh}}]{Gerber97}
\bibinfo{author}{\bibnamefont{Gerber}, \bibfnamefont{A.}},
  \bibinfo{author}{\bibfnamefont{A.}~\bibnamefont{Milner}},
  \bibinfo{author}{\bibfnamefont{G.}~\bibnamefont{Deutscher}},
  \bibinfo{author}{\bibfnamefont{M.}~\bibnamefont{Karpovsky}}, and
  \bibinfo{author}{\bibfnamefont{A.}~\bibnamefont{Gladkikh}},
  \bibinfo{year}{1997}, \bibinfo{journal}{Phys. Rev. Lett.}
  \textbf{\bibinfo{volume}{78}}, \bibinfo{pages}{4277}.

\bibitem[{\citenamefont{Goldman and Markovi\'c}(1998)}]{Goldman98}
\bibinfo{author}{\bibnamefont{Goldman}, \bibfnamefont{A.~M.}}, and
  \bibinfo{author}{\bibfnamefont{N.}~\bibnamefont{Markovi\'c}},
  \bibinfo{year}{1998}, \bibinfo{journal}{Physics Today}
  \textbf{\bibinfo{volume}{51}}, \bibinfo{pages}{39}.

\bibitem[{\citenamefont{Gordon}(2000)}]{gordon}
\bibinfo{author}{\bibnamefont{Gordon}, \bibfnamefont{R.}},
  \bibinfo{year}{2000}, \bibinfo{journal}{MRS Bull.}
  \textbf{\bibinfo{volume}{25}}, \bibinfo{pages}{52}.

\bibitem[{\citenamefont{Gorkov} \emph{et~al.}(1979)\citenamefont{Gorkov,
  Larkin, and Khmel'nitskii}}]{Khmelnitskii79}
\bibinfo{author}{\bibnamefont{Gorkov}, \bibfnamefont{L.~P.}},
  \bibinfo{author}{\bibfnamefont{A.~L.} \bibnamefont{Larkin}}, and
  \bibinfo{author}{\bibfnamefont{D.~E.} \bibnamefont{Khmel'nitskii}},
  \bibinfo{year}{1979}, \bibinfo{journal}{Pis'ma Zh. Eksp. Teor. Fiz.}
  \textbf{\bibinfo{volume}{30}}, \bibinfo{pages}{248}, \bibinfo{note}{[Sov.
  Phys. JETP Lett. {\bf 30}, 228 (1979)]}.

\bibitem[{\citenamefont{Guinea and Sch\"{o}n}(1986)}]{guinea}
\bibinfo{author}{\bibnamefont{Guinea}, \bibfnamefont{F.}}, and
  \bibinfo{author}{\bibfnamefont{G.}~\bibnamefont{Sch\"{o}n}},
  \bibinfo{year}{1986}, \bibinfo{journal}{Europhys. Lett.}
  \textbf{\bibinfo{volume}{1}}, \bibinfo{pages}{585}.

\bibitem[{\citenamefont{Hadacek} \emph{et~al.}(2004)\citenamefont{Hadacek,
  Sankquer, and Vill\'egier}}]{Hadacek04}
\bibinfo{author}{\bibnamefont{Hadacek}, \bibfnamefont{N.}},
  \bibinfo{author}{\bibfnamefont{M.}~\bibnamefont{Sankquer}}, and
  \bibinfo{author}{\bibfnamefont{J.~C.} \bibnamefont{Vill\'egier}},
  \bibinfo{year}{2004}, \bibinfo{journal}{Phys. Rev. B}
  \textbf{\bibinfo{volume}{69}}, \bibinfo{pages}{024505}.

\bibitem[{\citenamefont{Halperin}(1986)}]{halperin}
\bibinfo{author}{\bibnamefont{Halperin}, \bibfnamefont{W.~P.}},
  \bibinfo{year}{1986}, \bibinfo{journal}{Rev. Mod. Phys.}
  \textbf{\bibinfo{volume}{58}}, \bibinfo{pages}{533}.

\bibitem[{\citenamefont{Ishida and Ikeda}(1998)}]{Ishida98}
\bibinfo{author}{\bibnamefont{Ishida}, \bibfnamefont{H.}}, and
  \bibinfo{author}{\bibfnamefont{R.}~\bibnamefont{Ikeda}},
  \bibinfo{year}{1998}, \bibinfo{journal}{J.~Phys.~Soc.~Jpn.}
  \textbf{\bibinfo{volume}{67}}, \bibinfo{pages}{983}.

\bibitem[{\citenamefont{Jaeger} \emph{et~al.}(1989)\citenamefont{Jaeger,
  Haviland, Orr, and Goldman}}]{Jaeger89}
\bibinfo{author}{\bibnamefont{Jaeger}, \bibfnamefont{H.}},
  \bibinfo{author}{\bibfnamefont{D.}~\bibnamefont{Haviland}},
  \bibinfo{author}{\bibfnamefont{B.}~\bibnamefont{Orr}}, and
  \bibinfo{author}{\bibfnamefont{A.}~\bibnamefont{Goldman}},
  \bibinfo{year}{1989}, \bibinfo{journal}{Phys Rev B}
  \textbf{\bibinfo{volume}{40}}, \bibinfo{pages}{182}.

\bibitem[{\citenamefont{Jaeger} \emph{et~al.}(1986)\citenamefont{Jaeger,
  Haviland, Goldman, and Orr}}]{Jaeger86}
\bibinfo{author}{\bibnamefont{Jaeger}, \bibfnamefont{H.~M.}},
  \bibinfo{author}{\bibfnamefont{D.~B.} \bibnamefont{Haviland}},
  \bibinfo{author}{\bibfnamefont{A.~M.} \bibnamefont{Goldman}}, and
  \bibinfo{author}{\bibfnamefont{B.~G.} \bibnamefont{Orr}},
  \bibinfo{year}{1986}, \bibinfo{journal}{Phys. Rev. B}
  \textbf{\bibinfo{volume}{34}}, \bibinfo{pages}{4920}.

\bibitem[{\citenamefont{Jha and Middleton}(2005)}]{Jha05}
\bibinfo{author}{\bibnamefont{Jha}, \bibfnamefont{S.}}, and
  \bibinfo{author}{\bibfnamefont{A.~A.} \bibnamefont{Middleton}},
  \bibinfo{year}{2005}, \bibinfo{journal}{cond-mat/0511094} .

\bibitem[{\citenamefont{Kamenev and Andreev}(1999)}]{Kamenev99}
\bibinfo{author}{\bibnamefont{Kamenev}, \bibfnamefont{A.}}, and
  \bibinfo{author}{\bibfnamefont{A.}~\bibnamefont{Andreev}},
  \bibinfo{year}{1999}, \bibinfo{journal}{Phys. Rev. B}
  \textbf{\bibinfo{volume}{60}}, \bibinfo{pages}{2218}.

\bibitem[{\citenamefont{Kee} \emph{et~al.}(1998)\citenamefont{Kee, Aleiner, and
  Altshuler}}]{Kee98}
\bibinfo{author}{\bibnamefont{Kee}, \bibfnamefont{H.~Y.}},
  \bibinfo{author}{\bibfnamefont{I.~L.} \bibnamefont{Aleiner}}, and
  \bibinfo{author}{\bibfnamefont{B.~L.} \bibnamefont{Altshuler}},
  \bibinfo{year}{1998}, \bibinfo{journal}{Phys Rev B}
  \textbf{\bibinfo{volume}{58}}, \bibinfo{pages}{5757}.

\bibitem[{\citenamefont{Kim} \emph{et~al.}(1998)\citenamefont{Kim,
  Granstr\"{o}m, Friend, Johansson, Salaneck, Daik, Feast, and Cacialli}}]{kim}
\bibinfo{author}{\bibnamefont{Kim}, \bibfnamefont{J.~S.}},
  \bibinfo{author}{\bibfnamefont{M.}~\bibnamefont{Granstr\"{o}m}},
  \bibinfo{author}{\bibfnamefont{R.}~\bibnamefont{Friend}},
  \bibinfo{author}{\bibfnamefont{N.}~\bibnamefont{Johansson}},
  \bibinfo{author}{\bibfnamefont{W.}~\bibnamefont{Salaneck}},
  \bibinfo{author}{\bibfnamefont{R.}~\bibnamefont{Daik}},
  \bibinfo{author}{\bibfnamefont{W.}~\bibnamefont{Feast}}, and
  \bibinfo{author}{\bibfnamefont{F.}~\bibnamefont{Cacialli}},
  \bibinfo{year}{1998}, \bibinfo{journal}{J. Appl. Phys.}
  \textbf{\bibinfo{volume}{84}}, \bibinfo{pages}{6859}.

\bibitem[{\citenamefont{Kopnin}(2001)}]{Kopninbook}
\bibinfo{author}{\bibnamefont{Kopnin}, \bibfnamefont{N.~B.}},
  \bibinfo{year}{2001}, \emph{\bibinfo{title}{Theory of Nonequilibrium
  Superconductivity}} (\bibinfo{publisher}{Oxford University Press, Oxford}).

\bibitem[{\citenamefont{Kosterlitz}(1976)}]{kosterlitz}
\bibinfo{author}{\bibnamefont{Kosterlitz}, \bibfnamefont{J.}},
  \bibinfo{year}{1976}, \bibinfo{journal}{Phys. Rev. Lett.}
  \textbf{\bibinfo{volume}{37}}, \bibinfo{pages}{1577}.

\bibitem[{\citenamefont{Kravchenko and Sarachik}(2004)}]{Kravchenko04}
\bibinfo{author}{\bibnamefont{Kravchenko}, \bibfnamefont{S.~V.}}, and
  \bibinfo{author}{\bibfnamefont{M.~P.} \bibnamefont{Sarachik}},
  \bibinfo{year}{2004}, \bibinfo{journal}{Rep. Prog. Phys.}
  \textbf{\bibinfo{volume}{67}}, \bibinfo{pages}{1}.

\bibitem[{\citenamefont{Lang} \emph{et~al.}(2002)\citenamefont{Lang, Lang,
  Madhavan, J.~E.~Hoffman, Eisaki, Uchida, and Davis}}]{Lang}
\bibinfo{author}{\bibnamefont{Lang}, \bibfnamefont{K.~M.}},
  \bibinfo{author}{\bibfnamefont{V.~K.~M.} \bibnamefont{Lang}},
  \bibinfo{author}{\bibfnamefont{V.}~\bibnamefont{Madhavan}},
  \bibinfo{author}{\bibfnamefont{E.~W.~H.} \bibnamefont{J.~E.~Hoffman}},
  \bibinfo{author}{\bibfnamefont{H.}~\bibnamefont{Eisaki}},
  \bibinfo{author}{\bibfnamefont{S.}~\bibnamefont{Uchida}}, and
  \bibinfo{author}{\bibfnamefont{J.~C.} \bibnamefont{Davis}},
  \bibinfo{year}{2002}, \bibinfo{journal}{Nature}
  \textbf{\bibinfo{volume}{415}}, \bibinfo{pages}{412}.

\bibitem[{\citenamefont{Larkin}(1965)}]{Larkin65}
\bibinfo{author}{\bibnamefont{Larkin}, \bibfnamefont{A.~I.}},
  \bibinfo{year}{1965}, \bibinfo{journal}{Zh. Eksp. Teor. Fiz.}
  \textbf{\bibinfo{volume}{48}}, \bibinfo{pages}{232}, \bibinfo{note}{[Sov.
  Phys. JETP {\bf 21}, 153 (1965)]}.

\bibitem[{\citenamefont{Larkin}(1999)}]{Larkin99}
\bibinfo{author}{\bibnamefont{Larkin}, \bibfnamefont{A.~I.}},
  \bibinfo{year}{1999}, \bibinfo{journal}{Ann. Phys.}
  \textbf{\bibinfo{volume}{8}}, \bibinfo{pages}{785}.

\bibitem[{\citenamefont{Larkin and Ovchinnikov}(1983)}]{Larkin83}
\bibinfo{author}{\bibnamefont{Larkin}, \bibfnamefont{A.~I.}}, and
  \bibinfo{author}{\bibfnamefont{Y.~N.} \bibnamefont{Ovchinnikov}},
  \bibinfo{year}{1983}, \bibinfo{journal}{Phys. Rev. B}
  \textbf{\bibinfo{volume}{28}}, \bibinfo{pages}{6281}.

\bibitem[{\citenamefont{Larkin and Varlamov}(2005)}]{Larkinbook}
\bibinfo{author}{\bibnamefont{Larkin}, \bibfnamefont{A.~I.}}, and
  \bibinfo{author}{\bibfnamefont{A.}~\bibnamefont{Varlamov}},
  \bibinfo{year}{2005}, \emph{\bibinfo{title}{Theory of fluctuations in
  superconductors}} (\bibinfo{publisher}{Oxford University Press, New York}).

\bibitem[{\citenamefont{Lee and Ramakrishnan}(1985)}]{Lee_review}
\bibinfo{author}{\bibnamefont{Lee}, \bibfnamefont{P.~A.}}, and
  \bibinfo{author}{\bibfnamefont{T.~V.} \bibnamefont{Ramakrishnan}},
  \bibinfo{year}{1985}, \bibinfo{journal}{Rev. Mod. Phys.}
  \textbf{\bibinfo{volume}{57}}, \bibinfo{pages}{287}.

\bibitem[{\citenamefont{Liao} \emph{et~al.}(2005)\citenamefont{Liao, Xun, and
  Yu}}]{Liao05}
\bibinfo{author}{\bibnamefont{Liao}, \bibfnamefont{Z.~M.}},
  \bibinfo{author}{\bibfnamefont{J.}~\bibnamefont{Xun}}, and
  \bibinfo{author}{\bibfnamefont{D.~P.} \bibnamefont{Yu}},
  \bibinfo{year}{2005}, \bibinfo{journal}{Phys. Lett. A}
  \textbf{\bibinfo{volume}{345}}, \bibinfo{pages}{386}.

\bibitem[{\citenamefont{Lin} \emph{et~al.}(2001)\citenamefont{Lin, Jaeger,
  Sorensen, and Klabunde}}]{Lin2001}
\bibinfo{author}{\bibnamefont{Lin}, \bibfnamefont{X.}},
  \bibinfo{author}{\bibfnamefont{H.~M.} \bibnamefont{Jaeger}},
  \bibinfo{author}{\bibfnamefont{C.}~\bibnamefont{Sorensen}}, and
  \bibinfo{author}{\bibfnamefont{K.}~\bibnamefont{Klabunde}},
  \bibinfo{year}{2001}, \bibinfo{journal}{J. Phys. Chem. B}
  \textbf{\bibinfo{volume}{105}}, \bibinfo{pages}{3353}.

\bibitem[{\citenamefont{Liu} \emph{et~al.}(1993)\citenamefont{Liu, Haviland,
  Nease, and Goldman}}]{Liu93}
\bibinfo{author}{\bibnamefont{Liu}, \bibfnamefont{Y.}},
  \bibinfo{author}{\bibfnamefont{D.~B.} \bibnamefont{Haviland}},
  \bibinfo{author}{\bibfnamefont{B.}~\bibnamefont{Nease}}, and
  \bibinfo{author}{\bibfnamefont{A.}~\bibnamefont{Goldman}},
  \bibinfo{year}{1993}, \bibinfo{journal}{Phys. Rev B}
  \textbf{\bibinfo{volume}{47}}, \bibinfo{pages}{5931}.

\bibitem[{\citenamefont{Loh and Tripathi}(2006)}]{Loh05}
\bibinfo{author}{\bibnamefont{Loh}, \bibfnamefont{Y.~L.}}, and
  \bibinfo{author}{\bibfnamefont{V.}~\bibnamefont{Tripathi}},
  \bibinfo{year}{2006}, \bibinfo{journal}{Phys. Rev. Lett.}
  \textbf{\bibinfo{volume}{96}}, \bibinfo{pages}{046805}.

\bibitem[{\citenamefont{Loh} \emph{et~al.}(2005)\citenamefont{Loh, Tripathi,
  and Turlakov}}]{Loh}
\bibinfo{author}{\bibnamefont{Loh}, \bibfnamefont{Y.~L.}},
  \bibinfo{author}{\bibfnamefont{V.}~\bibnamefont{Tripathi}}, and
  \bibinfo{author}{\bibfnamefont{M.}~\bibnamefont{Turlakov}},
  \bibinfo{year}{2005}, \bibinfo{journal}{Phys. Rev. B}
  \textbf{\bibinfo{volume}{72}}, \bibinfo{pages}{233404}.

\bibitem[{\citenamefont{Lukyanov and Zamolodchikov}(2004)}]{zamolodchikov}
\bibinfo{author}{\bibnamefont{Lukyanov}, \bibfnamefont{S.}}, and
  \bibinfo{author}{\bibfnamefont{A.}~\bibnamefont{Zamolodchikov}},
  \bibinfo{year}{2004}, \bibinfo{journal}{Journal of Stat. Mechanics} ,
  \bibinfo{pages}{P05003}.

\bibitem[{\citenamefont{Maekawa} \emph{et~al.}(1983)\citenamefont{Maekawa,
  Ebisawa, and Fukuyama}}]{Maekawa83}
\bibinfo{author}{\bibnamefont{Maekawa}, \bibfnamefont{S.}},
  \bibinfo{author}{\bibfnamefont{H.}~\bibnamefont{Ebisawa}}, and
  \bibinfo{author}{\bibfnamefont{H.}~\bibnamefont{Fukuyama}},
  \bibinfo{year}{1983}, \bibinfo{journal}{J. Phys. Soc. Jpn.}
  \textbf{\bibinfo{volume}{52}}, \bibinfo{pages}{1352}.

\bibitem[{\citenamefont{Maekawa and Fukuyama}(1982)}]{Fukuyama81}
\bibinfo{author}{\bibnamefont{Maekawa}, \bibfnamefont{S.}}, and
  \bibinfo{author}{\bibfnamefont{H.}~\bibnamefont{Fukuyama}},
  \bibinfo{year}{1982}, \bibinfo{journal}{J. Phys. Soc. Jpn.}
  \textbf{\bibinfo{volume}{51}}, \bibinfo{pages}{1380}.

\bibitem[{\citenamefont{Maki}(1968{\natexlab{a}})}]{Maki68}
\bibinfo{author}{\bibnamefont{Maki}, \bibfnamefont{K.}},
  \bibinfo{year}{1968}{\natexlab{a}}, \bibinfo{journal}{Prog. Theor. Phys.}
  \textbf{\bibinfo{volume}{39}}, \bibinfo{pages}{897}.

\bibitem[{\citenamefont{Maki}(1968{\natexlab{b}})}]{Maki68b}
\bibinfo{author}{\bibnamefont{Maki}, \bibfnamefont{K.}},
  \bibinfo{year}{1968}{\natexlab{b}}, \bibinfo{journal}{Prog. Theor. Phys.}
  \textbf{\bibinfo{volume}{40}}, \bibinfo{pages}{193}.

\bibitem[{\citenamefont{Markovi\'{c}}
  \emph{et~al.}(1999)\citenamefont{Markovi\'{c}, Christiansen, Mack, Huber, and
  Goldman}}]{Markovic99}
\bibinfo{author}{\bibnamefont{Markovi\'{c}}, \bibfnamefont{N.}},
  \bibinfo{author}{\bibfnamefont{C.}~\bibnamefont{Christiansen}},
  \bibinfo{author}{\bibfnamefont{A.~M.} \bibnamefont{Mack}},
  \bibinfo{author}{\bibfnamefont{W.~H.} \bibnamefont{Huber}}, and
  \bibinfo{author}{\bibfnamefont{A.~M.} \bibnamefont{Goldman}},
  \bibinfo{year}{1999}, \bibinfo{journal}{Phys. Rev. B}
  \textbf{\bibinfo{volume}{60}}, \bibinfo{pages}{4320}.

\bibitem[{\citenamefont{Matveev and Larkin}(1997)}]{Matveev97}
\bibinfo{author}{\bibnamefont{Matveev}, \bibfnamefont{K.~A.}}, and
  \bibinfo{author}{\bibfnamefont{A.~I.} \bibnamefont{Larkin}},
  \bibinfo{year}{1997}, \bibinfo{journal}{Phys. Rev. Lett.}
  \textbf{\bibinfo{volume}{78}}, \bibinfo{pages}{3749}.

\bibitem[{\citenamefont{McLean and Stephen}(1979)}]{McLean79}
\bibinfo{author}{\bibnamefont{McLean}, \bibfnamefont{W.}}, and
  \bibinfo{author}{\bibfnamefont{M.}~\bibnamefont{Stephen}},
  \bibinfo{year}{1979}, \bibinfo{journal}{Phys. Rev. B}
  \textbf{\bibinfo{volume}{19}}.

\bibitem[{\citenamefont{Mehta}(1991)}]{mehta}
\bibinfo{author}{\bibnamefont{Mehta}, \bibfnamefont{M.}}, \bibinfo{year}{1991},
  \emph{\bibinfo{title}{Random Matrices}} (\bibinfo{publisher}{Academic Press,
  San Diego}).

\bibitem[{\citenamefont{Meyer} \emph{et~al.}(2004)\citenamefont{Meyer, Kamenev,
  and Glazman}}]{Jul'ka}
\bibinfo{author}{\bibnamefont{Meyer}, \bibfnamefont{J.~S.}},
  \bibinfo{author}{\bibfnamefont{A.}~\bibnamefont{Kamenev}}, and
  \bibinfo{author}{\bibfnamefont{L.~I.} \bibnamefont{Glazman}},
  \bibinfo{year}{2004}, \bibinfo{journal}{Phys. Rev. B}
  \textbf{\bibinfo{volume}{70}}, \bibinfo{pages}{045310}.

\bibitem[{\citenamefont{Middleton and Wingreen}(1993)}]{Middleton93}
\bibinfo{author}{\bibnamefont{Middleton}, \bibfnamefont{A.~A.}}, and
  \bibinfo{author}{\bibfnamefont{N.~S.} \bibnamefont{Wingreen}},
  \bibinfo{year}{1993}, \bibinfo{journal}{Phys. Rev. Lett}
  \textbf{\bibinfo{volume}{71}}, \bibinfo{pages}{3198}.

\bibitem[{\citenamefont{Mowbray and Skolnick}(2005)}]{Mow05}
\bibinfo{author}{\bibnamefont{Mowbray}, \bibfnamefont{D.~J.}}, and
  \bibinfo{author}{\bibfnamefont{M.~S.} \bibnamefont{Skolnick}},
  \bibinfo{year}{2005}, \bibinfo{journal}{J. Phys. D-Applied Physics}
  \textbf{\bibinfo{volume}{38}}, \bibinfo{pages}{2059}.

\bibitem[{\citenamefont{Muller and Ioffe}(2004)}]{Ioffe}
\bibinfo{author}{\bibnamefont{Muller}, \bibfnamefont{M.}}, and
  \bibinfo{author}{\bibfnamefont{L.~B.} \bibnamefont{Ioffe}},
  \bibinfo{year}{2004}, \bibinfo{journal}{Phys. Rev. Lett.}
  \textbf{\bibinfo{volume}{93}}, \bibinfo{pages}{256403}.

\bibitem[{\citenamefont{Murray} \emph{et~al.}(2000)\citenamefont{Murray, Kagan,
  and Bawendi}}]{Murray2000}
\bibinfo{author}{\bibnamefont{Murray}, \bibfnamefont{C.}},
  \bibinfo{author}{\bibfnamefont{C.}~\bibnamefont{Kagan}}, and
  \bibinfo{author}{\bibfnamefont{M.}~\bibnamefont{Bawendi}},
  \bibinfo{year}{2000}, \bibinfo{journal}{Annual review of Materials Science}
  \textbf{\bibinfo{volume}{30}}, \bibinfo{pages}{545}.

\bibitem[{\citenamefont{Murray} \emph{et~al.}(1993)\citenamefont{Murray,
  Norris, and Bawendi}}]{Murray93}
\bibinfo{author}{\bibnamefont{Murray}, \bibfnamefont{C.~B.}},
  \bibinfo{author}{\bibfnamefont{D.~J.} \bibnamefont{Norris}}, and
  \bibinfo{author}{\bibfnamefont{M.~G.} \bibnamefont{Bawendi}},
  \bibinfo{year}{1993}, \bibinfo{journal}{J. Am. Chem. Soc.}
  \textbf{\bibinfo{volume}{115}}, \bibinfo{pages}{8706}.

\bibitem[{\citenamefont{Nagaev}(1992)}]{nagaev}
\bibinfo{author}{\bibnamefont{Nagaev}, \bibfnamefont{E.}},
  \bibinfo{year}{1992}, \bibinfo{journal}{Physics Reports,}
  \textbf{\bibinfo{volume}{222}}, \bibinfo{pages}{199}.

\bibitem[{\citenamefont{Narayanan} \emph{et~al.}(2004)\citenamefont{Narayanan,
  Wang, and Lin}}]{Narayanan2004}
\bibinfo{author}{\bibnamefont{Narayanan}, \bibfnamefont{S.}},
  \bibinfo{author}{\bibfnamefont{J.}~\bibnamefont{Wang}}, and
  \bibinfo{author}{\bibfnamefont{X.}~\bibnamefont{Lin}}, \bibinfo{year}{2004},
  \bibinfo{journal}{Phys. Rev. Lett.} \textbf{\bibinfo{volume}{93}},
  \bibinfo{pages}{135503}.

\bibitem[{\citenamefont{Orr} \emph{et~al.}(1985)\citenamefont{Orr, Jaeger, and
  Goldman}}]{Orr_PRB85}
\bibinfo{author}{\bibnamefont{Orr}, \bibfnamefont{B.~G.}},
  \bibinfo{author}{\bibfnamefont{H.~M.} \bibnamefont{Jaeger}}, and
  \bibinfo{author}{\bibfnamefont{A.~M.} \bibnamefont{Goldman}},
  \bibinfo{year}{1985}, \bibinfo{journal}{Phys. Rev. B}
  \textbf{\bibinfo{volume}{32}}, \bibinfo{pages}{7586}.

\bibitem[{\citenamefont{Orr} \emph{et~al.}(1986)\citenamefont{Orr, Jaeger,
  Goldman, and Kuper}}]{Orr86}
\bibinfo{author}{\bibnamefont{Orr}, \bibfnamefont{B.~G.}},
  \bibinfo{author}{\bibfnamefont{H.~M.} \bibnamefont{Jaeger}},
  \bibinfo{author}{\bibfnamefont{A.~M.} \bibnamefont{Goldman}}, and
  \bibinfo{author}{\bibfnamefont{C.~G.} \bibnamefont{Kuper}},
  \bibinfo{year}{1986}, \bibinfo{journal}{Phys. Rev. Lett}
  \textbf{\bibinfo{volume}{56}}, \bibinfo{pages}{378}.

\bibitem[{\citenamefont{Ovchinnikov}(1973)}]{Ovchinnikov73}
\bibinfo{author}{\bibnamefont{Ovchinnikov}, \bibfnamefont{Y.~N.}},
  \bibinfo{year}{1973}, \bibinfo{journal}{Zh. Eksp. Teor. Fiz.}
  \textbf{\bibinfo{volume}{64}}, \bibinfo{pages}{719}, \bibinfo{note}{[Sov.
  Phys. JETP, {\bf 37}, 366 (1973)]}.

\bibitem[{\citenamefont{Ovchinnikov and
  Kresin}(2005{\natexlab{a}})}]{Ovchinnikov05a}
\bibinfo{author}{\bibnamefont{Ovchinnikov}, \bibfnamefont{Y.~N.}}, and
  \bibinfo{author}{\bibfnamefont{V.~Z.} \bibnamefont{Kresin}},
  \bibinfo{year}{2005}{\natexlab{a}}, \bibinfo{journal}{Eur. Phys. J B}
  \textbf{\bibinfo{volume}{45}}, \bibinfo{pages}{5}.

\bibitem[{\citenamefont{Ovchinnikov and
  Kresin}(2005{\natexlab{b}})}]{Ovchinnikov05b}
\bibinfo{author}{\bibnamefont{Ovchinnikov}, \bibfnamefont{Y.~N.}}, and
  \bibinfo{author}{\bibfnamefont{V.~Z.} \bibnamefont{Kresin}},
  \bibinfo{year}{2005}{\natexlab{b}}, \bibinfo{journal}{Eur. Phys. J B}
  \textbf{\bibinfo{volume}{47}}, \bibinfo{pages}{333}.

\bibitem[{\citenamefont{Pankov and Dobrosavlijevic}(2005)}]{Pankov}
\bibinfo{author}{\bibnamefont{Pankov}, \bibfnamefont{S.}}, and
  \bibinfo{author}{\bibfnamefont{V.}~\bibnamefont{Dobrosavlijevic}},
  \bibinfo{year}{2005}, \bibinfo{journal}{Phys. Rev. Lett.}
  \textbf{\bibinfo{volume}{94}}.

\bibitem[{\citenamefont{Parthasarathy}
  \emph{et~al.}(2004)\citenamefont{Parthasarathy, Lin, Elteto, Rosenbaum, and
  Jaeger}}]{Parthasarathy04}
\bibinfo{author}{\bibnamefont{Parthasarathy}, \bibfnamefont{R.}},
  \bibinfo{author}{\bibfnamefont{X.-M.} \bibnamefont{Lin}},
  \bibinfo{author}{\bibfnamefont{K.}~\bibnamefont{Elteto}},
  \bibinfo{author}{\bibfnamefont{T.~F.} \bibnamefont{Rosenbaum}}, and
  \bibinfo{author}{\bibfnamefont{H.~M.} \bibnamefont{Jaeger}},
  \bibinfo{year}{2004}, \bibinfo{journal}{Phys. Rev. Lett.}
  \textbf{\bibinfo{volume}{92}}, \bibinfo{pages}{076801}.

\bibitem[{\citenamefont{Parthasarathy}
  \emph{et~al.}(2001)\citenamefont{Parthasarathy, Lin, and Jaeger}}]{partha01}
\bibinfo{author}{\bibnamefont{Parthasarathy}, \bibfnamefont{R.}},
  \bibinfo{author}{\bibfnamefont{X.-M.} \bibnamefont{Lin}}, and
  \bibinfo{author}{\bibfnamefont{H.}~\bibnamefont{Jaeger}},
  \bibinfo{year}{2001}, \bibinfo{journal}{Phys. Rev. Lett.}
  \textbf{\bibinfo{volume}{87}}, \bibinfo{pages}{186807}.

\bibitem[{\citenamefont{Pollak and Adkins}(1992)}]{pollak}
\bibinfo{author}{\bibnamefont{Pollak}, \bibfnamefont{M.}}, and
  \bibinfo{author}{\bibfnamefont{C.~J.} \bibnamefont{Adkins}},
  \bibinfo{year}{1992}, \bibinfo{journal}{Phil. Mag. B}
  \textbf{\bibinfo{volume}{65}}, \bibinfo{pages}{855}.

\bibitem[{\citenamefont{Punnoose and Finkelstein}(2005)}]{Finkelsteinscience}
\bibinfo{author}{\bibnamefont{Punnoose}, \bibfnamefont{A.}}, and
  \bibinfo{author}{\bibfnamefont{A.~M.} \bibnamefont{Finkelstein}},
  \bibinfo{year}{2005}, \bibinfo{journal}{Science}
  \textbf{\bibinfo{volume}{310}}, \bibinfo{pages}{289}.

\bibitem[{\citenamefont{Radhakrishnan}
  \emph{et~al.}(1990)\citenamefont{Radhakrishnan, Subramaniam,
  Sankaranaranyanan, Rao, and Srinivasan}}]{radha}
\bibinfo{author}{\bibnamefont{Radhakrishnan}, \bibfnamefont{V.}},
  \bibinfo{author}{\bibfnamefont{C.~K.} \bibnamefont{Subramaniam}},
  \bibinfo{author}{\bibfnamefont{S.}~\bibnamefont{Sankaranaranyanan}},
  \bibinfo{author}{\bibfnamefont{G.~V.~S.} \bibnamefont{Rao}}, and
  \bibinfo{author}{\bibfnamefont{R.}~\bibnamefont{Srinivasan}},
  \bibinfo{year}{1990}, \bibinfo{journal}{Physica C}
  \textbf{\bibinfo{volume}{167}}, \bibinfo{pages}{53}.

\bibitem[{\citenamefont{Ralph} \emph{et~al.}(1995)\citenamefont{Ralph, Black,
  and Tinkham}}]{Ralph95}
\bibinfo{author}{\bibnamefont{Ralph}, \bibfnamefont{D.}},
  \bibinfo{author}{\bibfnamefont{C.}~\bibnamefont{Black}}, and
  \bibinfo{author}{\bibfnamefont{M.}~\bibnamefont{Tinkham}},
  \bibinfo{year}{1995}, \bibinfo{journal}{Phys. Rev. Lett.}
  \textbf{\bibinfo{volume}{74}}, \bibinfo{pages}{3241}.

\bibitem[{\citenamefont{Romero and Drndic}(2005)}]{Romero05}
\bibinfo{author}{\bibnamefont{Romero}, \bibfnamefont{H.}}, and
  \bibinfo{author}{\bibfnamefont{M.}~\bibnamefont{Drndic}},
  \bibinfo{year}{2005}, \bibinfo{journal}{Phys. Rev. Lett}
  \textbf{\bibinfo{volume}{95}}, \bibinfo{pages}{156801}.

\bibitem[{\citenamefont{Rotkina} \emph{et~al.}(2005)\citenamefont{Rotkina, Oh,
  Eckstein, and Rotkin}}]{rotkina}
\bibinfo{author}{\bibnamefont{Rotkina}, \bibfnamefont{L.}},
  \bibinfo{author}{\bibfnamefont{S.}~\bibnamefont{Oh}},
  \bibinfo{author}{\bibfnamefont{J.}~\bibnamefont{Eckstein}}, and
  \bibinfo{author}{\bibfnamefont{S.}~\bibnamefont{Rotkin}},
  \bibinfo{year}{2005}, \bibinfo{journal}{Phys. Rev. B}
  \textbf{\bibinfo{volume}{72}}, \bibinfo{pages}{233407}.

\bibitem[{\citenamefont{Sachdev}(2001)}]{Sachdevbook}
\bibinfo{author}{\bibnamefont{Sachdev}, \bibfnamefont{S.}},
  \bibinfo{year}{2001}, \emph{\bibinfo{title}{Quantum phase transitions}}
  (\bibinfo{publisher}{Cambridge university press}).

\bibitem[{\citenamefont{Schmid}(1983)}]{Schmid83}
\bibinfo{author}{\bibnamefont{Schmid}, \bibfnamefont{A.}},
  \bibinfo{year}{1983}, \bibinfo{journal}{Phys. Rev. Lett.}
  \textbf{\bibinfo{volume}{51}}, \bibinfo{pages}{1506}.

\bibitem[{\citenamefont{Sch\"{o}n and Zaikin}(1990)}]{Zaikin}
\bibinfo{author}{\bibnamefont{Sch\"{o}n}, \bibfnamefont{G.}}, and
  \bibinfo{author}{\bibfnamefont{A.}~\bibnamefont{Zaikin}},
  \bibinfo{year}{1990}, \bibinfo{journal}{Phys. Rep.}
  \textbf{\bibinfo{volume}{198}}, \bibinfo{pages}{237}.

\bibitem[{\citenamefont{Shapira and Deutscher}(1983)}]{Shapira83}
\bibinfo{author}{\bibnamefont{Shapira}, \bibfnamefont{Y.}}, and
  \bibinfo{author}{\bibfnamefont{G.}~\bibnamefont{Deutscher}},
  \bibinfo{year}{1983}, \bibinfo{journal}{Phys. Rev. B}
  \textbf{\bibinfo{volume}{27}}, \bibinfo{pages}{4463}.

\bibitem[{\citenamefont{Shklovskii}(1973)}]{Shklovskii73}
\bibinfo{author}{\bibnamefont{Shklovskii}, \bibfnamefont{B.~I.}},
  \bibinfo{year}{1973}, \bibinfo{journal}{Fiz. Tekh. Poluprovodn.
  (S.-Petersburg)} \textbf{\bibinfo{volume}{6}}, \bibinfo{pages}{2335}.

\bibitem[{\citenamefont{Shklovskii and Efros}(1988)}]{Shklovskii}
\bibinfo{author}{\bibnamefont{Shklovskii}, \bibfnamefont{B.~I.}}, and
  \bibinfo{author}{\bibfnamefont{A.~L.} \bibnamefont{Efros}},
  \bibinfo{year}{1988}, \emph{\bibinfo{title}{Electronic properties of Doped
  Semiconductors}} (\bibinfo{publisher}{Springer-Verlag, New York}).

\bibitem[{\citenamefont{Simanek}(1979)}]{Simanek79}
\bibinfo{author}{\bibnamefont{Simanek}, \bibfnamefont{E.}},
  \bibinfo{year}{1979}, \bibinfo{journal}{Sol. State. Comm.}
  \textbf{\bibinfo{volume}{31}}, \bibinfo{pages}{419}.

\bibitem[{\citenamefont{Simanek}(1994)}]{Simanekbook}
\bibinfo{author}{\bibnamefont{Simanek}, \bibfnamefont{E.}},
  \bibinfo{year}{1994}, \emph{\bibinfo{title}{Nonhomogeneous Superconductors}}
  (\bibinfo{publisher}{Oxford University Press, Oxford}).

\bibitem[{\citenamefont{Sim\'{a}nek and Brown}(1986)}]{Simanek86}
\bibinfo{author}{\bibnamefont{Sim\'{a}nek}, \bibfnamefont{E.}}, and
  \bibinfo{author}{\bibfnamefont{R.}~\bibnamefont{Brown}},
  \bibinfo{year}{1986}, \bibinfo{journal}{Phys. Rev. B}
  \textbf{\bibinfo{volume}{34}}, \bibinfo{pages}{3495}.

\bibitem[{\citenamefont{Simon} \emph{et~al.}(1987)\citenamefont{Simon,
  Dalrymple, Vechten, Fuller, and Wolf}}]{Simon}
\bibinfo{author}{\bibnamefont{Simon}, \bibfnamefont{R.~W.}},
  \bibinfo{author}{\bibfnamefont{B.~J.} \bibnamefont{Dalrymple}},
  \bibinfo{author}{\bibfnamefont{D.~V.} \bibnamefont{Vechten}},
  \bibinfo{author}{\bibfnamefont{W.~W.} \bibnamefont{Fuller}}, and
  \bibinfo{author}{\bibfnamefont{S.~A.} \bibnamefont{Wolf}},
  \bibinfo{year}{1987}, \bibinfo{journal}{Phys. Rev. B}
  \textbf{\bibinfo{volume}{36}}, \bibinfo{pages}{1962}.

\bibitem[{\citenamefont{Smith and Ambegaokar}(1996)}]{Smith96}
\bibinfo{author}{\bibnamefont{Smith}, \bibfnamefont{R.~A.}}, and
  \bibinfo{author}{\bibfnamefont{V.}~\bibnamefont{Ambegaokar}},
  \bibinfo{year}{1996}, \bibinfo{journal}{Phys. Rev. Lett.}
  \textbf{\bibinfo{volume}{77}}, \bibinfo{pages}{4962}.

\bibitem[{\citenamefont{Sondhi} \emph{et~al.}(1997)\citenamefont{Sondhi,
  Girvin, Carini, and Shahar}}]{Sondhi97}
\bibinfo{author}{\bibnamefont{Sondhi}, \bibfnamefont{S.}},
  \bibinfo{author}{\bibfnamefont{S.}~\bibnamefont{Girvin}},
  \bibinfo{author}{\bibfnamefont{J.}~\bibnamefont{Carini}}, and
  \bibinfo{author}{\bibfnamefont{D.}~\bibnamefont{Shahar}},
  \bibinfo{year}{1997}, \bibinfo{journal}{Rev. Mod. Phys.}
  \textbf{\bibinfo{volume}{69}}, \bibinfo{pages}{315}.

\bibitem[{\citenamefont{Thompson}(1970)}]{Thompson70}
\bibinfo{author}{\bibnamefont{Thompson}, \bibfnamefont{R.~S.}},
  \bibinfo{year}{1970}, \bibinfo{journal}{Phys. Rev. B}
  \textbf{\bibinfo{volume}{1}}, \bibinfo{pages}{327}.

\bibitem[{\citenamefont{Tinkham}(1996)}]{Tinkham96}
\bibinfo{author}{\bibnamefont{Tinkham}, \bibfnamefont{M.}},
  \bibinfo{year}{1996}, \emph{\bibinfo{title}{Introduction to
  Superconductivity}} (\bibinfo{publisher}{McGraw-Hill, New York}).

\bibitem[{\citenamefont{Tran} \emph{et~al.}(2005)\citenamefont{Tran,
  Beloborodov, Lin, Vinokur, and Jaeger}}]{Tran05}
\bibinfo{author}{\bibnamefont{Tran}, \bibfnamefont{T.}},
  \bibinfo{author}{\bibfnamefont{I.~S.} \bibnamefont{Beloborodov}},
  \bibinfo{author}{\bibfnamefont{X.~M.} \bibnamefont{Lin}},
  \bibinfo{author}{\bibfnamefont{V.~M.} \bibnamefont{Vinokur}}, and
  \bibinfo{author}{\bibfnamefont{H.~M.} \bibnamefont{Jaeger}},
  \bibinfo{year}{2005}, \bibinfo{journal}{Phys. Rev. Lett.}
  \textbf{\bibinfo{volume}{95}}, \bibinfo{pages}{076806}.

\bibitem[{\citenamefont{Usadel}(1970)}]{Usadel70}
\bibinfo{author}{\bibnamefont{Usadel}, \bibfnamefont{K.~D.}},
  \bibinfo{year}{1970}, \bibinfo{journal}{Phys. Rev. Lett.}
  \textbf{\bibinfo{volume}{25}}, \bibinfo{pages}{507}.

\bibitem[{\citenamefont{Wegner}(1979)}]{Wegner79}
\bibinfo{author}{\bibnamefont{Wegner}, \bibfnamefont{F.}},
  \bibinfo{year}{1979}, \bibinfo{journal}{Z. Phys. B}
  \textbf{\bibinfo{volume}{35}}, \bibinfo{pages}{207}.

\bibitem[{\citenamefont{Wehrenberg}
  \emph{et~al.}(2002)\citenamefont{Wehrenberg, Wang, and
  Guyot-Sionnest}}]{Wehrenberg02}
\bibinfo{author}{\bibnamefont{Wehrenberg}, \bibfnamefont{B.~L.}},
  \bibinfo{author}{\bibfnamefont{C.}~\bibnamefont{Wang}}, and
  \bibinfo{author}{\bibfnamefont{P.~J.} \bibnamefont{Guyot-Sionnest}},
  \bibinfo{year}{2002}, \bibinfo{journal}{Phys. Chem. B}
  \textbf{\bibinfo{volume}{106}}, \bibinfo{pages}{10634}.

\bibitem[{\citenamefont{Yakimov} \emph{et~al.}(2003)\citenamefont{Yakimov,
  Dvurechenskii, Nikiforov, and Bloshkin}}]{Yakimov03}
\bibinfo{author}{\bibnamefont{Yakimov}, \bibfnamefont{A.~I.}},
  \bibinfo{author}{\bibfnamefont{A.~V.} \bibnamefont{Dvurechenskii}},
  \bibinfo{author}{\bibfnamefont{A.~I.} \bibnamefont{Nikiforov}}, and
  \bibinfo{author}{\bibfnamefont{A.~A.} \bibnamefont{Bloshkin}},
  \bibinfo{year}{2003}, \textbf{\bibinfo{volume}{77}}, \bibinfo{pages}{445},
  \bibinfo{note}{[Sov. Phys. JETP Lett. {\bf 77}, 376 (2003)]}.

\bibitem[{\citenamefont{Yu} \emph{et~al.}(2004)\citenamefont{Yu, Wang,
  Wehrenberg, and Guyot-Sionnest}}]{Yu04}
\bibinfo{author}{\bibnamefont{Yu}, \bibfnamefont{D.}},
  \bibinfo{author}{\bibfnamefont{C.}~\bibnamefont{Wang}},
  \bibinfo{author}{\bibfnamefont{B.~L.} \bibnamefont{Wehrenberg}}, and
  \bibinfo{author}{\bibfnamefont{P.}~\bibnamefont{Guyot-Sionnest}},
  \bibinfo{year}{2004}, \bibinfo{journal}{Phys. Rev. Lett.}
  \textbf{\bibinfo{volume}{92}}, \bibinfo{pages}{216802}.

\bibitem[{\citenamefont{Yu} \emph{et~al.}(1991)\citenamefont{Yu, Duxbury,
  Jeffers, and Dubson}}]{Yu91}
\bibinfo{author}{\bibnamefont{Yu}, \bibfnamefont{X.}},
  \bibinfo{author}{\bibfnamefont{P.~M.} \bibnamefont{Duxbury}},
  \bibinfo{author}{\bibfnamefont{G.}~\bibnamefont{Jeffers}}, and
  \bibinfo{author}{\bibfnamefont{M.~A.} \bibnamefont{Dubson}},
  \bibinfo{year}{1991}, \bibinfo{journal}{Phys. Rev. B}
  \textbf{\bibinfo{volume}{44}}, \bibinfo{pages}{13163}.

\bibitem[{\citenamefont{van~der Zant} \emph{et~al.}(1992)\citenamefont{van~der
  Zant, Fritschy, Elion, Geerlings, and Mooij}}]{Zant92}
\bibinfo{author}{\bibnamefont{van~der Zant}, \bibfnamefont{H.~S.~J.}},
  \bibinfo{author}{\bibfnamefont{F.~C.} \bibnamefont{Fritschy}},
  \bibinfo{author}{\bibfnamefont{W.~J.} \bibnamefont{Elion}},
  \bibinfo{author}{\bibfnamefont{L.~J.} \bibnamefont{Geerlings}}, and
  \bibinfo{author}{\bibfnamefont{J.~E.} \bibnamefont{Mooij}},
  \bibinfo{year}{1992}, \bibinfo{journal}{Phys. Rev. Lett}
  \textbf{\bibinfo{volume}{69}}, \bibinfo{pages}{2971}.

\bibitem[{\citenamefont{Zhang and Shklovskii}(2004)}]{Shklovskii04}
\bibinfo{author}{\bibnamefont{Zhang}, \bibfnamefont{J.}}, and
  \bibinfo{author}{\bibfnamefont{B.~I.} \bibnamefont{Shklovskii}},
  \bibinfo{year}{2004}, \bibinfo{journal}{Phys. Rev. B}
  \textbf{\bibinfo{volume}{70}}, \bibinfo{pages}{115317}.

\bibitem[{\citenamefont{Zhu} \emph{et~al.}(2000)\citenamefont{Zhu, Zhang,
  Guenter, and Jin}}]{zhu}
\bibinfo{author}{\bibnamefont{Zhu}, \bibfnamefont{F.}},
  \bibinfo{author}{\bibfnamefont{K.}~\bibnamefont{Zhang}},
  \bibinfo{author}{\bibfnamefont{E.}~\bibnamefont{Guenter}}, and
  \bibinfo{author}{\bibfnamefont{C.}~\bibnamefont{Jin}}, \bibinfo{year}{2000},
  \bibinfo{journal}{Thin Solid Films} \textbf{\bibinfo{volume}{363}},
  \bibinfo{pages}{314}.

\end{thebibliography}

\end{document}